\documentstyle[12pt,epsfig,rotating,a4p] {article}

\newcommand {\zf}{$_{Z\!F}\!I\!^{\textstyle T}\!\!T\!\!_{{\textstyle E}\!R}$}

\begin{document}
\newcommand{\eel} {\mbox{$\epsilon_{\mathrm{e}^+ \mathrm{e}^-}$}}
\newcommand{\emu} {\mbox{$\epsilon_{\mu^+ \mu^-}$}}
\newcommand{\nprode}{\mbox{$N^{\mathrm{e}^+ \mathrm{e}^-}_{prod}$}}
\newcommand{\nprodm}{\mbox{$N^{\mu^+ \mu^-}_{prod}$}}
\newcommand{\nexpe}{\mbox{$N_{exp}^{\mathrm{e}^+ \mathrm{e}^-}$}}
\newcommand{\nexpm}{\mbox{$N_{exp}^{\mu^+ \mu^-}$}}
\newcommand{\nexpt}{\mbox{$N_{exp}^{total}$}}
\newcommand{\Z}{\mbox{$\mathrm{Z}^{0}$}}
\newcommand{\W}{\mbox{$\mathrm{W}^{\pm}$}}
\newcommand{\bZo}{{\bf \mbox{$\mathrm{Z}^{0}$}}}
\newcommand{\Zg}{\mbox{$\mathrm{Z}^{0}\gamma$}}
\newcommand{\ZZ}{\mbox{$\mathrm{Z}^{0}\mathrm{Z}^{0}$}}
\newcommand{\WW}{\mbox{$\mathrm{W}^{+}\mathrm{W}^{-}$}}
\newcommand{\Zs}{\mbox{$\mathrm{Z}^{*}$}}
\newcommand{\h}{\mbox{$\mathrm{h}^{0}$}}
\newcommand{\Ho}{\mbox{$\mathrm{H}^{0}$}}
\newcommand{\A}{\mbox{$\mathrm{A}^{0}$}}
\newcommand{\Hpm}{\mbox{$\mathrm{H}^{\pm}$}}
\newcommand{\Hsm}{\mbox{$\mathrm{H}^{0}_{SM}$}}
\newcommand{\mHsm}{\mbox{$m_{\mathrm{H}^0_{SM}}$}}
\newcommand{\sctop}{\mbox{$\tilde{\mathrm{t}}$}}
\newcommand{\mstop}{\mbox{$m_{\tilde{\mathrm{t}}}$}}
\newcommand{\mstopL}{\mbox{$m_{\tilde{\mathrm{t}}_{\mathrm{L}}}$}}
\newcommand{\mstopR}{\mbox{$m_{\tilde{\mathrm{t}}_{\mathrm{R}}}$}}
\newcommand{\X}{\mbox{${\tilde{\chi}^0}$}}
\newcommand{\ko}{\mbox{${\tilde{\chi}^0}$}}
\newcommand{\kol}{\mbox{${\tilde{\chi}_1^0}$}}
\newcommand{\bho}{\mbox{$\boldmath{\mathrm{H}^{0}}$}}
\newcommand{\ee}{\mbox{$\mathrm{e}^{+}\mathrm{e}^{-}$}}
\newcommand{\bee}{\mbox{$\boldmath {\mathrm{e}^{+}\mathrm{e}^{-}} $}}
\newcommand{\mm}{\mbox{$\mu^{+}\mu^{-}$}}
\newcommand{\bmm}{\mbox{$\boldmath {\mu^{+}\mu^{-}} $}}
\newcommand{\nn}{\mbox{$\nu \bar{\nu}$}}
\newcommand{\bnn}{\mbox{$\boldmath {\nu \bar{\nu}} $}}
\newcommand{\qq}{\mbox{$\mathrm{q} \bar{\mathrm{q}}$}}
\newcommand{\ff}{\mbox{$\mathrm{f} \bar{\mathrm{f}}$}}
\newcommand{\bqq}{\mbox{$\boldmath {\mathrm{q} \bar{\mathrm{q}}} $}}
\newcommand{\pb}{\mbox{$\mathrm{pb}^{-1}$}}
\newcommand{\ra}{\mbox{$\rightarrow$}}
\newcommand{\br}{\mbox{$\boldmath {\rightarrow}$}}
\newcommand{\erh}{\mbox{$\mathrm{e}^+\mathrm{e}^-\rightarrow\mathrm{hadrons}$}}
\newcommand{\tautau}{\mbox{$\tau^{+}\tau^{-}$}}
\newcommand{\gamgam}{\mbox{$\gamma\gamma$}}
\newcommand{\tpm}{\mbox{$\tau^{\pm}$}}
\newcommand{\uu}{\mbox{$\mathrm{u} \bar{\mathrm{u}}$}}
\newcommand{\dd}{\mbox{$\mathrm{d} \bar{\mathrm{d}}$}}
\newcommand{\bb}{\mbox{$\mathrm{b} \bar{\mathrm{b}}$}}
\newcommand{\cc}{\mbox{$\mathrm{c} \bar{\mathrm{c}}$}}
\newcommand{\nunu}{\mbox{$\nu \bar{\nu}$}}
\newcommand{\mQ}{\mbox{$m_{\mathrm{Q}}$}}
\newcommand{\mZ}{\mbox{$m_{\mathrm{Z}}$}}
\newcommand{\mH}{\mbox{$m_{\mathrm{H}}$}}
\newcommand{\mh}{\mbox{$m_{\mathrm{h}}$}}
\newcommand{\mA}{\mbox{$m_{\mathrm{A}}$}}
\newcommand{\mHpm}{\mbox{$m_{\mathrm{H}^{\pm}}$}}
\newcommand{\mW}{\mbox{$m_{\mathrm{W}^{\pm}}$}}
\newcommand{\mt}{\mbox{$m_{\mathrm{t}}$}}
\newcommand{\mb}{\mbox{$m_{\mathrm{b}}$}}
\newcommand{\lpm}{\mbox{$\ell ^+ \ell^-$}}
\newcommand{\G}{\mbox{$\mathrm{GeV}$}}
\newcommand{\Gc}{\mbox{$\mathrm{GeV}$}}
\newcommand{\Gcs}{\mbox{$\mathrm{GeV}$}}
\newcommand{\epsnn}{\mbox{$\epsilon^{\nu\bar{\nu}}$(\%)}}
\newcommand{\Nnn}{\mbox{$N^{\nu \bar{\nu}}_{exp}$}}
\newcommand{\epsll}{\mbox{$\epsilon^{\ell^{+}\ell^{-}}$(\%)}}
\newcommand{\Nll}{\mbox{$N^{\ell^+\ell^-}_{exp}$}}
\newcommand{\Nexp}{\mbox{$N^{total}_{exp}$}}
\newcommand{\kl}{\mbox{$\mathrm{K_{L}}$}}
\newcommand{\dedx}{\mbox{d$E$/d$x$}}
\newcommand{\etal}{\mbox{$et$ $al.$}}
\newcommand{\ie}{\mbox{$i.e.$}}
\newcommand{\sba}{\mbox{$\sin ^2 (\beta -\alpha)$}}
\newcommand{\cba}{\mbox{$\cos ^2 (\beta -\alpha)$}}
\newcommand{\tanb}{\mbox{$\tan \beta$}}
\newcommand{\sqrts}{\mbox{$\sqrt {s}$}}
\newcommand{\sqrtsp}{\mbox{$\sqrt {s'}$}}
\newcommand{\ltapprox}{\stackrel{<}{\sim}}
\newcommand{\gtapprox}{\stackrel{>}{\sim}}
\renewcommand{\arraystretch}{1.4}

\begin{titlepage}
\vspace*{-1cm}
\begin{center}{\Large   EUROPEAN LABORATORY FOR PARTICLE PHYSICS}
\end{center}
\begin{flushright}
\large
  CERN-EP/98-029 \\
  20th February 1998 \\
\end{flushright}
\bigskip
\begin{center}{\huge\bf A Search for Neutral Higgs Bosons \\
                          in the MSSM and Models with     \\[4mm]
                            Two Scalar Field Doublets     }
\end{center}\bigskip
\bigskip\begin{center}{{\Large The OPAL Collaboration} \\[10mm]}
\end{center}\bigskip
\begin{center}{\large\bf  Abstract}\end{center}
A search is described for the neutral Higgs bosons \h\ and \A\
predicted by models with two scalar field doublets and, in particular,
the Minimal Supersymmetric Standard Model (MSSM).
The search in the \Z\h\ and \h\A\ production channels is 
based on data corresponding to an integrated luminosity of
25 \pb\ from \ee\ collisions at centre-of-mass
energies between 130 and 172~GeV collected with the OPAL detector at LEP.
The observation of a number of candidates consistent with Standard Model
background expectations is used in combination
with earlier results from data collected at the \Z\ resonance
to set limits on \mh\ and \mA\ in 
general models with two scalar field doublets and in the MSSM.
For example, in the MSSM, for $\tanb>1$, minimal and maximal scalar top quark
mixing and soft SUSY-breaking masses of 1~TeV, the 95\% confidence level limits
$\mh>59.0$~GeV and $\mA>59.5$~GeV are obtained.
For the first time, the MSSM parameter space is explored in a detailed scan.

\vspace*{15mm}
\begin{center} Submitted to {European Physics Journal C} \end{center}
\vspace*{1 cm}
\begin{center}
\end{center}
\end{titlepage}
%
\begin{center}{\Large        The OPAL Collaboration
}\end{center}\bigskip
\begin{center}{
K.\thinspace Ackerstaff$^{  8}$,
G.\thinspace Alexander$^{ 23}$,
J.\thinspace Allison$^{ 16}$,
N.\thinspace Altekamp$^{  5}$,
K.J.\thinspace Anderson$^{  9}$,
S.\thinspace Anderson$^{ 12}$,
S.\thinspace Arcelli$^{  2}$,
S.\thinspace Asai$^{ 24}$,
S.F.\thinspace Ashby$^{  1}$,
D.\thinspace Axen$^{ 29}$
% ...
}\end{center}\bigskip
\bigskip
$^{  1}$School of Physics and Astronomy, University of Birmingham,
Birmingham B15 2TT, UK
\newline
$^{  2}$Dipartimento di Fisica dell' Universit\`a di Bologna and INFN,
I-40126 Bologna, Italy
\newline
% ...
$^{  5}$Cavendish Laboratory, Cambridge CB3 0HE, UK
\newline
% ...
$^{  8}$CERN, European Organisation for Particle Physics,
CH-1211 Geneva 23, Switzerland
\newline
$^{  9}$Enrico Fermi Institute and Department of Physics,
University of Chicago, Chicago IL 60637, USA
\newline
% ...
$^{ 12}$Indiana University, Department of Physics,
Swain Hall West 117, Bloomington IN 47405, USA
\newline
% ...
$^{ 16}$Department of Physics, Schuster Laboratory, The University,
Manchester M13 9PL, UK
\newline
% ...
$^{ 23}$Department of Physics and Astronomy, Tel Aviv University,
Tel Aviv 69978, Israel
\newline
$^{ 24}$International Centre for Elementary Particle Physics and
Department of Physics, University of Tokyo, Tokyo 113, and
Kobe University, Kobe 657, Japan
\newline
% ...
$^{ 29}$University of British Columbia, Department of Physics,
Vancouver BC V6T 1Z1, Canada
\newline
% ...
$^{  b}$ and Royal Society University Research Fellow
\newline
$^{  c}$ and Institute of Nuclear Research, Debrecen, Hungary
\newline
$^{  d}$ and Department of Experimental Physics, Lajos Kossuth
University, Debrecen, Hungary
\newline
$^{  e}$ and Department of Physics, New York University, NY 1003, USA
\newline
%
\newpage
\section{Introduction}
The data accumulated with the OPAL detector
at centre-of-mass energies of $130-172$~GeV,
corresponding to approximately 25 \pb\ of integrated luminosity, have opened up
new kinematic domains for particle searches.
We describe searches performed in these high energy data
for neutral Higgs bosons which decay to hadrons or tau-leptons.

In the Standard Model (SM)~\cite{sm}, spontaneous symmetry-breaking is 
effected 
by the self-interaction of one scalar (Higgs) field doublet~\cite{higgs}.
The model predicts one Higgs boson, \Hsm, the mass of which 
is not specified.
The OPAL search for \Hsm\ at $\sqrts=130-172$~GeV
has resulted in a lower bound on its mass of $\mHsm>69.4$~GeV
at the 95\% confidence level (CL)~\cite{smpaper}.
 
Supersymmetry (SUSY)~\cite{susy} is considered an attractive
possible extension of the SM,
since it provides a solution to one of the outstanding
problems of the SM, that of the ``hierarchy" of energy scales~\cite{hierarchy}.
The implementation of SUSY requires at least two Higgs field doublets.
There are exactly two in the Minimal Supersymmetric extension of the
Standard Model (MSSM)~\cite{mssm}.
These fields couple separately to up-type quarks for the first doublet,
and to down-type quarks and charged leptons for the second doublet,
and have vacuum expectation values $v_1$ and $v_2$, respectively. 
Scalar field doublets that couple in this manner may
exist more generally and the class of such models is known as
Type II Two Higgs Doublet Model (2HDM)~\cite{higgshunter}.
Here, 2HDM models will be understood to have no extra particles besides those
of the SM and the two scalar doublets,
and there is no mass relation between the different
neutral and charged scalar field particles.
In any two Higgs field doublet model,
the Higgs sector comprises five physical Higgs bosons: two neutral CP-even
scalars \h\ and \Ho\ (with masses satisfying $\mh<\mH$ by definition), one
CP-odd scalar \A\ and two charged scalars \Hpm. 
In this paper the searches are restricted to the neutral Higgs bosons \h\ and
\A.
The heaviest neutral Higgs boson \Ho\ is likely to have a mass beyond the
reach of LEP, and OPAL searches for \Hpm\ bosons have been
published separately~\cite{charged}.

At the current \ee\ centre-of-mass energies (\sqrts) accessed by
LEP, the \h\ and \A\  
bosons are expected to be produced predominantly via two processes: 
the ``Higgs-strahlung"
process \ee\ra\h\Z\ (where the \Z\ boson is on-shell)
and the ``pair production" process \ee\ra\h\A. Contributions from the
\WW\ and \Z\Z\ fusion processes account for a small part of the total
production, except close to the kinematic limit of the
\ee\ra\h\Z\ process~\cite{bosoncs}.
For these two principal processes, the cross-sections 
$\sigma_{\mathrm{hZ}}$ and $\sigma_{\mathrm{hA}}$
are related at tree-level 
to the SM cross-sections~\cite{lep2higgs}: 
\begin{eqnarray}
\ee\ra\h\Z\;:&&
\sigma_{\mathrm{hZ}}=\sin^2(\beta -\alpha)~\sigma^{\mathrm{SM}}_{\mathrm{HZ}},
\label{equation:xsec_zh} \\
\ee\ra\h\A\;:&&
\sigma_{\mathrm{hA}}=
\cos^2(\beta-\alpha)~\bar{\lambda}~\sigma^{\mathrm{SM}}_{\nn},
\label{equation:xsec_ah}
\end{eqnarray} 
where $\sigma^{\mathrm{SM}}_{\mathrm{HZ}}$ and
$\sigma^{\mathrm{SM}}_{\nn}$ are the cross-sections for the SM processes
\ee\ra\Hsm\Z\ and \ee\ra\nn, and
$\bar{\lambda}$ is a kinematic factor, depending 
on \mh, \mA\ and \sqrts, typically having values between 0.5 and 0.7 for the
centre-of-mass energies under consideration.
The angle $\beta$ is defined in terms of the
vacuum expectation values of the two scalar fields, $\tanb=v_2/v_1$, and
$\alpha$ is the mixing angle of the CP-even (\h,\Ho) fields.
The coefficients \sba\ and \cba\ indicate complementarity
in the cross-sections of the two processes,
a feature which is exploited in deriving bounds for
Higgs boson masses and other model parameters.

In the MSSM at tree level the following mass relations are predicted:
$\mh\leq\mZ$, $\mA\leq\mH$, $\mZ\leq\mH$ and $\mHpm\geq\mW$~\cite{higgshunter}.
Loop corrections, dominantly from the top and scalar top quarks (\sctop), 
strongly modify
these mass relations and also have some moderate impact on the Higgs boson
couplings~\cite{carena}. 
The shift in \mh, approximately proportional to 
$m_{\rm t}^2\times~\log(\mstop^2/m_{\rm t}^2)$,
can be several tens of GeV. While the top~quark mass has been measured to be
$m_{\rm t}=(175{\pm}5)$~GeV~\cite{mtop},
the mass of the scalar top quark \mstop\ depends
on the mixing in the \sctop\ sector which, in turn, depends on 
several other parameters of the MSSM. However, 
even for a choice of SUSY parameters which maximises
the mass shift, \mh\ should not exceed approximately 135~GeV~\cite{lep2higgs}.
Although this bound is beyond the ultimate reach of the LEP
collider, a substantial fraction of this mass range is accessible.

In this work we undertake a more detailed examination of the MSSM parameter 
space than has been done in the past.
The coefficients \sba\ and \cba\ which appear
in Eqs.~(\ref{equation:xsec_zh}) and~(\ref{equation:xsec_ah})
depend on a number of MSSM parameters which enter
via mixing in the \sctop\ sector and which will be summarised in a
later section. We perform detailed scans over broad ranges of these parameters.
Each of these scans is considered as an independent
``model'' within the MSSM, and results are provided for each.
In increasing order of generality, they include:
(A) Particular choices of parameters for ``minimal'' and ``maximal'' 
\sctop-mixing and soft SUSY-breaking masses fixed to be large
as defined in~\cite{lep2higgs} which are considered as the ``benchmark case''
and adopted by most search groups,
(B) varying the parameters over a wide range, but keeping relations between
some of them corresponding to minimal and maximal mixing, and
(C) a ``general'' scan where the MSSM parameters are allowed to vary
independently within wide, but reasonable, ranges.  

The final-state topologies of the processes (\ref{equation:xsec_zh}) and
(\ref{equation:xsec_ah}) are determined by the decays of the \Z,
\h\ and \A\ bosons. Since the Higgs bosons couple to fermions with a strength
proportional to the fermion mass, the Higgs
dominantly decays into pairs of the most massive
fermions which are allowed by the kinematics, most notably
b~quarks and tau-leptons for the LEP mass range.
For particular choices of the model parameters (e.g. for
$\tanb<1$) decays into \cc\ may be enhanced. For $2\mA\leq\mh$, the process 
\h\ra\A\A\ is also allowed and may even be the dominant decay, leading to more 
complex final states than 
those from direct decays into fermion pairs. In the MSSM, Higgs bosons may
also decay into SUSY particles if allowed by kinematics. In particular,
the decay into pairs of neutralinos (\ko) may lead to ``invisible" Higgs decay 
channels\footnote{In the MSSM R-parity is conserved and throughout this paper 
we assume that the lightest supersymmetric particle
is the lightest neutralino, \kol.}
which must be considered in a full treatment of the model.

In searching for the process \ee\ra\h\Z, 
the fact is exploited that in most of the MSSM parameter space
with $\tanb>0.7$
the decay properties of the \h\ boson are essentially those of the
SM Higgs boson. Thus, earlier OPAL searches for the
\Hsm\ boson~\cite{smpaper}, including those performed at
energies above the \Z\ mass~\cite{opal94,opal96},
are interpreted here as searches for 
\ee\ra\h\Z. The reduction of the search sensitivity due to \sba\ 
in Eq. (1) is taken into account.
Dedicated searches for ``invisible" final states at
$\sqrts\approx\mZ$~\cite{topol} are also included.
Since these searches are published, we only summarise the results
which are relevant for the present purpose.
These searches for \Hsm\ are also efficient for
\ee\ra\h\Z\ra\A\A\Z, sometimes after small modifications to the search.

In searching for the process \ee\ra\h\A, the following final states are most
important:
(\h\ra\bb)(\A\ra\bb),
(\h\ra\tautau)(\A\ra\qq) and (\h\ra\qq)(\A\ra\tautau).
The searches in these channels using the data accumulated by OPAL between
$\sqrts=130$ and 172~GeV have not been published.
Therefore they are described in greater detail.
The search for \h\A\ra\A\A\A\ra\bb\bb\bb\ is important when the decay
\h\ra\A\A\ is
kinematically allowed and is also presented here for the first time for
data taken above the \Z\ energy.

This paper starts in Section~\ref{sect:detector}
with a short description of the OPAL detector, the data 
samples used and the various Monte Carlo
simulations used to obtain the detection efficiencies and to estimate
the backgrounds from SM processes.
This is followed by a description of the event selections for the
various \h\Z\ and \h\A\ channels in Sections~\ref{sect:searches}
and~\ref{sect:hasearches}.

A new statistical method~\cite{bock},
summarised in Section~\ref{section:combchan},
has been used to combine the results
from different search channels and data sets taken at different
centre-of-mass energies.

The model-independent and 2HDM results are summarised in
Section~\ref{section:modindep},
followed by those interpreted within the MSSM in
Section~\ref{section:mssminterpretation}.

Previous OPAL searches for the \h\ and \A\ bosons, based on data
collected at \sqrts$\approx$\mZ\ are described in~\cite{opal94,opal96}. 
The relevant publications from the other LEP 
collaborations describing neutral SUSY Higgs boson searches
are listed in~\cite{mssmall}.

\section{Experimental considerations}
\label{sect:detector}
%
The present search includes data collected with the OPAL 
detector~\cite{detector}
in 1995 at $\sqrts=130-136$~GeV (5.2 \pb), in 1996 at $\sqrts=161$~GeV 
(10.0 \pb),
and at $170-172$~GeV (10.4~\pb). The results are combined with those from 
earlier
searches~\cite{opal96}
which are based on the analysis of up to 4.5 million hadronic \Z\ decays.
 
The OPAL detector is an apparatus
with nearly complete solid angle coverage and excellent hermeticity.
The central tracking detector consists of a high-resolution
silicon microstrip vertex detector ($\mu$VTX)~\cite{simvtx}
with polar angle\footnote{
OPAL uses a right-handed
coordinate system where the $+z$ direction is along the electron beam and
where $+x$ points to the centre of the LEP ring.  
The polar angle, $\theta$, is
defined with respect to the $+z$ direction and the azimuthal angle, $\phi$,
with respect to the horizontal, $+x$ direction.}
coverage $|\cos\theta|<0.9$, which immediately
surrounds the beam-pipe, followed by a high-precision vertex drift chamber,
a large-volume jet chamber, and $z$-chambers,
all in a uniform  
0.435~T axial magnetic field. A lead-glass electromagnetic calorimeter
is located outside the magnet coil, which, in combination with
the forward calorimeter, gamma catcher, and silicon-tungsten
luminometer~\cite{sw}, complete the geometrical acceptance
down to 33~mrad from the beam direction.  The silicon-tungsten luminometer
serves to measure the integrated luminosity using small-angle Bhabha
scattering events~\cite{lumino}.
The magnet return yoke is instrumented with streamer tubes
for hadron calorimetry and is surrounded by several layers of muon chambers.

Events are reconstructed from charged-particle tracks and
energy deposits (``clusters") in the electromagnetic and hadronic calorimeters.
The tracks and clusters must pass a set of quality requirements
similar to those used in
previous OPAL Higgs boson searches~\cite{higgsold}.
In calculating the total visible energies and momenta, $E_{\rm vis}$
and $\vec{P}_{\rm vis}$, of events and of
individual jets, corrections are applied against 
double-counting of energy in the case of tracks and associated
clusters~\cite{lep2neutralino,opalhiggs1990}. For the analysis presented here, 
charged particles and
neutral clusters are grouped into jets using the Durham algorithm~\cite{durham}.

\label{btag}
The tagging of jets originating from b quarks plays an important role in
Higgs boson searches, since both \h\ and \A\ decay preferentially to \bb\
over large domains of the two-field-doublet and MSSM parameter spaces.
Primary and secondary vertices are reconstructed in three dimensions
following two algorithms, using only tracks which pass an additional
set of quality requirements.
The first method (BTAG1)~\cite{btag1}
considers all such tracks in a jet and 
attempts to fit them
to a common vertex. Tracks are discarded from the vertex by an iterative
procedure which drops the track with the largest $\chi^2$ contribution to the
vertex fit, until the largest $\chi^2$ contribution is less than 4, with at
least two tracks remaining.
In the second method (BTAG2)~\cite{btag2},
the intersection of all pairs of
such tracks in a jet having impact parameter significance $b/\sigma_b$
(where $b$ is the impact parameter\footnote{
The impact parameter is taken to be
positive if in the two-dimensional projection the track path 
crosses the jet
axis in the direction of the flight direction; otherwise it is negative.
}
and $\sigma_b$ its error) greater
than 2.5 are considered as seed vertices. The other tracks
in the jet are added to the seed vertex in the order which results in 
the greatest vertex probability after each addition. The process continues
until either all tracks in the jet are added or the resulting vertex 
probability falls below 1\%, in which case the last track 
to have been added  is dropped. If more than one acceptable vertex
per jet is found via this algorithm, the best is chosen according to
a set of criteria~\cite{btag1} involving track multiplicity and the vertex 
decay length significance $S\equiv L/\sigma_L$, where $L$ is the vertex decay 
length\footnote{
The vertex decay length is the projection onto the jet direction of
the distance between the primary vertex, as reconstructed for the event
(see~\cite{btag1} for the algorithm), and
the secondary vertex, as reconstructed for the jet.
The decay length is taken to be positive if the vector that connects the primary
to the secondary vertex is at an angle of less than 90$^{\circ}$ from
the direction of the associated jet, and negative otherwise.
}
and $\sigma_L$ its error.
In both algorithms, $S$ is then used to
distinguish between b-flavoured hadron decays and background.
The methods are found to not be fully correlated, and using them both
adds discriminating power.
Charged track impact parameters are also used 
to complement
the secondary vertex algorithms
via the forward multiplicity,
defined as the number of tracks in a jet with 
$b/\sigma_b>2.5$.
Finally, semi-leptonic b-hadron decays are exploited by first identifying
electrons via the method described in~\cite{nn5} and
muons with the algorithm described in~\cite{muon}, and then considering the
transverse momentum with respect to the corresponding jet axis.

The signal detection efficiencies and accepted background cross-sections
are estimated using a variety of Monte Carlo samples. 
The HZHA generator~\cite{hzha} is used to simulate Higgs boson
production processes. The detection efficiencies are determined at 
fixed values of the Higgs boson masses using sample sizes varying between 500
and 10,000 events. Efficiencies at arbitrary masses are evaluated using
spline fits in the (\mh,\mA) plane between these points.
The background processes are simulated primarily by
the following event generators:
PYTHIA~\cite{pythia} ((\Z$/\gamma)^*$\ra\qq($\gamma$)), 
EXCALIBUR~\cite{excalibur} and grc4f~\cite{grc4f} (four-fermion processes 
(4f)),
BHWIDE~\cite{bhwide} (\ee$(\gamma)$),
KORALZ~\cite{koralz} (\mm$(\gamma)$ and \tautau$(\gamma)$),
PHOJET~\cite{phojet} and
Vermaseren~\cite{vermaseren} (hadronic and leptonic two-photon processes
($\gamma\gamma$)).
The generated partons are hadronised using JETSET~\cite{pythia} and the
resulting particles are
processed through a full simulation~\cite{gopal} of the OPAL detector.

\section{\boldmath Searches for the process \ee\ra\h\Z}
\label{sect:searches}
The OPAL searches for
the SM process \ee\ra\Hsm\Z\ are interpreted as searches for the process
\ee\ra\h\Z\ via Eq.~(1). The searches for \Hsm\ using all data
recorded at center-of-mass energies up to 172 GeV are described
in~\cite{smpaper}. They make use of the
following final states:
\begin{itemize}
\item ``Four-jet": (\Hsm\ra\bb)(\Z\ra\qq) (q=u,d,s,c,b),
\item ``Missing energy": (\Hsm\ra\qq)(\Z\ra\nn) (q includes quarks and gluons),
\item ``Charged lepton": (\Hsm\ra\qq)(\Z\ra\ee, \mm), and
\item ``Tau-lepton": (\Hsm\ra\qq)(\Z\ra\tautau) and
                     (\Hsm\ra\tautau)(\Z\ra\qq).
\end{itemize}

The search in the missing energy channel is also sensitive to small
contributions to \Hsm\ production coming from the \WW\ fusion process 
\ee\ra\nn\Hsm\, while the search in
the charged lepton channel is sensitive to those from the \Z\Z\ fusion process
\ee\ra\ee\Hsm.
These contributions are taken into account in the corresponding channels.

The results from these published searches at $\sqrt{s}=161-172$ GeV are 
summarised in Table~\ref{smhiggs},
which lists the signal detection efficiencies for two Higgs boson masses
in the range of interest, the residual expected number of background events
and the number of selected
candidate events in each search channel.
As can be seen from the table, the selection criteria
applied at $\sqrts=161$~GeV select
one candidate in the missing energy channel, with a mass of  
$(39.3{\pm}4.9)$~GeV, while those applied at $170-172$~GeV select one candidate 
in the four-jet channel with a mass of $(75.6{\pm}3.0)$~GeV. The earlier OPAL 
searches applied to \Z\ boson decays~\cite{opal96} selected one candidate in 
the charged lepton 
channel \mm\Hsm\ with a mass of $(61.2{\pm}1.0)$~GeV (with 0.38$\pm$0.04
events expected
from background) and two candidates in the missing energy channel with masses
of $(6.3{\pm}0.8)$~GeV and $(24.8{\pm}3.0)$~GeV (with 2.3$\pm$0.4 events 
expected from background).
All these candidates are considered as possible Higgs boson events when limits
in the MSSM parameter space are computed. 
\begin{table}[tbp]
\begin{center}
\begin{tabular}{||l||c|c||cc||}
\hline\hline
\multicolumn{5}{||c||}{\bf\boldmath $\sqrts=161$~GeV} \\\hline\hline
Channel & {\large $\epsilon$}{\small (\mH=60~GeV)(\%)} &
          {\large $\epsilon$}{\small (\mH=70~GeV)(\%)} & Background & Data\\ 
\hline\hline
Four-jet                  & 30    & 31     & 0.75$\pm$0.08   & 0  \\
Missing energy            & 56    & 41     & 0.90$\pm$0.10   & 1  \\
Charged leptons \Z\ra\ee\ & 70    & 59     & 0.06$\pm$0.02   & 0  \\
Charged leptons \Z\ra\mm\ & 75    & 73     & 0.04$\pm$0.03   & 0  \\ 
Tau-lepton \Z\ra\tautau\  & 24    & 17     & 0.10$\pm$0.03   & 0  \\
Tau-lepton \Z\ra\qq\      & 19    & 17     & 0.06$\pm$0.03   & 0  \\[0.5ex]
\hline\hline
\multicolumn{5}{||c||}{\bf\boldmath $\sqrts=170-172$~GeV} \\\hline\hline
Channel & {\large $\epsilon$}{\small (\mH=60~GeV)(\%)} &
          {\large $\epsilon$}{\small (\mH=70~GeV)(\%)} & Background & Data\\ 
\hline\hline
Four-jet                  & 27    & 28      & 0.88$\pm$0.07   & 1  \\
Missing energy            & 47    & 41      & 0.55$\pm$0.05   & 0  \\
Charged leptons \Z\ra\ee\ & 64    & 65      & 0.08$\pm$0.02   & 0  \\
Charged leptons \Z\ra\mm\ & 69    & 71      & 0.06$\pm$0.03   & 0  \\ 
Tau-lepton \Z\ra\tautau\  & 27    & 22      & 0.41$\pm$0.03   & 0  \\
Tau-lepton \Z\ra\qq\      & 15    & 19      & 0.18$\pm$0.03   & 0  \\[0.5ex]
\hline\hline
\end{tabular}
\caption[]{\label{smhiggs}\sl
         Summary of the searches for the SM Higgs boson at 
         centre-of-mass energies of
         161 and $170-172$~GeV.  For each channel the signal detection 
         efficiencies
         for $\mHsm=60$ and 70~GeV, the number of expected background
         events, and the
         number of events selected are given.
         The statistical error on the efficiencies is typically 1-4\%.
}
\end{center}
\end{table}

The above searches are also sensitive to the 
process \ee\ra\h\Z\ followed by \h\ra\A\A.
The selection is slightly changed 
with respect to that described in~\cite{smpaper} in the case of the 
four-jet and missing
energy channels.
In the four-jet channel the likelihood discriminant, as described later in
Section~\ref{section:ah}, is reoptimised for
the \Z\A\A\ final state. This increases the efficiency from 23\% to 29\%,
while the background expectation remains approximately equal.
In the case of the missing energy channel at $\sqrts=161$~GeV a requirement
on the jet resolution parameter, $y_{23}<0.05$, limits the acceptance to
two-jet events only. Removing this requirement, the detection efficiency for
\h\ra\A\A\ ($\mh=60$~GeV, $\mA=30$~GeV)
increases from 38\% to 68\% while the
background increases from 0.9 to 1.1 events. For the decay of the \A\ 
boson, only the
predominant \bb\ final state is considered.
For the charged lepton and tau-lepton channels, 
Monte Carlo simulations have demonstrated that the
detection efficiencies for this two-stage process are close to those of the 
\h\ decay to fermion pairs.
The detection efficiencies for the particular case of $\mh=60$~GeV
and $\mA=30$~GeV, a point close to the boundary of the kinematically-allowed
region for \h\ra\A\A, are shown in Table~\ref{hAA}. 

\section{\boldmath Searches for the process \ee\ra\h\A}
\label{sect:hasearches}
\begin{table}[tbp]
\begin{center}
\begin{tabular}{||l|l|c||}
\hline\hline
\multicolumn{3}{||c||}{\bf\boldmath $\sqrts=161$~GeV} \\\hline\hline
SM search & applied to the process & {\large $\epsilon$}{\small (\%)}  \\  
\hline\hline
Four-jet        &(\A\A\ra\bb\bb)(\Z\ra\qq)     & 29 \\
Missing energy  &(\A\A\ra\qq\qq)(\Z\ra\nn)     & 68 \\
Charged leptons &(\A\A\ra\qq\qq)(\Z\ra\ee)     & 60 \\
Charged leptons &(\A\A\ra\qq\qq)(\Z\ra\mm)     & 74 \\
Tau-lepton      &(\A\A\ra\qq\qq)(\Z\ra\tautau) & 16 \\
\hline\hline
\multicolumn{3}{||c||}{\bf\boldmath $\sqrts=170-172$~GeV} \\\hline\hline
SM search & applied to the process & {\large $\epsilon$}{\small (\%)} \\  
\hline\hline
Four-jet        &(\A\A\ra\bb\bb)(\Z\ra\qq)     & 34 \\
Missing energy  &(\A\A\ra\bb\bb)(\Z\ra\nn)     & 59 \\
Charged leptons &(\A\A\ra\qq\qq)(\Z\ra\ee)     & 57 \\
Charged leptons &(\A\A\ra\qq\qq)(\Z\ra\mm)     & 73 \\
Tau-lepton      &(\A\A\ra\qq\qq)(\Z\ra\tautau) & 17 \\
\hline\hline
\end{tabular}
\caption[]{\label{hAA}\sl
         Signal detection efficiencies for the searches for the SM Higgs boson, 
         at centre-of-mass energies of 161 and $170-172$~GeV,
         applied to the processes with \h\ra\A\A\ followed by \A\ra\bb. 
         The efficiencies are quoted for $\mh=60$~GeV and
         $\mA=30$~GeV, with typical statistical errors of 1-4\%.
}
\end{center}
\end{table}
In this section the searches for the MSSM process \ee\ra\h\A\
for final states (\h\ra\bb)(\A\ra\bb), (\h\ra\tautau)(\A\ra\qq),
(\h\ra\qq)(\A\ra\tautau) and (\h\ra\A\A\ra\bb\bb)(\A\ra\bb) are summarised.
If the same branching ratios to \bb\ and \tautau\ are assumed 
for \h\ and \A\ as for \Hsm, then the first three of the above 
final states account for approximately 90\% of all \h\A\ decays,
in the portion of the
$(\mA,\mh)$ phase space where $\mA>\mh$. For points in this
phase space where $\mh\geq\mA$ and the decay \h\ra\A\A\ is
kinematically allowed, the last final state represents approximately
66\% of decays of this type.  

\subsection{\boldmath The channel \h\A\ra\bb\bb}
\label{section:ah}
The signature for events from the process \h\A\ra\bb\bb\ is  
four energetic
jets containing b-hadrons and a visible energy close to the 
centre-of-mass energy.
As \sqrts\ changes from 130~GeV to 172~GeV, the background changes considerably in size
and composition. At 130~GeV and 161~GeV the main background comes from
(\Z$/\gamma)^*$\ra\qq\
with or without initial state radiation
accompanied by hard gluon emission. Four-fermion processes, in particular
\ee\ra\WW, play a minor role since the threshold for these processes is
at most only marginally crossed.
At $170-172$~GeV this background becomes more important,
while the (\Z$/\gamma)^*$ background is reduced.
Two-photon processes have a large cross-section at all energies; however,
the selection requiring multihadronic final states with a visible energy close
to \sqrts\ reduces them to a negligible level.

The selection proceeds in two phases. First, a preselection is applied to
retain only those events which have some similarity to the signal.
The events remaining after preselection are then analysed using a likelihood
technique.

\begin{figure}[p]
\centerline{
\epsfig{file=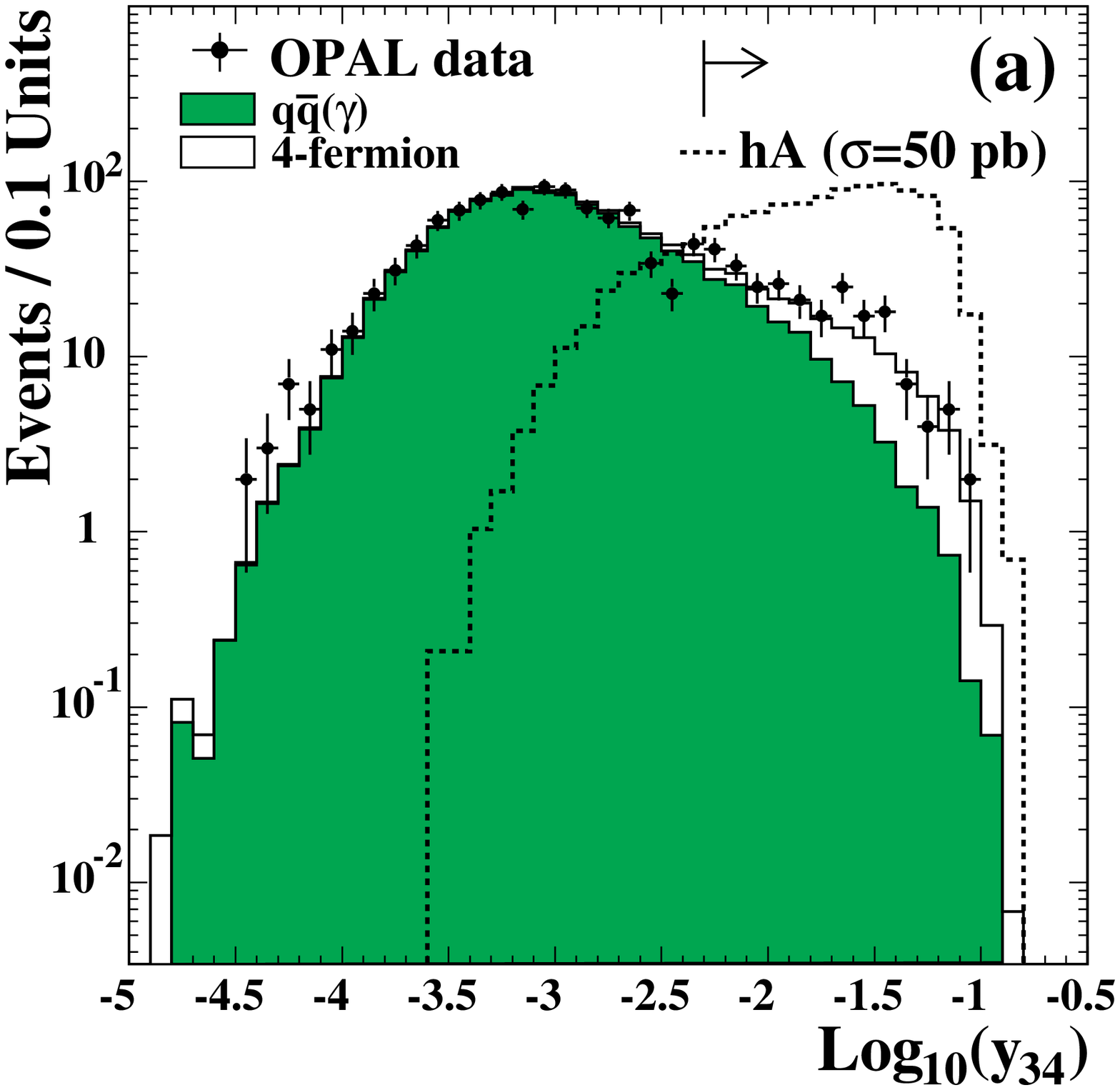,width=8.5cm}
\epsfig{file=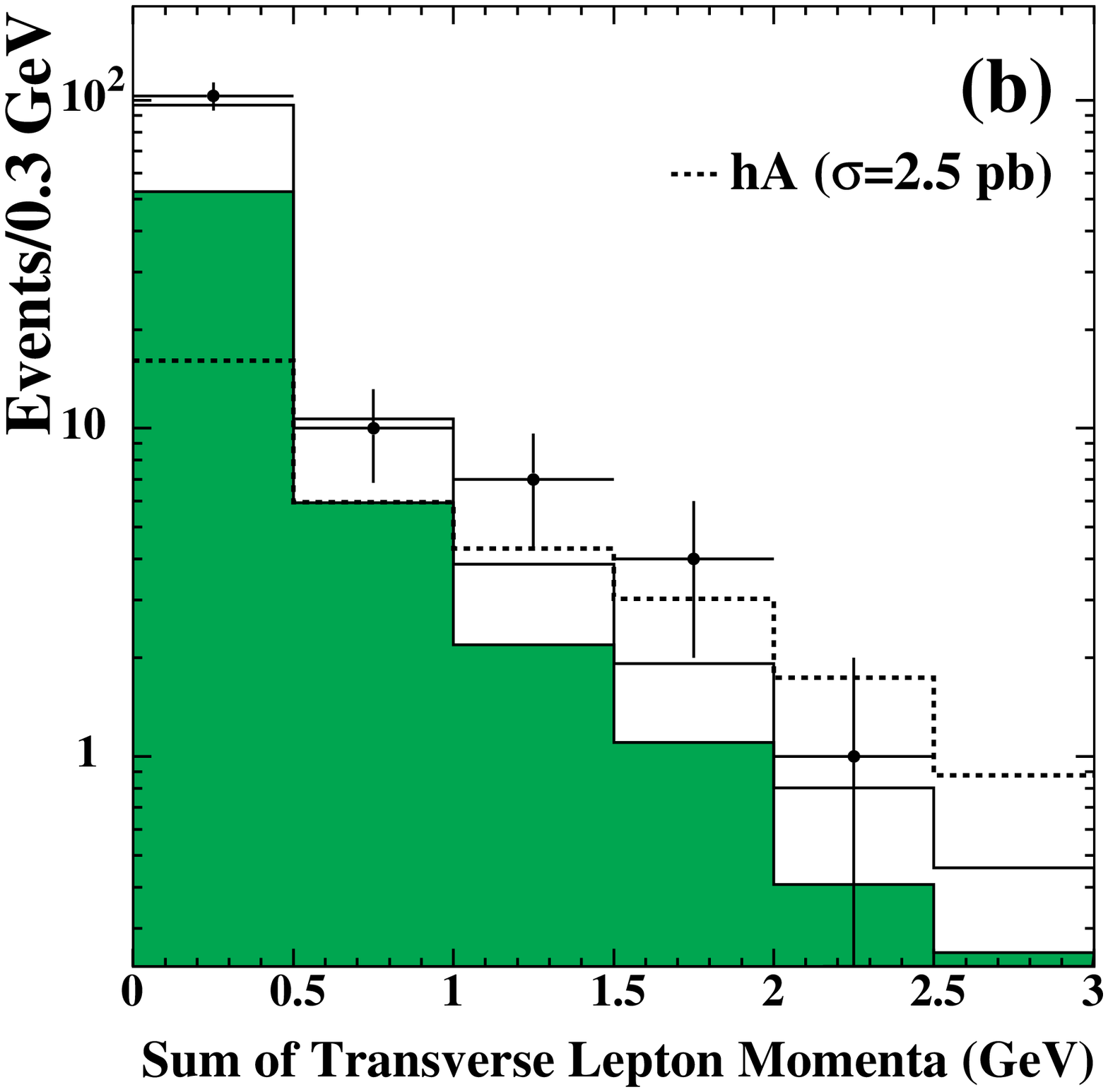,width=8.5cm}
} 
\centerline{
\epsfig{file=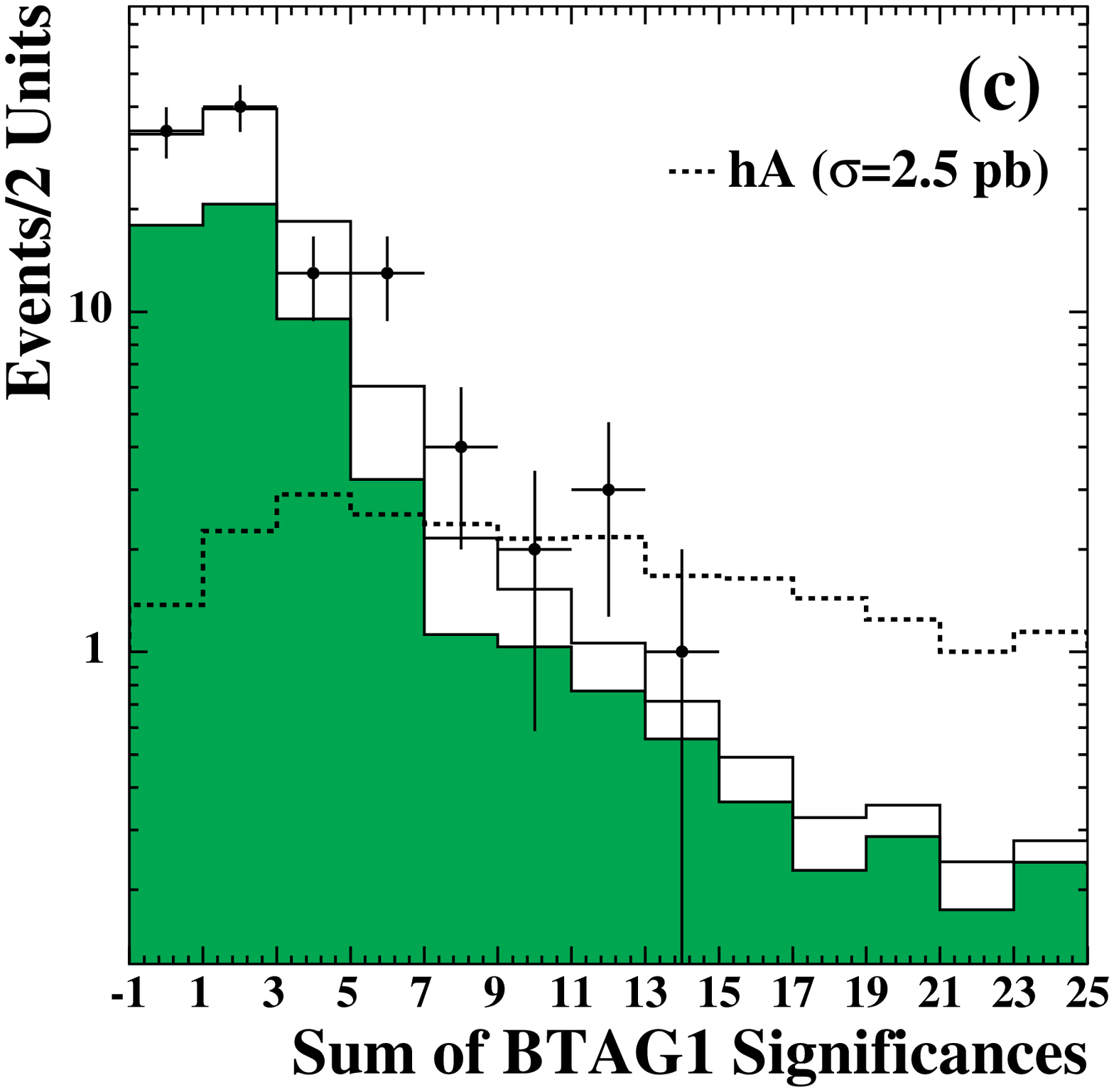,width=8.5cm}
\epsfig{file=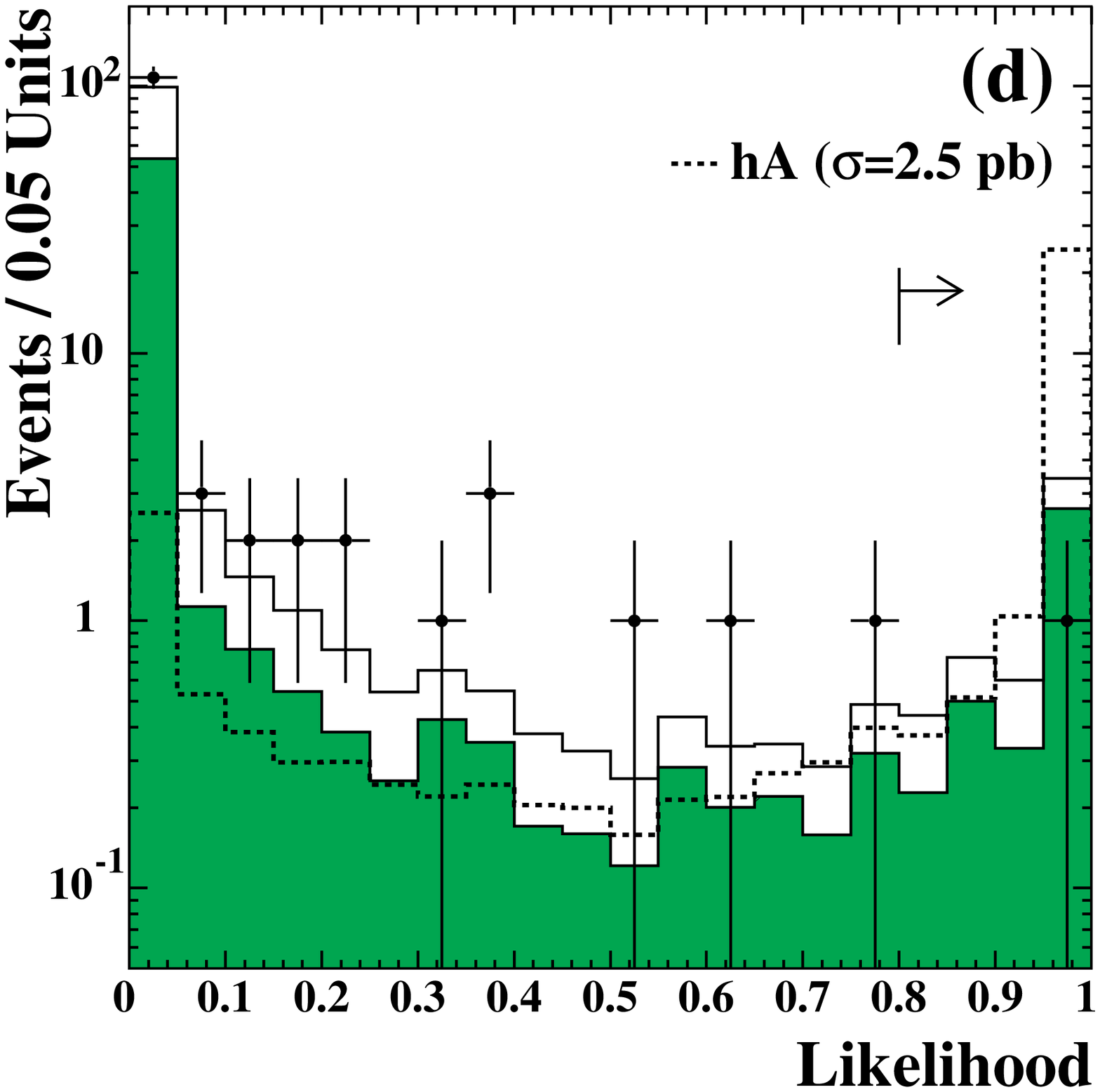,width=8.5cm}
}
\caption[]{\label{figure:ahbb_cuts}\sl
         Selection variables relevant 
         for the \h\A\ra\bb\bb\ analysis.
         (a) The logarithm of the jet resolution parameter 
         $y_{34}$ after preselection requirements (1) and (2).
         (b) The sum of the two largest lepton transverse momenta
         after the preselection.
         (c) The sum of the two largest secondary vertex decay length
         significances of the BTAG1 algorithm
         (defined in Section~\ref{btag}) after the preselection.
         (d) The \A\h\ likelihood distribution after the preselection.
         The three centre-of-mass energies are added for all histograms.
         The points represent the data.
         The shaded histograms show the $\qq$ background and the open
         histograms show the four-fermion background, normalised to 
         the integrated luminosity of the data. Two-photon processes are not
         included.
         The dashed line histogram shows the expectation for \A\h\ signal
         events with $\mA=\mh=55$~GeV,
         where the displayed production cross-sections
         have been chosen for visibility.}
\end{figure}

The preselection consists of the following requirements:
\begin{itemize}
\item[(1)]
      The events must qualify as being hadronic final states
      as described in~\cite{tkmh}.
\item[(2)]
      The radiative process \ee\ra(\Z$/\gamma)^*$\ra\qq$\gamma$ is largely
      eliminated by requiring that the effective centre-of-mass energy,
      \sqrtsp, obtained by discarding the radiative photon from the event 
      following~\cite{sprim}, is greater than 110, 140 and 150~GeV
      for $\sqrts=130-136$, 161, and $170-172$~GeV, respectively.
\item[(3)]
      The events are reconstructed into four jets using
      the Durham algorithm~\cite{durham}. The jet resolution parameter
      $y_{34}$, at which the number of jets changes from 3 to 4, is required 
      to be larger than 0.005.
      The distribution of $\log y_{34}$ is shown in
      Fig.~\ref{figure:ahbb_cuts}(a). 
\item[(4)]
      The value of the event shape $C$-parameter must be greater than 0.45.
      The $C$-parameter is defined as $C=3(e_1e_2+e_2e_3+e_3e_1)$, with
      $e_1$, $e_2$ and $e_3$, being the eigenvalues of the normalised momentum
      tensor of the event~\cite{cpar}. It ranges from $C=0$ for perfectly
      back-to-back two-jet events to $C=1$ for perfectly spherical events.
\item[(5)]
      Each of the four reconstructed jets must contain at least two charged
      tracks and at least two electromagnetic calorimeter clusters.
\item[(6)]
      A fit of the jet four-momenta, in which the energy and momentum of
      the final state is constrained to that of the initial \ee\ state,
      must yield a $\chi^2$-probability larger than 0.01.
\end{itemize}

The results of the preselection are listed in Table~\ref{table:ahbb_cuts}.
Except for the number of events retained after (1),
the agreement between observed
events and expected background predicted by the Monte Carlo simulation
is good.
The discrepancy after the first requirement can be explained by inaccurate 
modelling
of events that radiatively return to the \Z\ and by not including the 
two-photon events in the Monte Carlo prediction.
After a cut on \sqrtsp, these events are rejected and the background prediction
from Monte Carlo describes the data well.

Next, a likelihood technique is used to classify the remaining events as
either (\Z$/\gamma)^*$\ra\qq$(\gamma)$ (1), a four-fermion process (2), or
\A\h\ra\bb\bb\ (3).
Three kinematic and six b-tagging variables are input to the likelihood.
The kinematic quantities are: the smallest angle between
any jet pair, the logarithm of the probability for a fit of the jet momenta
with energy, momentum and equal di-jet mass constraint,
and the smallest di-jet mass difference after
the energy momentum conserving fit.
For the b-tagging quantities used here, all vertices
on which their calculation is based must contain at least two tracks, each
having two hits in the $r-\phi$ and $r-z$ ladders of the $\mu$VTX 
detector. 
The calculation of forward multiplicities takes into account only
tracks which have two hits in the $r-\phi$ and $r-z$ ladders of
the $\mu$VTX detector.
These quantities are (see Section~\ref{btag} for a description):
the sum of the two largest and the sum of the two smallest decay length 
significances in the four jets for
vertices reconstructed with algorithm BTAG1,
the sum of all decay length significances for vertices reconstructed 
with algorithm BTAG2,
the sum of the two largest and the sum of the two smallest forward 
multiplicities,
and the sum of the two largest transverse momenta with
respect to the corresponding jet axis for identified leptons.
For all of these quantities there is good agreement in their distributions
between data and Monte Carlo background predictions, and as examples the
distributions for the sum of the two largest lepton transverse momenta and
the sum of the two largest decay length significances for
the vertex algorithm BTAG1
are shown in Fig.~\ref{figure:ahbb_cuts}(b) and (c), respectively.

For each of these input variables (labelled by $i$)
a normalised histogram $f^j_i(x_i)$ is constructed from
the Monte Carlo for each of the two background classes and the signal
(labelled by $j=1,2,3$).
For a single variable, the probability for an event to belong to class $j$ is:
\[
p^j_i(x_i)=\frac{f^j_i(x_i)}{\sum^3_{k=1}f^k_i(x_i)},
\]
and the joint discriminating variable for class $j$ is defined as:
\[
{\cal P}^j(\vec{x})=\frac{\prod^9_{i=1}p^j_i(x_i)}{\sum^3_{k=1}\prod^9_{i=1}p^k_i(x_i)},
\]
where the product runs over the nine input variables.
The signal likelihood is defined as:
\[
{{\cal L}^{\mathrm{Ah}}}(\vec{x})=\frac{{\cal P}^3(\vec{x})}{\sum^3_{j=1}
{\cal P}^{j}(\vec{x})},
\]
and is required to be greater than 0.8 for the final selection.
The likelihood distribution is shown in Fig.~\ref{figure:ahbb_cuts}(d)
and over the entire range good agreement between the data and Monte Carlo
background prediction can be observed.
Some irreducible background from four-fermion processes also shows up as a
small peak near a likelihood of 1.
In total one event is selected, with a centre-of-mass energy of 172~GeV,
while 1.65, 1.61 and 1.95 events are expected from the background simulations
at $\sqrts=130-136$, 161, and $170-172$~GeV, respectively.
The Poisson probability to select one or fewer events when 5.2 are expected 
is 2.3\%.

\begin{figure}[p]
\centerline{
\epsfig{file=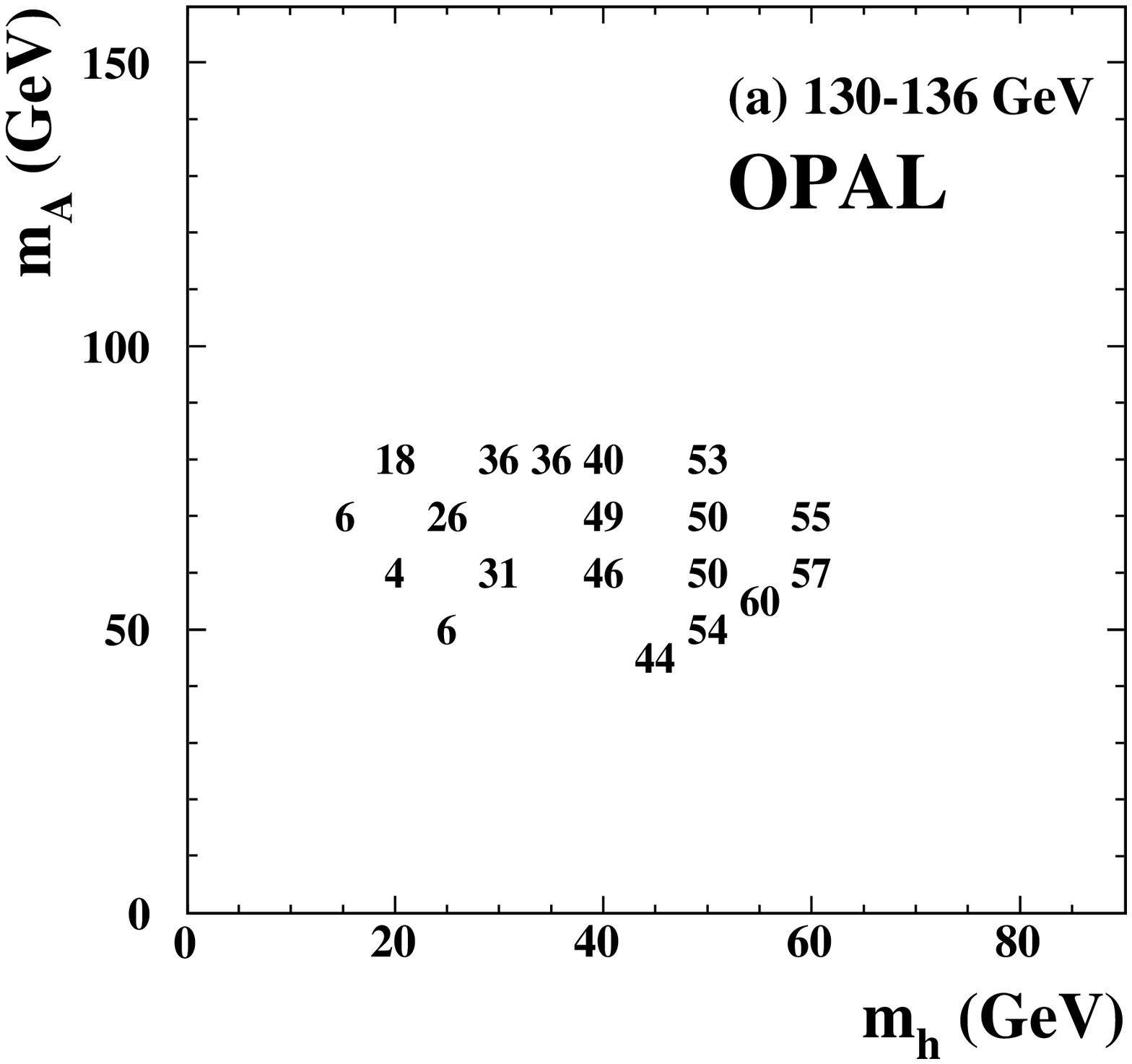,width=8.5cm}
\epsfig{file=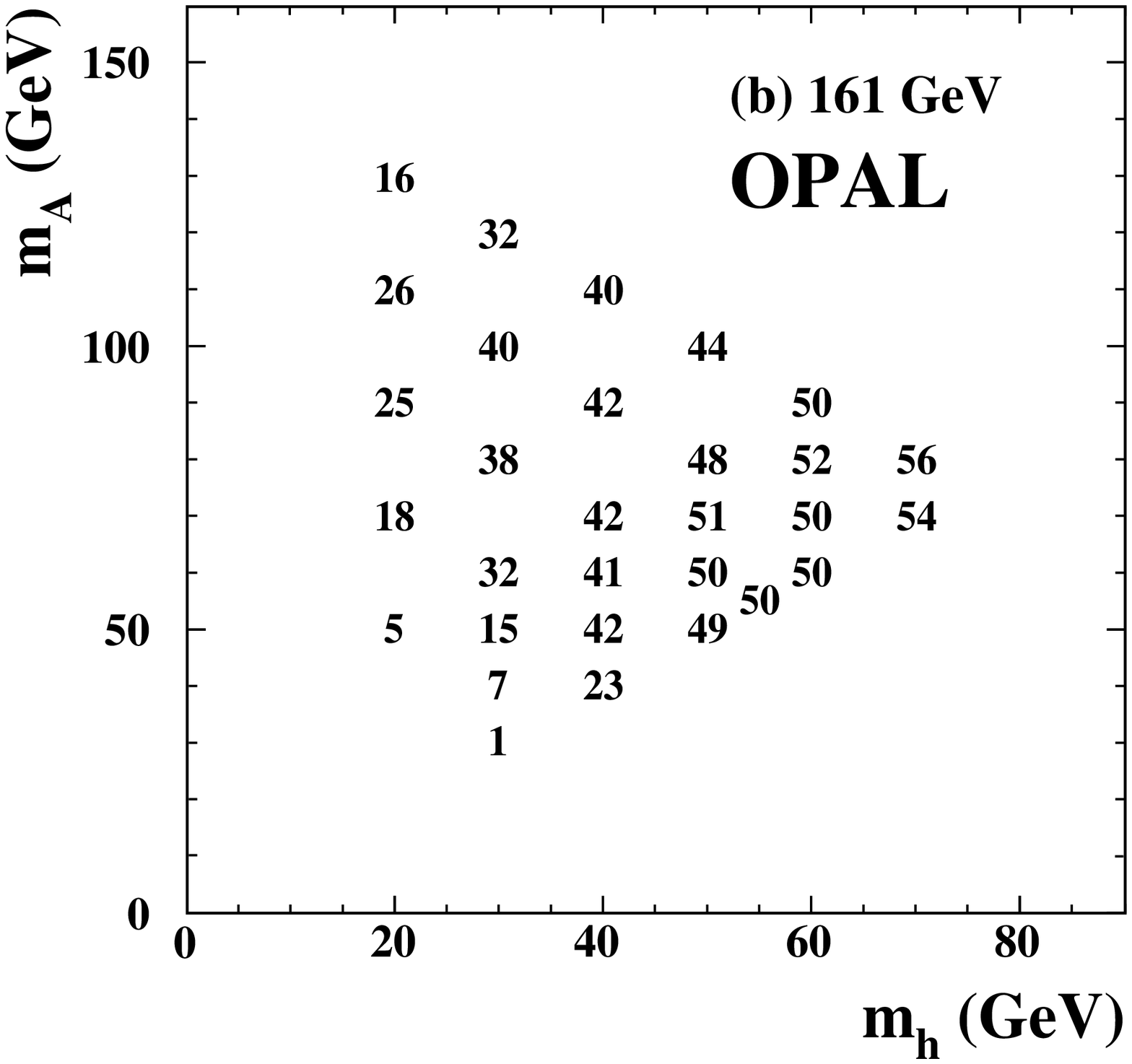,width=8.5cm}
} 
\centerline{
\epsfig{file=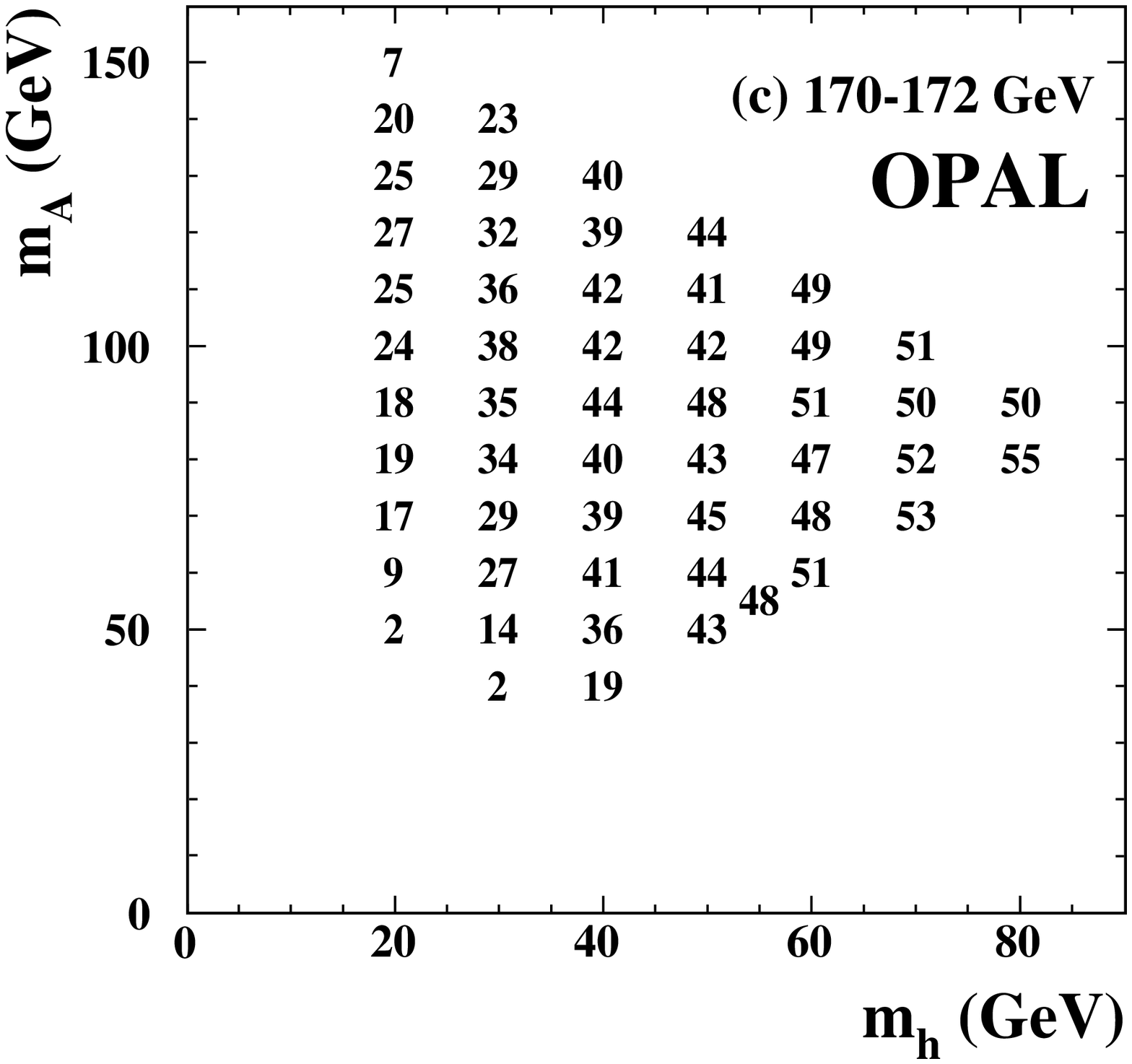,width=8.5cm}
\epsfig{file=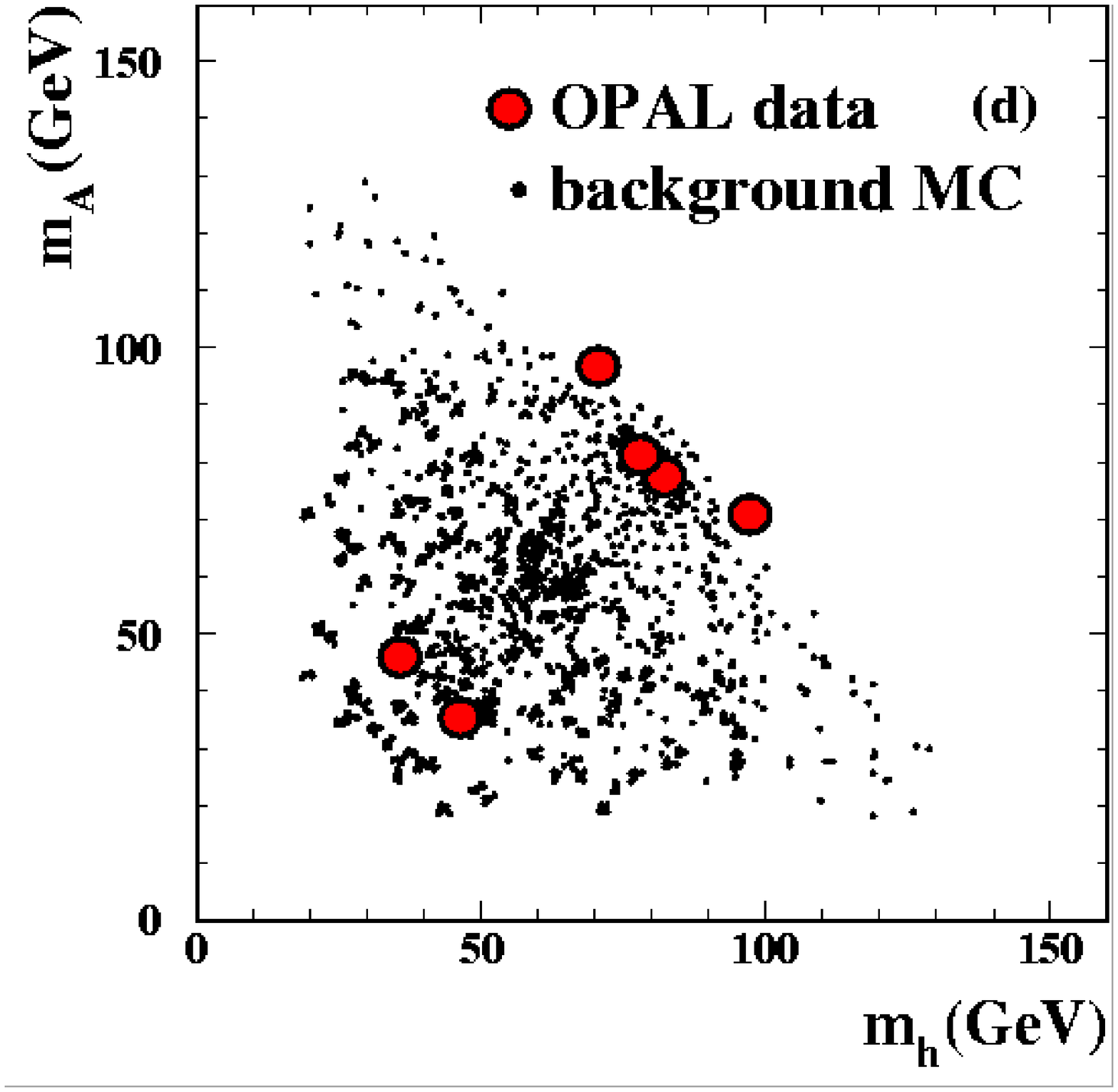,width=8.5cm,height=7.9cm}
}
\caption[]{\label{figure:ahbb_eff}\sl
         (a)--(c) The efficiencies (in \%) for the detection of 
         \h\A\ra\bb\bb\ at the three
         centre-of-mass energies in the (\mh,~\mA) plane. 
         The numbers are shown only for $\mA\geq\mh$ due to the symmetry 
         in \mh\ and \mA.
         (d) The position in the (\mh,~\mA) plane of the six possible
         mass combinations of the candidate event
         superimposed on the expected Standard Model background
         (scaled up by a factor of 100)
         for all center-of-mass energies. 
}
\end{figure}
\begin{figure}[p]
\centerline{
\epsfig{file=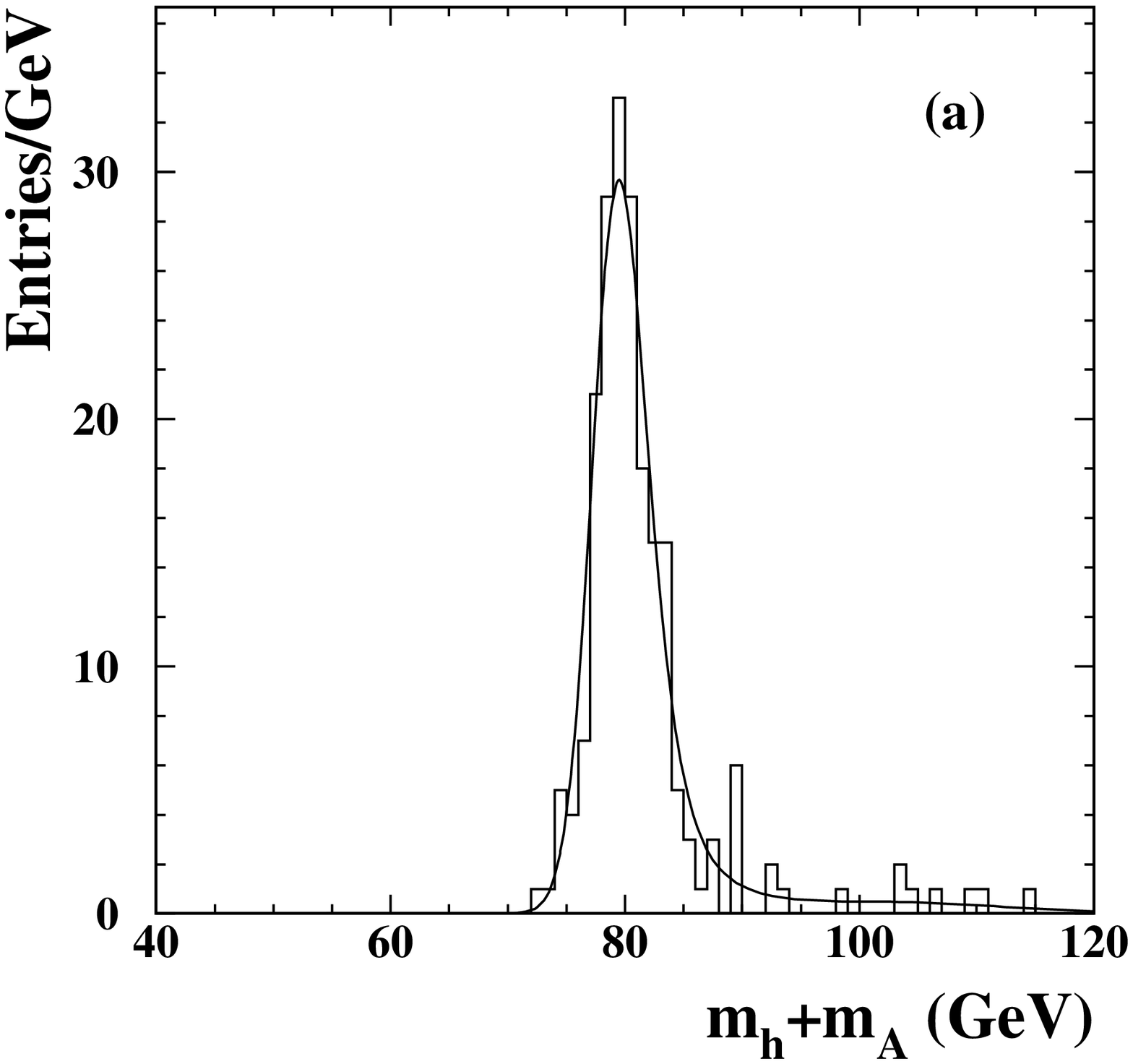,width=8.5cm}
\epsfig{file=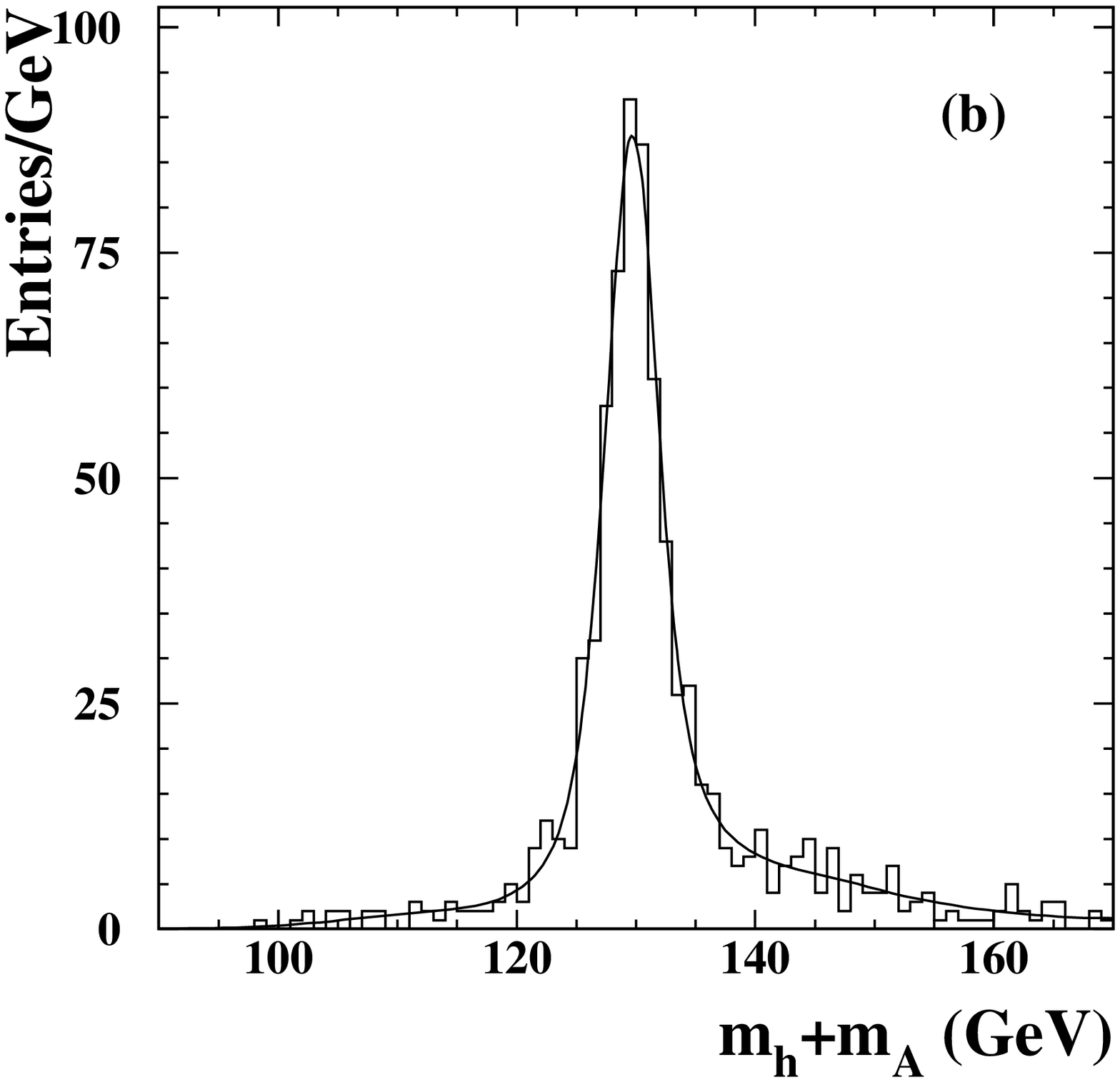,width=8.5cm}
} 
\centerline{
\epsfig{file=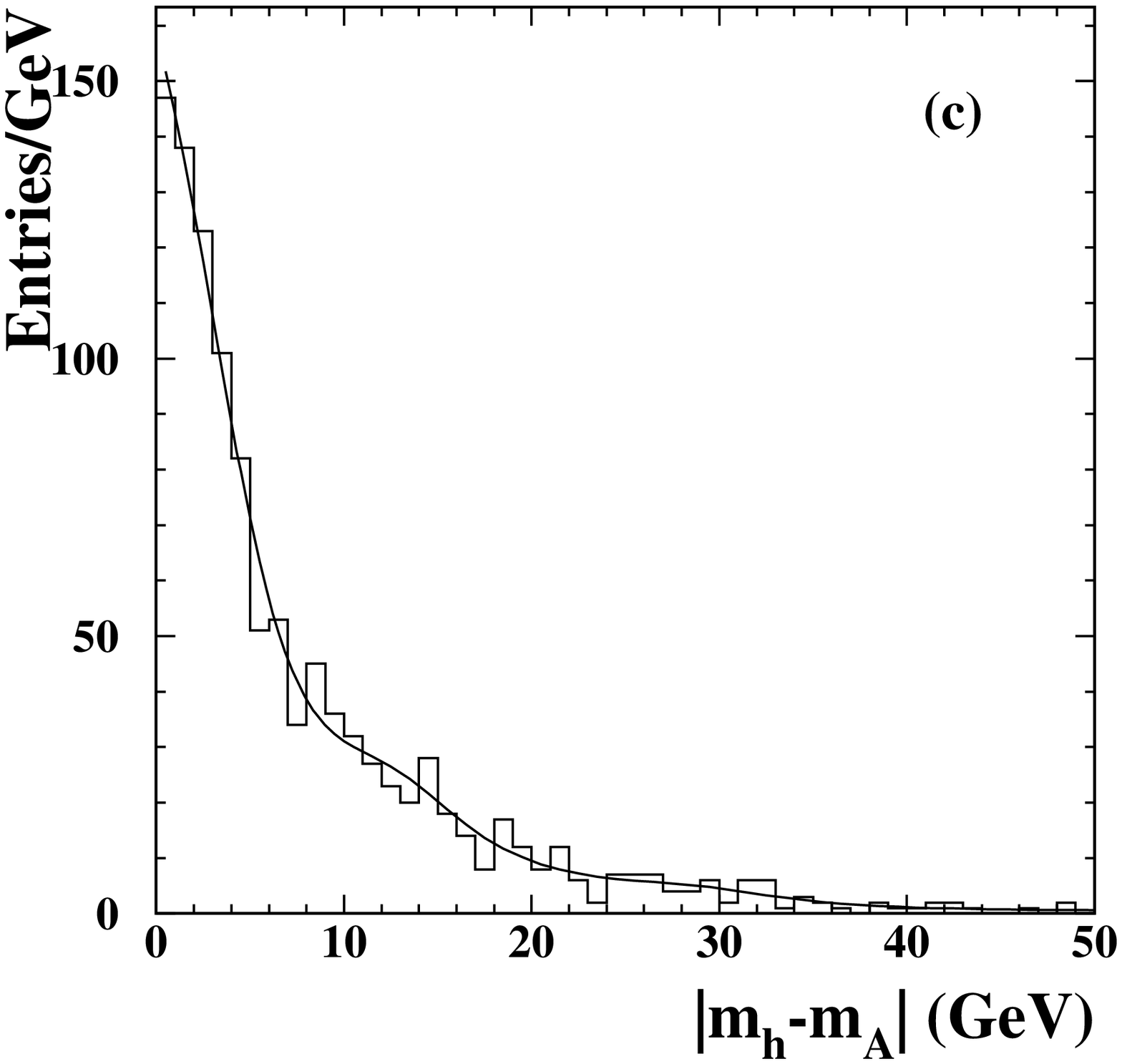,width=8.5cm}
\epsfig{file=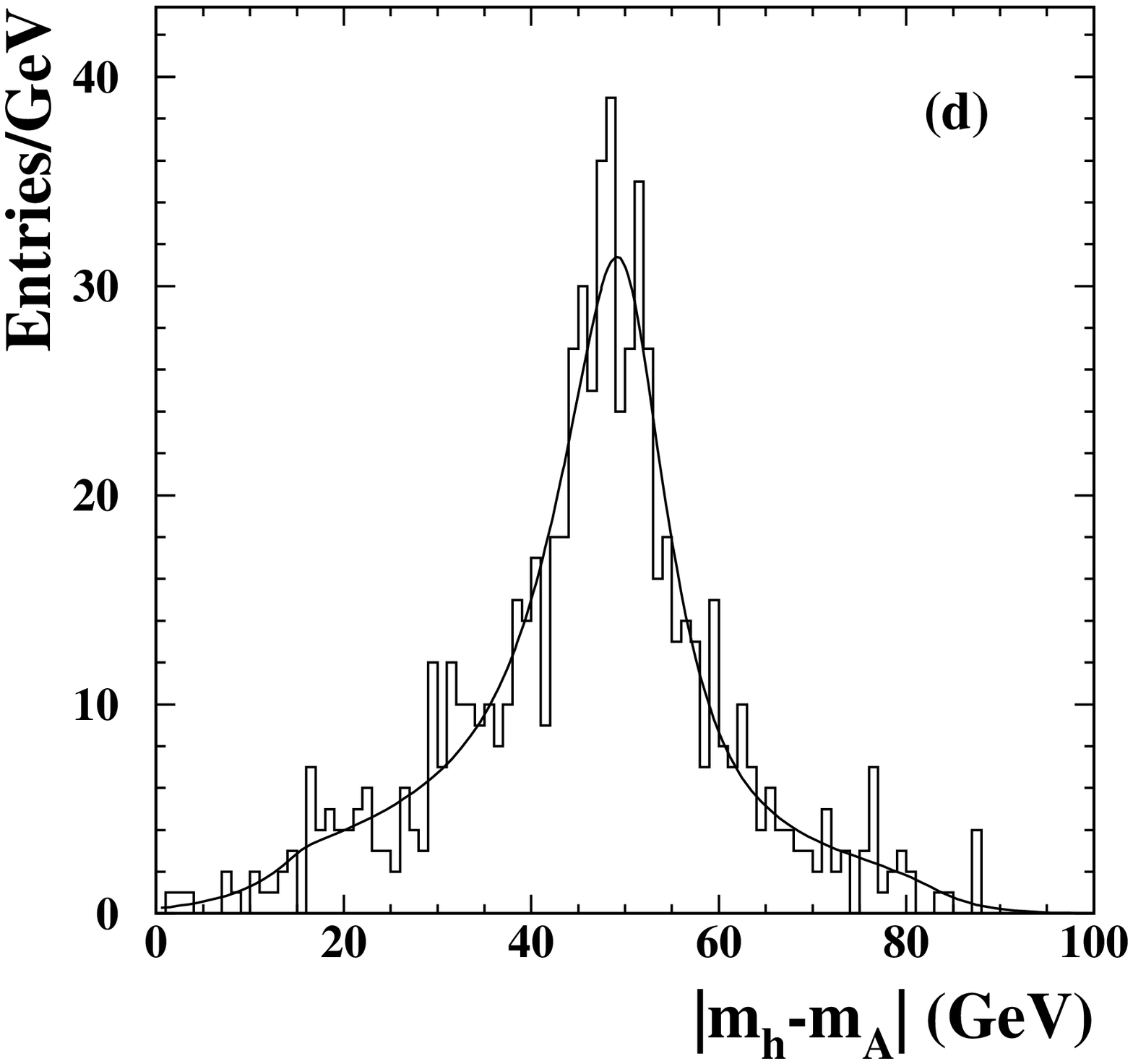,width=8.5cm}
}
\caption[]{\label{figure:ahbb_masshistos}\sl
          Mass resolution curves for \h\A\ra\bb\bb.
          (a) The mass sum of
          \mh\ and \mA\ for $\mh+\mA=80$~GeV.
          (b) The mass sum for $\mh+\mA=130$~GeV.
          (c) The mass difference of \mh\ and \mA\ for $\mh-\mA=0$.
          (d) The mass difference for $|\mh-\mA|=50$~GeV.
          Only the combination with the smallest difference between measured
          and true mass sum or difference is plotted.
          The histograms show the simulated
          distributions and the solid lines represent
          smooth fitted functions. 
}
\end{figure}

\begin{table}[tbp]
\begin{center}
\begin{tabular}{||c||c||c||c|c||c||}
\hline\hline
\multicolumn{6}{||c||}{\bf\boldmath $\sqrts=130-136$~GeV} \\\hline\hline
Cut & Data & Total Bkg. & $\qq(\gamma)$ & 4f & {\large $\epsilon$}{\small(\%)} 
\\
\hline\hline
(1)            & 1536   & 1489    & 1478    &  11     & 100 \\ 
(2)            &  445   &  496    &  489    &   7     &  97 \\
(3)            &   83   &   65.5  &   62.9  &   2.6   &  84 \\
(4)            &   64   &   48.1  &   45.8  &   2.3   &  83 \\
(5)            &   41   &   34.4  &   33.0  &   1.4   &  79 \\
(6)            &   28   &   28.2  &   26.9  &   1.3   &  75 \\
${\cal L}$ cut &    0   &    1.65$\pm$0.33
                                &    1.57 &   0.08  &  60 \\
\hline\hline
\multicolumn{6}{||c||}{\bf\boldmath $\sqrts=161$~GeV} \\\hline\hline
Cut & Data & Total Bkg. & $\qq(\gamma)$ & 4f & {\large $\epsilon$}{\small(\%)} 
\\
\hline\hline
(1)            & 1499   & 1399    & 1346    &  53     & 100 \\ 
(2)            &  394   &  378    &  352    &  26     &  90 \\
(3)            &   62   &   54.1  &   37.2  &  16.1   &  72 \\
(4)            &   49   &   40.6  &   26.1  &  14.4   &  71 \\
(5)            &   40   &   33.2  &   21.3  &  11.7   &  69 \\
(6)            &   33   &   30.5  &   19.6  &  10.9   &  63 \\
${\cal L}$ cut &    0   &    1.61$\pm$0.11
                                &    1.20 &   0.41  &  50 \\
\hline\hline
\multicolumn{6}{||c||}{\bf\boldmath $\sqrts=170-172$~GeV} \\\hline\hline
Cut & Data & Total Bkg. & $\qq(\gamma)$ & 4f & {\large $\epsilon$}{\small(\%)} 
\\
\hline\hline
(1)            & 1403   & 1254    & 1137    & 117     & 100 \\ 
(2)            &  369   &  367    &  299    &  68     &  88 \\
(3)            &   92   &   81.4  &   31.7  &  49.7   &  70 \\
(4)            &   77   &   69.7  &   21.8  &  47.9   &  69 \\
(5)            &   69   &   60.1  &   18.1  &  42.0   &  67 \\
(6)            &   64   &   56.3  &   16.3  &  40.0   &  60 \\
${\cal L}$ cut &    1   &    1.95$\pm$0.10
                                &    0.95 &   1.00  &  48 \\
\hline\hline
\end{tabular}
\caption[]{\label{table:ahbb_cuts}\sl
         Effect of the selection criteria on data,
         background (normalised to the integrated luminosity of the data) 
         and signal simulation ($\mh=\mA=55$~GeV) at the three centre-of-mass 
         energies for the signal channel $\h\A\ra\bb\bb$. The quoted errors 
         on the
         background are statistical.
}
\end{center}
\end{table}
The signal detection efficiencies are affected
by the following main uncertainties, all expressed as relative percentages.
The error from Monte Carlo statistics is typically $4-10$\%.
However, near the region where the limit is set larger Monte Carlo samples
were generated, and a fit is made through the grid of efficiencies
in the (\mh,~\mA) plane, so the efficiency at a certain mass point is
effectively based on higher statistics. This results in a statistical error
of approximately 1\% in the region near the limit.
The preselection requirements on \sqrtsp, $y_{34}$, and $C$
were varied by amounts equivalent to
the difference between the mean value of data and
Monte Carlo. This results in an uncertainty estimate for the modelling
of the preselection variables 
ranging from 3.5\% to 7.6\%, depending on the centre-of-mass energy.
The uncertainty associated with the requirement on $\chi^2$-probability
was determined to be 2\%.
To assess the systematic uncertainty from the b-tagging,
the impact parameter resolution was varied by 10\%,
and the b-hadron charged decay multiplicity was varied by
0.35 units~\cite{bmult}.
The resulting errors range from 3.6\% to 4.5\% depending on the
detector configuration (data taken in 1995 versus data taken in 1996.)
The effect of binning in the likelihood was investigated by using a linear
bin-to-bin interpolation. This gives rise to an uncertainty ranging from 0.0\%
to 0.4\%.
The theoretical uncertainty on the cross-section is estimated to be 1\%.
The systematic error on the integrated luminosity is 0.6\%~\cite{lumino}.
The total systematic uncertainty is calculated by adding the above
uncertainties in quadrature.
This gives uncertainties ranging from 5\% to 9\%
depending on the centre-of-mass energy.

The residual background estimate has a statistical error of 20\%, 7\%, 5\% at
the $130-136$, 161, $170-172$~GeV centre-of-mass energies, respectively.
The uncertainty from modelling the preselection was estimated in the same manner
as for the efficiency, giving uncertainties ranging from 3.3\% to 16\%.
The effects of impact parameter resolution and
binning in the likelihood
were also investigated in the same manner as for the efficiency.
Uncertainties in the impact parameter resolution
give errors between 15\% and 23\%.
The effect of modelling the variables used in the likelihood selection is
cross-checked by reweighting the Monte Carlo events in such a manner as to 
better approximate
the data distributions within the ability of the Monte Carlo samples to
represent the data. It should be noted that the agreement in these
distributions before the reweighting procedure already gives a good
$\chi^2$ probability.
This procedure yields errors ranging from 0.4\% to 22\%, and the corresponding
error estimates on the signal detection efficiencies are negligible.
In all cases errors due to the tracking resolution uncertainty
were larger than those due to the
reweighting procedure, and therefore the former are
taken as the systematic errors for these two checks.
The effect of binning on the likelihood for the background was estimated to
range from 0.1\% to 4.0\%.
The effect of modelling the Standard Model physics was investigated by
comparing different Monte Carlo event generators.
For the \Z/$\gamma^*$\ra\qq\ background the results using the PYTHIA generator
were compared to those using the HERWIG generator~\cite{herwig}.
For the four-fermion background the EXCALIBUR and grc4f generators were
compared.
The uncertainties for the physics modelling range from 3.6\% to 14\%.
The Monte Carlo generators have an uncertainty on the calculated 
cross-sections of 0.5\%.
Including the error on the integrated luminosity of 0.6\%~\cite{lumino}, 
total relative uncertainties of 27\%, 28\%, 24\% are assigned to the residual
background estimates at center-of-mass energies of $130-136$, 161, 
$170-172$~GeV, respectively. 

To make use of the mass information in the calculation of exclusion limits, the
hypothetical Higgs masses have to be determined. The four jets can be 
combined in three ways
into two jet-pairs.
The invariant masses of all jet-pair combinations of an event are
considered.
They are calculated using a kinematic fit assuming energy and momentum
conservation
(4C-fit).
The mass distributions have a non-Gaussian shape.
Since \h\ and \A\ cannot be distinguished, the mass
difference $\Delta M=|\mA-\mh|$ and the sum $M=\mA+\mh$
are considered instead of the masses themselves.
Figure~\ref{figure:ahbb_masshistos} shows examples of
these mass distributions at $\sqrts=172$~GeV.
The resolution in $M$ is roughly 3~GeV and does not vary
with mass. The tails of the distribution of $M$, caused by
the reconstructed jets not corresponding to the parton final states,
exist largely above the mass peak for $M=80$~GeV, whereas for $M=130$~GeV
they are present both above and below the nominal mass value.
The shape of $\Delta M$ is independent of the mass sum.
Its resolution is also approximately 3~GeV and the
distribution shows substantial tails.
Smooth functions fitted to these mass distribution histograms are used
for exclusion limit calculations instead of the histograms themselves.

\subsection{\boldmath The channel \h\A\ra\tautau\qq}
The \tautau\qq\ final state can be produced via the processes 
\ee\ra\h\A\ra\tautau\qq\ and \qq\tautau.
These processes are characterised by a pair of tau-leptons and
a pair of energetic hadronic jets.
The backgrounds are predominantly from 
(\Z$/\gamma)^*$\ra\qq$(\gamma)$ and four-fermion processes.
The search in these channels was restricted to the data recorded at the 
centre-of-mass energies $\sqrts=161$ and $170-172$~GeV.

The selection begins with the identification of tau-leptons, identical
to that in~\cite{smpaper}, using three algorithms which 
classify each tau-lepton candidates as decaying into an electron, a muon, or
hadrons.

\begin{figure}[p]
\centerline{\hfill
\epsfig{file=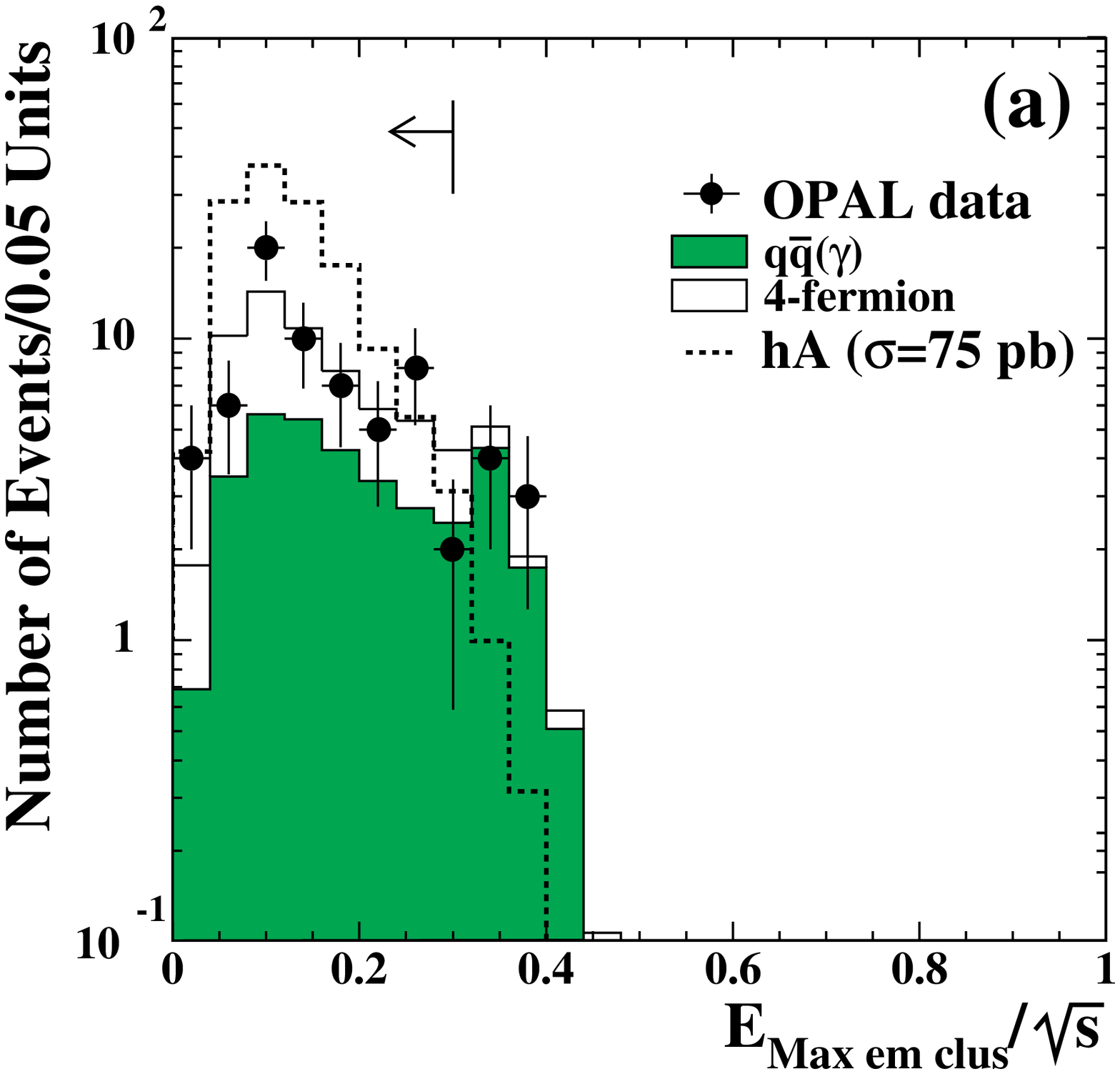,width=8.5cm}\hfill
\epsfig{file=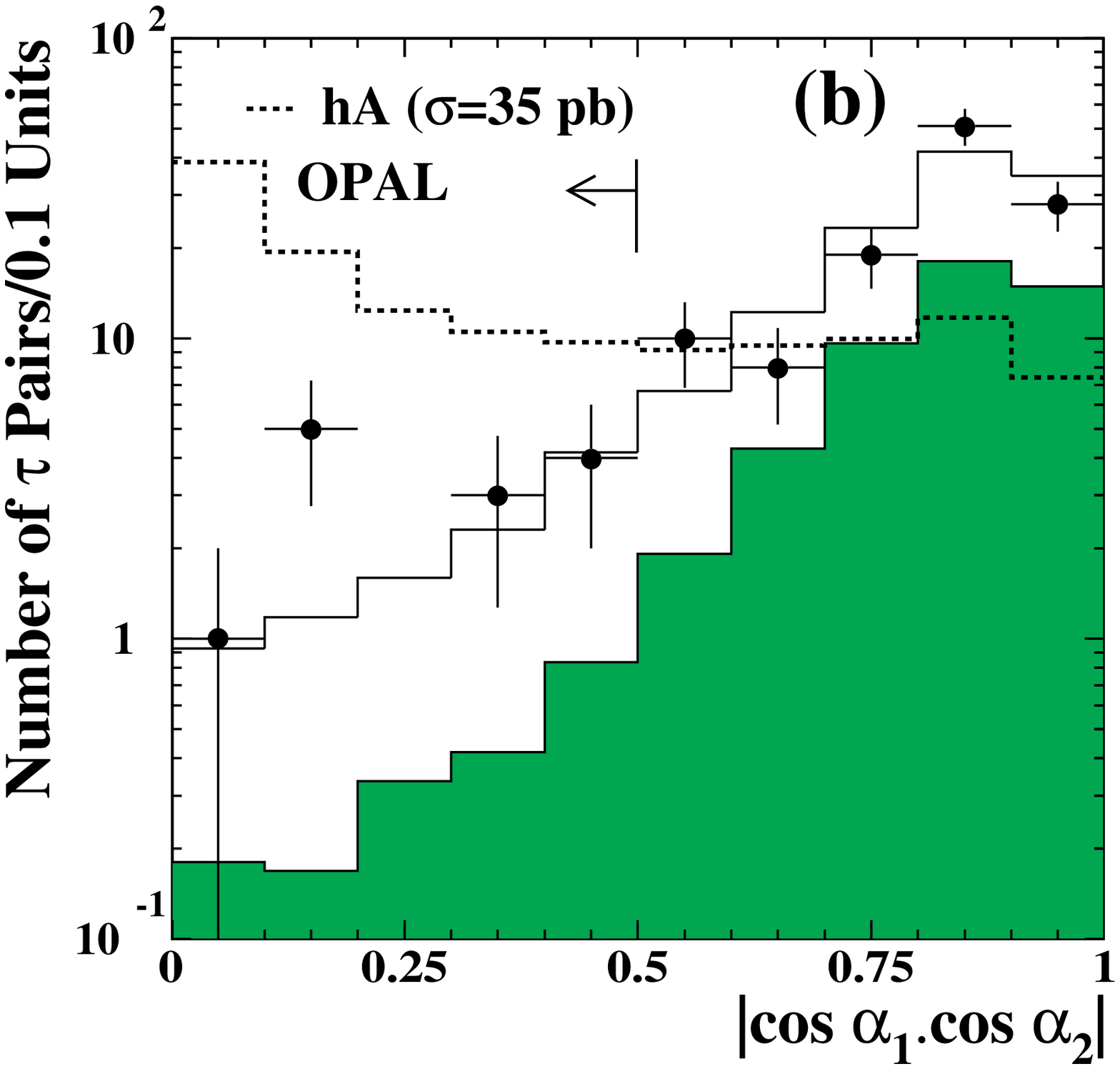,width=8.5cm}\hfill
}
\caption[]{\label{tautau:f1}\sl
         Selection variables relevant for the \h\A\ra\qq\tautau\ analysis.
         (a) The energy of the most energetic electromagnetic
              cluster scaled by $\sqrt{s}$ after cut (3).
         (b) The pairwise isolation parameter (see text) after cut (4).
         The $\sqrts=161$ and $170-172$~GeV data are added together.
         The points represent the data.
         The shaded histograms show the $\qq$ background and the open
         histograms show the four-fermion background, normalised to 
         the integrated luminosity of the data. Two-photon processes are not
         included.
         The dashed histograms are simulated signals for
         $\mh=\mA=55$~GeV, where the displayed production cross-sections
          have been chosen for visibility.
          The background simulations are normalised to the integrated
         luminosity of the data.
         Arrows indicate domains accepted by the selection.
}
\end{figure}

In the selection that follows, the tau-lepton direction is approximated
by that of the visible decay products.
If two tau-lepton candidates 
have momentum vectors separated by less than 23$^\circ$,
one being identified as a leptonic (electron or muon) decay and one as hadronic,
the leptonic decay is chosen. 
The following selection,
which is identical to that in~\cite{smpaper} up to and including (4),
was made:
\begin{itemize}
\item[(1)]  
Events are required to have at least two tau-lepton
candidates, each with electric charge $|q|=1$, and at least nine charged tracks.

\item[(2)]
  Most of the two-photon and \ee\ra(\Z$/\gamma)^*$ background events are
  eliminated by requiring that 
  the energy in the forward detector, gamma catcher, and 
  silicon-tungsten luminometer be
  less  than 4, 10, and 10~GeV, respectively,
  that $|\cos\theta_{\rm miss}| < 0.97$ and that $P^T_{\rm vis} > 3$~GeV,
  where $\theta_{\rm miss}$ is the polar angle of the missing momentum 
  vector and
  $P^T_{\rm vis}$ is the total transverse momentum of the event.
  In addition, the scalar sum of all track and cluster transverse momenta is
  required to be larger than 40~GeV.
  Accelerator-related backgrounds in the forward detectors which have not
  been fully simulated are taken into account via small corrections to the
  signal detection efficiencies.
  
\item[(3)]
The remaining (\Z$/\gamma)^*$\ra\qq$(\gamma)$ background is partially
suppressed by
requiring that events contain at least four jets, reconstructed using
the cone algorithm as in~\cite{smpaper}, where single 
electrons and muons from tau-lepton decays are allowed to be recognised 
as low-multiplicity ``jets". 
Events with an energetic isolated photon are removed, where an energetic 
isolated
photon is defined as an electromagnetic cluster
with energy larger than 15~GeV and no track within a cone
of $30^\circ$ half-angle.

\item[(4)]
To suppress the process \WW\ra$\ell\nu$\qq$^{\prime}$,
events are rejected if they contain any track or cluster with 
energy exceeding $0.3\sqrt{s}$. Fig.~\ref{tautau:f1}(a) shows
the distribution of the energy of the most energetic electromagnetic cluster
scaled by $\sqrt{s}$, prior to this cut, for the data recorded at 
$\sqrts=161-172$~GeV, the expected backgrounds, and
a simulated Higgs boson signal with $\mh=\mA=55$~GeV.

\item[(5)]
The three tau-lepton identification algorithms
identify 2.3 $\tau$ candidates per signal event on average. 
Fake candidate pairs are removed by requiring that the sum of the track charges
be zero and that the candidates satisfy a pairwise isolation requirement,
$|\cos\alpha_i\cdot\cos\alpha_j| < 0.5$, where $\alpha_{i}$ is the
angle between the direction of the $i$-th $\tau$ candidate and that of the
nearest track not associated with it.
The indices $i,j$ run over all $\tau$ candidates with $i\neq j$.
Fig.~\ref{tautau:f1}(b) shows the distribution of this variable for pairs
of tau-lepton candidates, for the data recorded at $\sqrts=161-172$~GeV,
the expected backgrounds, and
a simulated Higgs boson signal with $\mh=\mA=55$~GeV.
In those instances where
more than one candidate pair passes the selection,
the pair whose members have the lowest track multiplicity is chosen.
If no distinction can be made, the candidate pair is chosen whose members have
the highest value of the isolation parameter $R_{em+cd}^{11/30}$. 
Here $R_{em+cd}^{11/30}$ is the ratio of the sum of the electromagnetic cluster
energies and charged track momenta within a cone of $11^\circ$ half-angle
centered on the $\tau$ candidate axis
to that within a $30^\circ$ half-angle cone.
\end{itemize}
 
The hadronic part of the event, obtained by 
excluding the tracks and clusters from the selected $\tau$ candidate pair,
is then split into two jets using the Durham algorithm~\cite{durham}. 
The invariant
masses of the tau-lepton pair, $m_{\tau\tau}$, and of the  hadron jet-pair,
$m_{\rm had}$, are calculated using only the tau-lepton and jet 
directions and 
requiring energy and momentum conservation. 
The resolutions of the mass distributions, later used in the
calculation of exclusion limits (see Section~\ref{section:combchan}),
are determined from signal events by fitting a Gaussian 
distribution
in an interval which excludes non-Gaussian tails, resulting in a typical 
error of 6 GeV on the reconstructed masses. 

\begin{figure}[p]
\centerline{
\epsfig{file=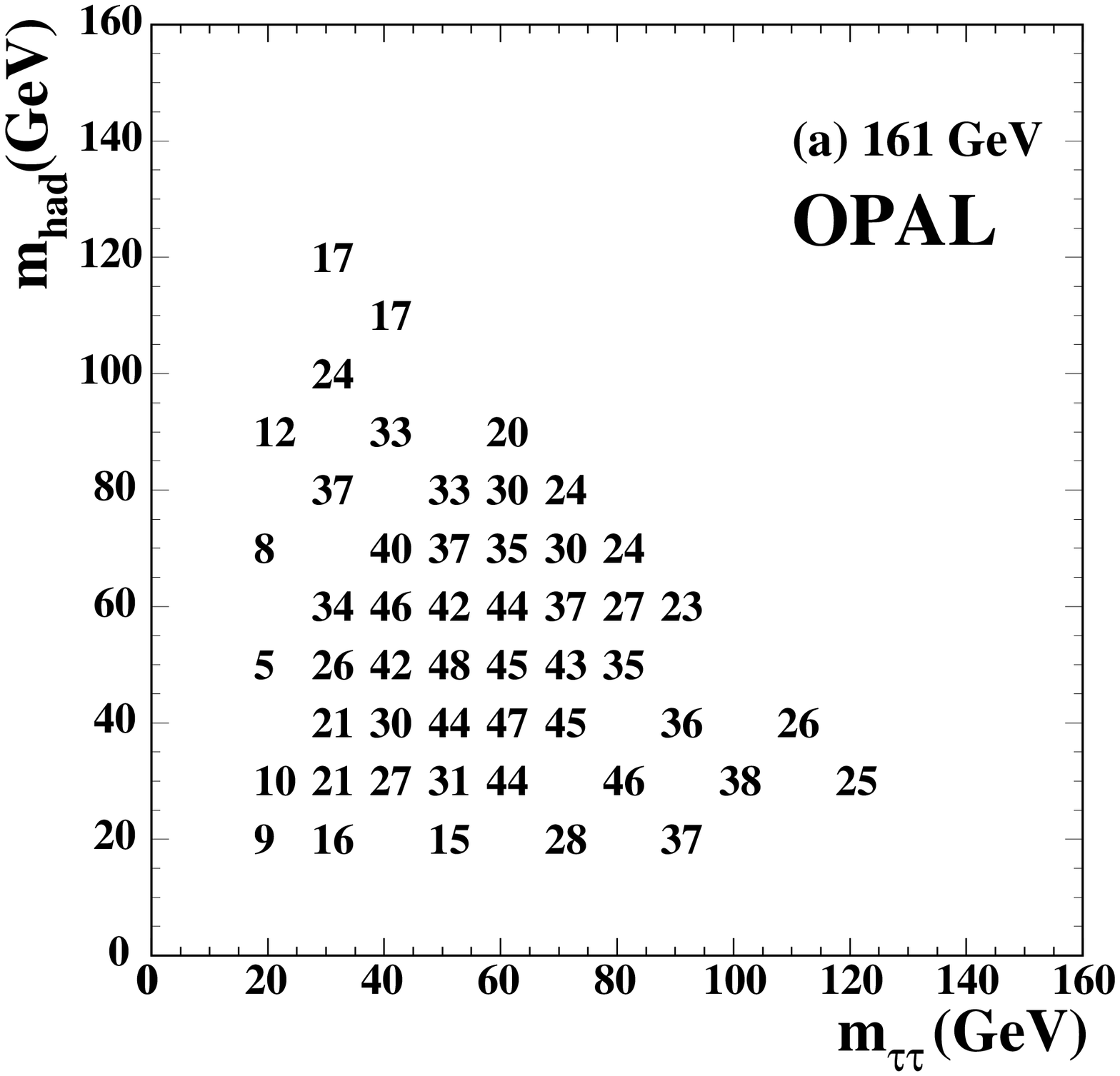,width=8.5cm}\hfill
\epsfig{file=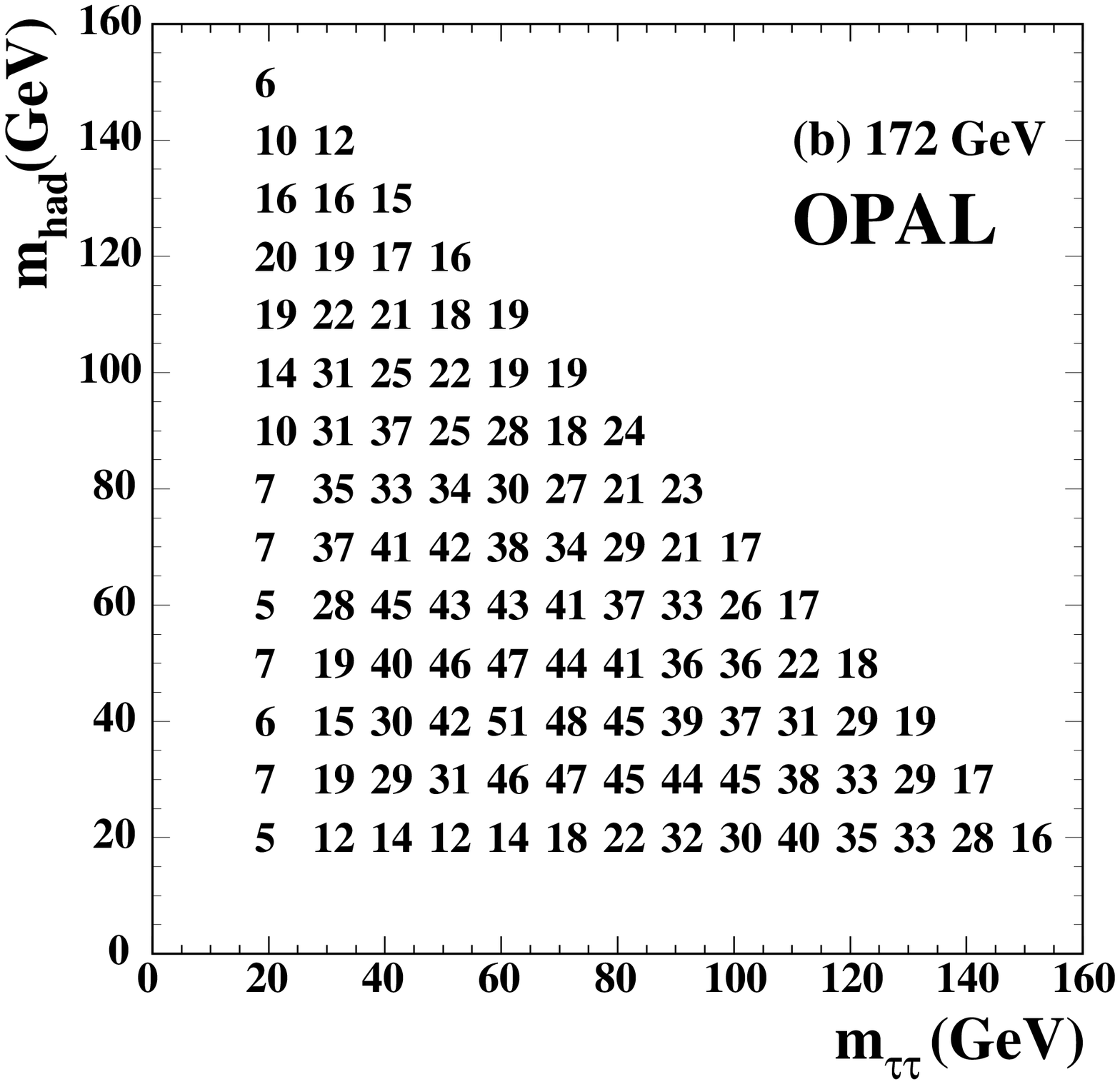,width=8.5cm}
}
\caption[]{\label{tautau:f2}\sl
         Efficiencies in percent for the signal process \h\A\ra\qq\tautau\ 
         in the ($m_{\tau\tau}$,$m_{\rm had}$)
         plane at $\sqrt{s}=161$ and 172 GeV, where $m_{\tau\tau}$ and
         $m_{\rm had}$ are the invariant masses of the tau-lepton pair
         and hadron jet-pair, respectively.
}
\end{figure}

\begin{figure}[htbp]
\centerline{
\epsfig{file=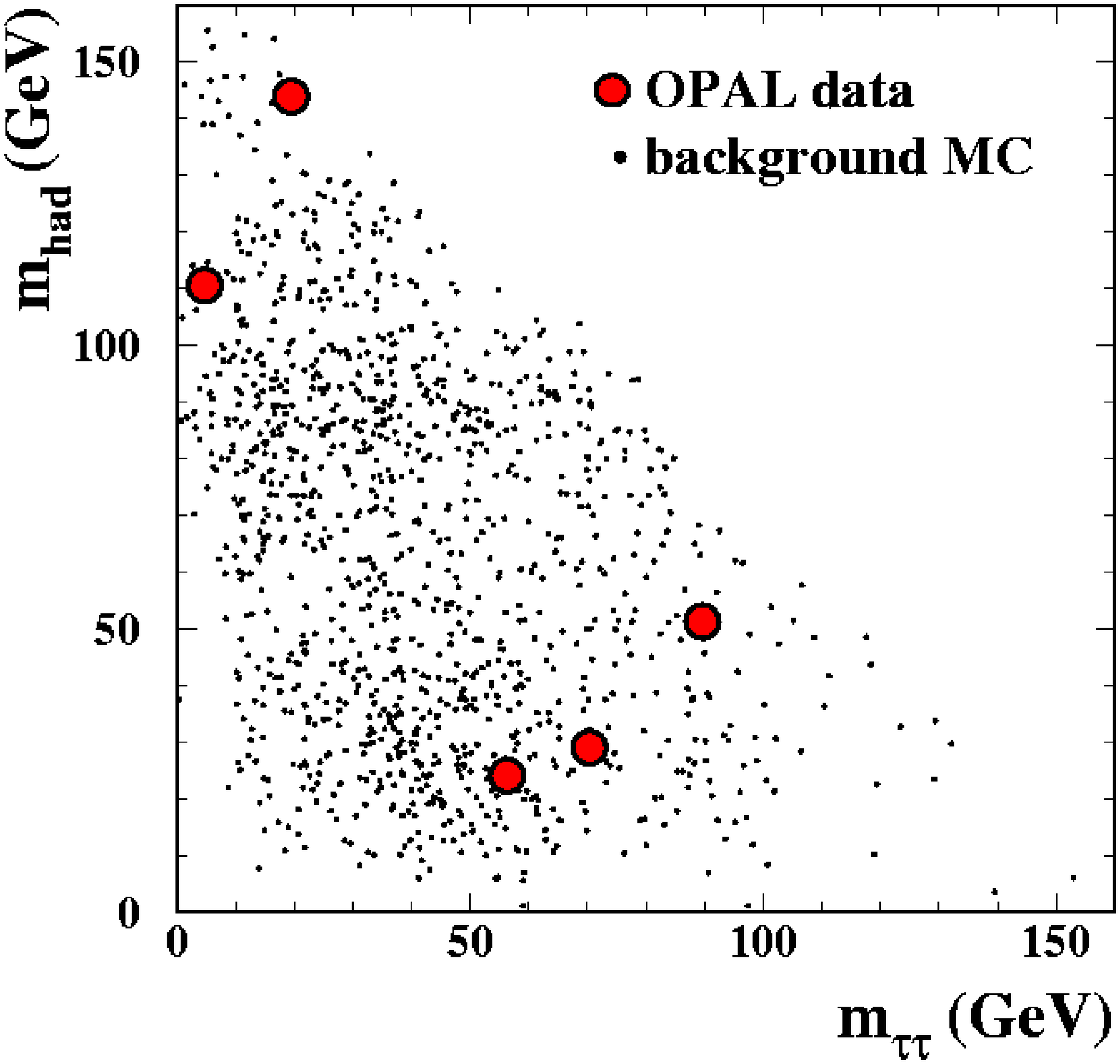,width=8.5cm}
}
\caption[]{\label{tautau:f4}\sl
         The position in the ($m_{\tau\tau}$,$m_{\rm had}$) plane of
         the \h\A\ra\tautau\qq\ channel candidates,
         superimposed on that of the arbitrarily normalised
         expected Standard Model background, for $\sqrt{s}=161$ and
         $170-172$ GeV.
}
\end{figure}

The numbers of observed
and expected 
events after each stage of the selection are given in Table~\ref{tautau_t1}
for $\sqrts=161$~GeV and $170-172$~GeV.
The agreement between the data and the expected background within the
limited statistics demonstrates the adequate modelling of the selection
criteria.
The detection efficiency for a Higgs boson signal with $\mh=\mA=55$~GeV
is also given. Five events survive the selection while the background is 
estimated to be
1.29 events at $\sqrts=161$~GeV and 2.54 events at $\sqrts=170-172$~GeV. 
Figure~\ref{tautau:f4} shows the positions of the surviving events in the
($m_{\tau\tau}$,$m_{\rm had}$) plane superimposed
on the expected background. 
\begin{table}[tp]
\begin{center}
\begin{tabular}{||c||c||c||c|c|c||c||}
\hline\hline
\multicolumn{7}{||c||}{\bf\boldmath $\sqrts=161$~GeV} \\\hline\hline
Cut & Data & Total Bkg. & $\qq$ & 4f & $\gamma\gamma$ &
{\large $\epsilon$}{\small (\%)} \\
\hline\hline
(1) & 402  & 398.0  & 92.8   & 18.3  & 286.9   & 67 \\
(2) &  45  & 44.7   & 30.9   & 13.3  & 0.5     & 62  \\
(3) &  32  & 30.1   & 19.6   & 10.1  & 0.4     & 61 \\
(4) & 26   & 22.7   & 14.3   & 8.0   & 0.4     & 57  \\
(5) & 0 & 1.29$\pm$0.24 & 0.41 & 0.88 & $<$0.21 & 46  \\
\hline\hline
\multicolumn{7}{||c||}{\bf\boldmath $\sqrts=170-172$~GeV} \\\hline\hline
Cut & Data & Total Bkg. & $\qq$ & 4f & $\gamma\gamma$ &
{\large $\epsilon$}{\small (\%)} \\
\hline\hline
(1) & 358  & 306.9  & 75.2   & 36.8  & 194.9    & 67 \\
(2) &  50  & 55.1   & 23.6   & 31.2  & 0.3     & 63  \\
(3) &  37  & 40.1   & 15.2   & 24.7  & 0.2     & 61 \\
(4) &  31  & 32.3   & 11.0   & 21.3  & $<$0.22 & 57 \\
(5) & 5    & 2.54$\pm$0.24  & 0.21   & 2.33 & $<$0.22 & 46  \\
\hline\hline
\end{tabular}
\caption[]{\label{tautau_t1}\sl
         Effect of the selection criteria on data,
         background (normalised to the integrated luminosity of the data) 
         and signal simulation ($\mh=\mA=55$~GeV) at the two centre-of-mass 
         energies for the signal channel \h\A\ra\tautau\qq. 
         The quoted errors on the background are statistical.}
\end{center}
\end{table}
The detection efficiencies for various values of $m_{\tau\tau}$ and 
$m_{\rm had}$, where $m_{\tau\tau}$ is the mass of the object (\h\ or
\A) decaying into the $\tau$-pair and $m_{\rm had}$ is the mass of the 
object decaying into the jet-pair, are given in
Fig.~\ref{tautau:f2}. 

The efficiencies are affected by the following
uncertainties:
Monte Carlo statistics, typically 2.2\%;
uncertainty in the tau-lepton identification efficiency, 3.6\%;
uncertainties due to modelling of selection variables excluding
the tau-lepton identification, 6.3\%;
uncertainties in the modelling of fragmentation and hadronisation, 2.4\%;
and uncertainty on the integrated luminosity, 0.6\%~\cite{lumino}.
Taking these uncertainties as independent and adding them in quadrature
results in a total systematic uncertainty  of 7.9\%  (relative errors).
The uncertainties due to the modelling of selection variables, including
those used in tau-lepton identification, were estimated by displacing the
cut values by an amount corresponding to the difference between the means
of the data and background Monte Carlo distributions. Using the same techniques,
the uncertainty in the number of expected background events was 
estimated to be 32\%, dominated by the systematic uncertainty associated with
the requirement on $|\cos\alpha_i\cdot\cos\alpha_j|$, which is steeply 
falling for
background at the position of the cut.
Because of this predominance of one single selection variable in the
uncertainty on the number of expected background events, 
the background is not subtracted when computing limits.

\subsection{\boldmath The channel \h\A\ra\bb\bb\bb}
\label{section:ah6b}
When $2\mA\leq\mh$ the decay  
\h\ra\A\A\ is allowed and may be dominant. In
these cases the process \ee\ra\h\A\ra\A\A\A\
can have a large branching ratio in the final state \bb\bb\bb.
Due to the presence of six b
quarks in the expected signature, the events are characterised by a
large number of jets and a large charged track multiplicity.
To reduce backgrounds b-tagging plays a crucial role.
The main background is from (\Z$/\gamma)^*$\ra{\bb}g($\gamma$) 
with hard gluon emission. At 161 and $170-172$~GeV
four-fermion processes also result in a small
contribution. Backgrounds from two-photon
processes are reduced to a negligible level by the event selection.

\begin{figure}[p]
\centerline{
\epsfig{file=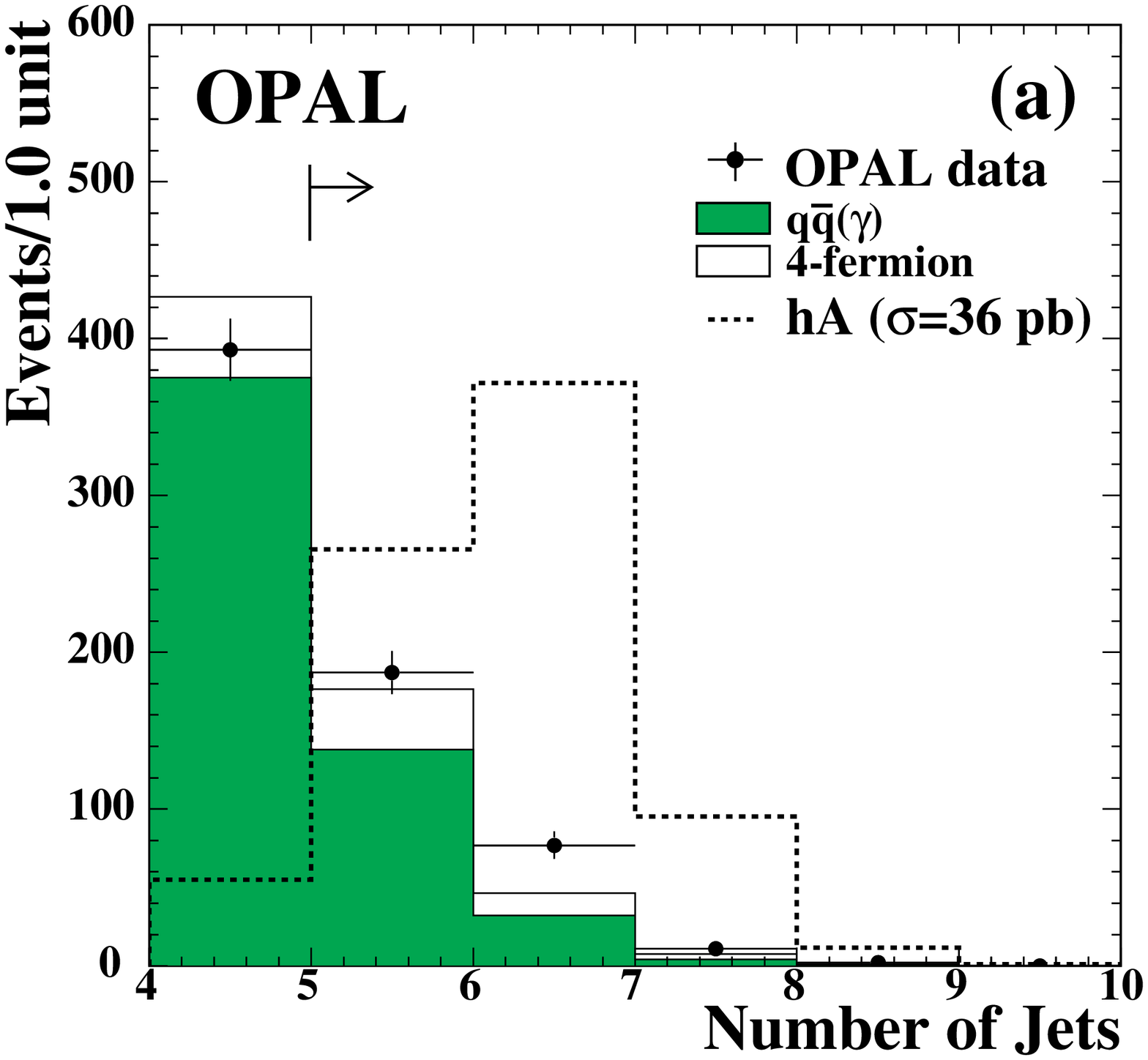,width=8.5cm} \hfill
\epsfig{file=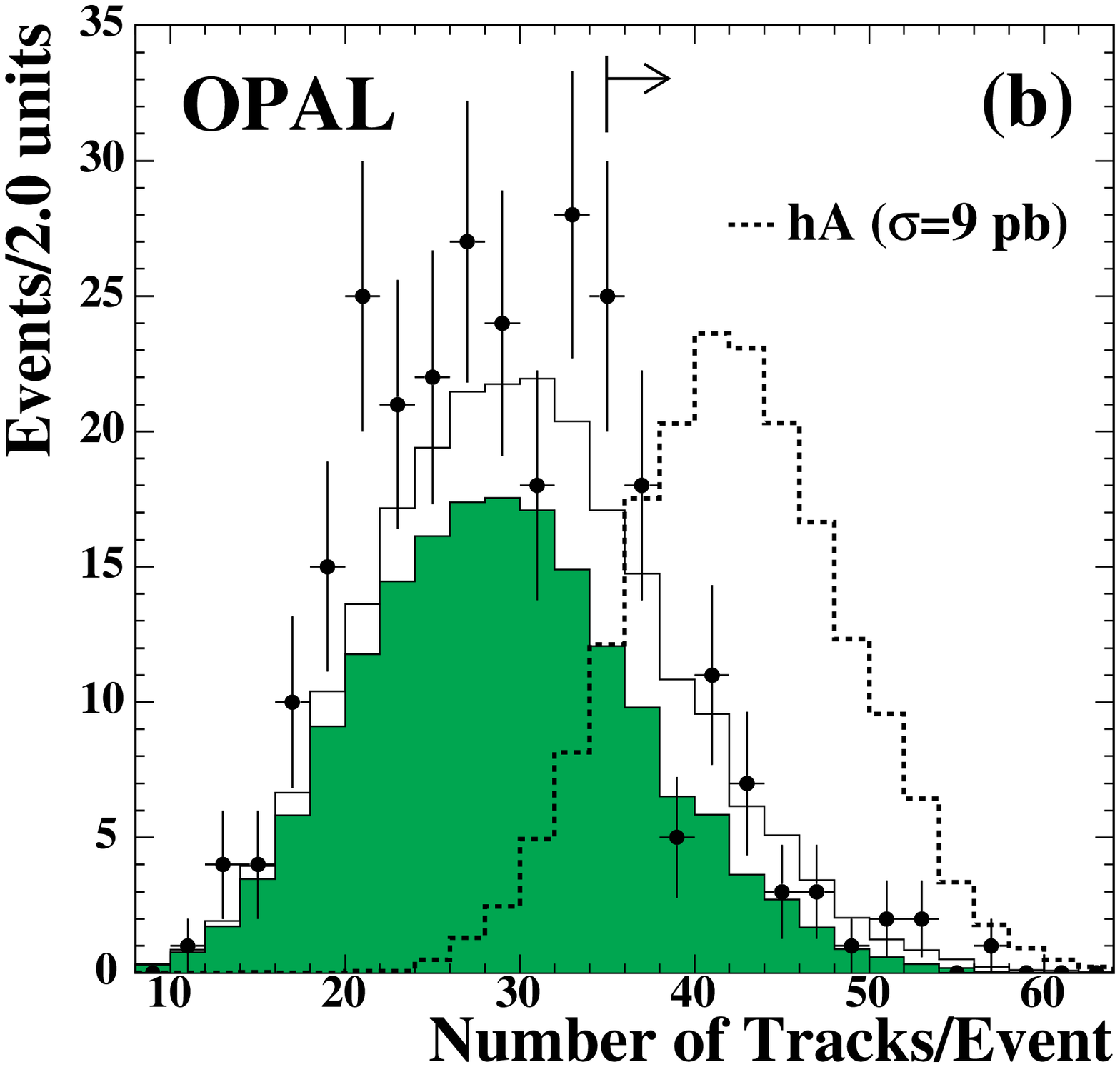,width=8.5cm}
}
\centerline{
\epsfig{file=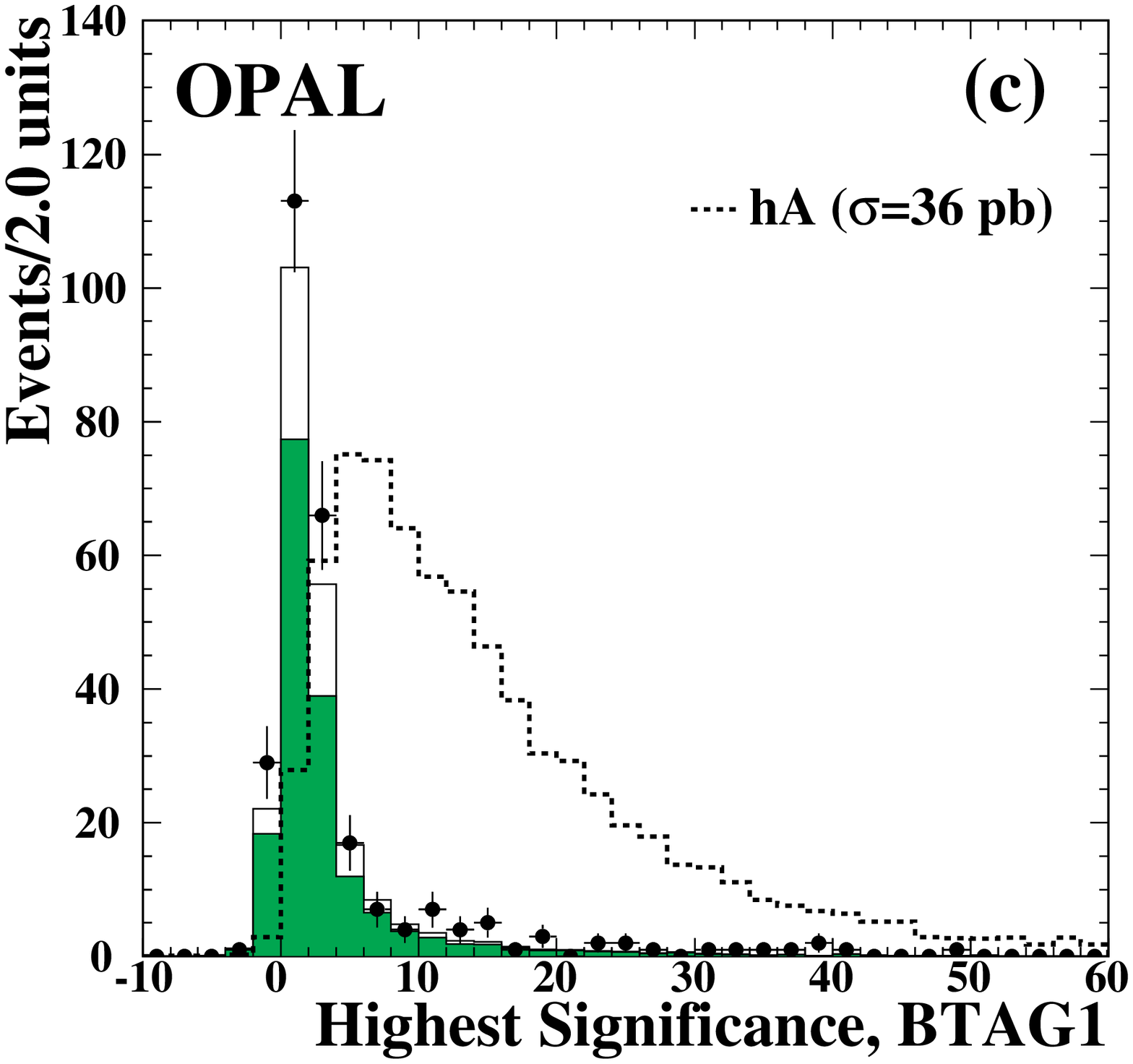,width=8.5cm} \hfill
\epsfig{file=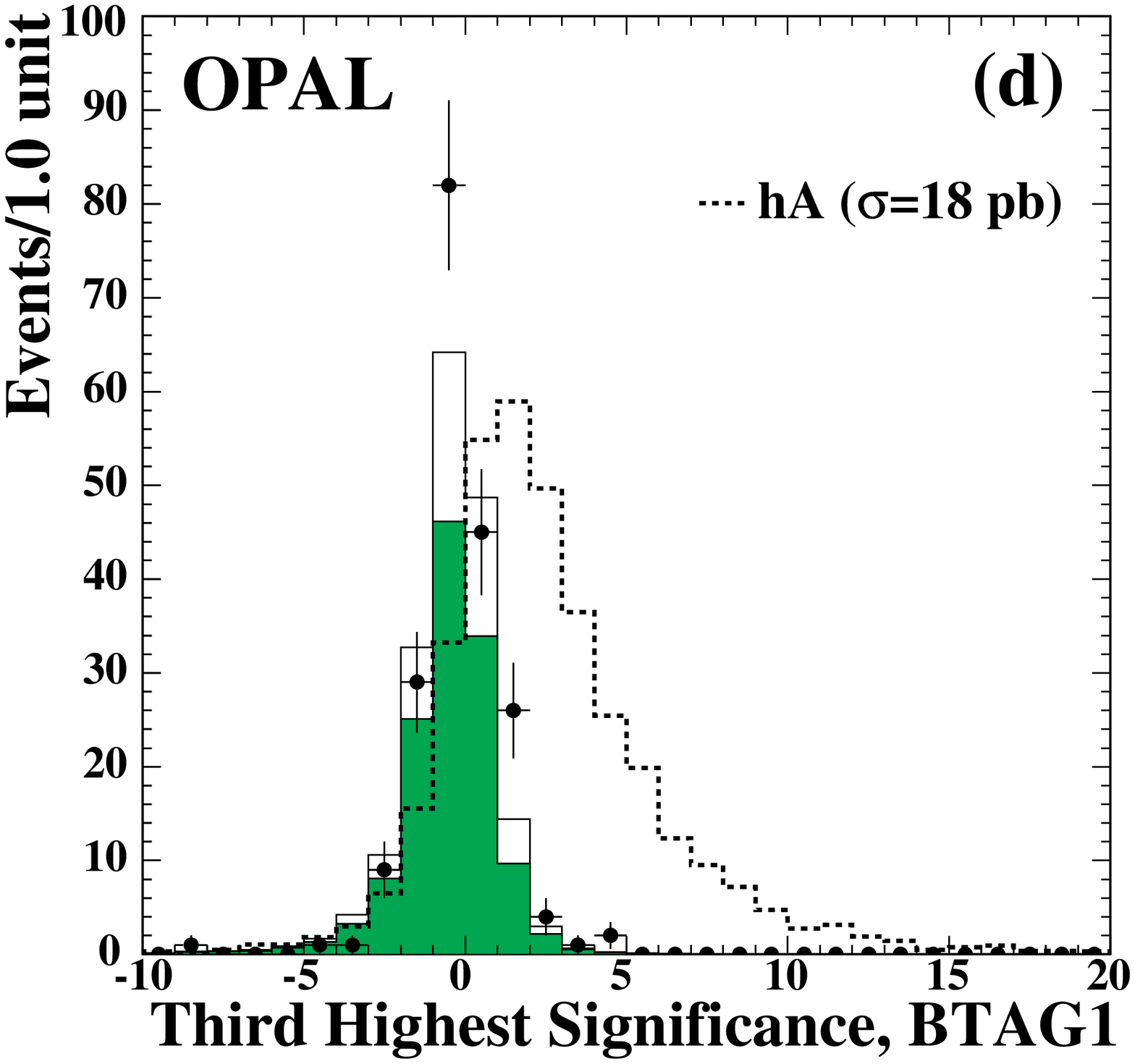,width=8.5cm}
}
\caption[]{\label{figure:ah6b_cuts}\sl
         Selection variables relevant 
         for the \h\A\ra\bb\bb\bb\ analysis.
         (a) The number of reconstructed jets using the Durham
         algorithm with $y_{\mathrm{cut}}=0.0015$ after multihadronic event
         selection and application of cut (3),
         for events with four or more jets.
         (b) The charged track multiplicity after cut (3).
         (c) The highest secondary vertex significance for the BTAG1
         algorithm (see Section~\ref{btag} for definition) after cut (3).
         (d) The third-highest secondary vertex significance 
         for the BTAG1 algorithm after cut (3).
         The distributions  are added for the centre-of-mass energies
         $130-136$~GeV, 161~GeV and $170-172$~GeV. Data are indicated
         by points.
         The shaded histograms show the $\qq$($\gamma$) background, 
         and the open
         histograms show the four-fermion background, normalised to 
         the integrated luminosity of the data. Two-photon processes are not
         included.
         The dashed lines represent a simulated signal with $\mh=60$~GeV,
         $\mA=30$~GeV for $\sqrts=171$~GeV, where the displayed production 
         cross-sections have been chosen for visibility.
         Arrows indicate domains accepted by the selection.
}
\end{figure}

The initial event selection follows that for \h\A\ra\bb\bb\, described
in Section~\ref{section:ah}. The data samples from $\sqrts=130-172$~GeV
are all used. The following requirements are made:

\begin{itemize}
\item[(1)]
      The events must qualify as hadronic final states
      as described in~\cite{tkmh}.

\item[(2)]
      Jets are reconstructed using the Durham~\cite{durham} 
      algorithm with $y_{\mathrm{cut}}=0.0015$. 
      Events having five or more jets are retained.

\item[(3)]
      As in the \h\A\ra\bb\bb\  analysis, the radiative process 
      \ee\ra(\Z/$\gamma)^*$\ra\qq$\gamma$ is largely
      eliminated by a requirement on the effective centre-of-mass energy,
      in this case $\sqrtsp>110$~GeV.
      The distribution of the number of jets for events with four or 
      more jets after application of this requirement is shown in 
      Fig.~\ref{figure:ah6b_cuts}(a) for data and simulated background and
      signal.

\item[(4)] 
      The number of charged tracks for the signal process is quite
      large, but the backgrounds from (\Z$/\gamma)^*$\ra\qq($\gamma$) 
      and four-fermion processes have long tails 
      extending to high multiplicities (see
      Fig.~\ref{figure:ah6b_cuts}(b)). 
      Candidate events are required to have more than 35 charged tracks.

\item[(5)]
      Three or more jets are required to show evidence for b~quark
      flavour, using the b-tagging algorithms discussed in 
      Section~\ref{btag} (BTAG1~\cite{btag1} and BTAG2~\cite{btag2}). 
      The secondary vertices used in these methods are in addition required to
      have at least two tracks each with two $\mu$VTX $r$-$\phi$ hits
      assigned~\cite{smpaper}.
      The decay length significances $S$, ordered in decreasing
      significance, must be successively $S>$ 8, 4, 3 for BTAG1 
      vertices and  $S>$ 8, 5, 3 for BTAG2 vertices. Events must 
      pass either the BTAG1 requirements or the BTAG2 requirements. The 
      distributions of the significance of the most significant and
      the third most significant vertex for the BTAG1 algorithm are shown in
      Fig.~\ref{figure:ah6b_cuts}(c) and (d) for events passing requirement (3).

\end{itemize}

Distributions of the variables relevant for the selection are shown in
Fig.~\ref{figure:ah6b_cuts}(a)--(d) for the $130-136$~GeV, 161~GeV 
and $170-172$~GeV data combined. Within the limited statistics,
the agreement of the data with the Monte Carlo simulations is
reasonable, except that it was found that the PYTHIA simulation of
\ee\ra(\Z/$\gamma)^*$\ra\qq$\gamma$ underestimates the number of events with
five or more jets. This discrepancy arises only in the
\Z\ radiative return peak. The loose requirement on \sqrtsp\ was made to 
decrease the number of \Z\ radiative return events in order to 
reduce the
effect of this uncertainty on the final selection, while maintaining a high
signal efficiency. This discrepancy, which would be eliminated with a harder 
requirement on \sqrtsp, vanishes subsequently with the application of the
remainder of the selection. 

\begin{table}[p]
\begin{center}
\begin{tabular}{||c||c||c||c|c||c||}
\hline\hline
\multicolumn{6}{||c||}{\bf\boldmath $\sqrts=130-136$~GeV} \\\hline\hline
Cut & Data & Total Bkg. & $\qq(\gamma)$ & 4f & {\large $\epsilon$}{\small(\%)} \\
\hline\hline
(1) & 1536  & 1489              & 1478     & 11    & 98 \\
(2) &  278  &  232              &  229     &  2.8  & 90 \\
(3) &   83  &   65.2            &   63.1   &  2.1  & 77 \\
(4) &    7  &   10.4            &   10.1   &  0.3  & 63 \\
(5) &    0  &   $<$0.08         &    0     &  0    & 22 \\
\hline\hline
\multicolumn{6}{||c||}{\bf\boldmath $\sqrts=161$~GeV} \\\hline\hline
Cut & Data & Total Bkg. & $\qq(\gamma)$ & 4f & {\large $\epsilon$}{\small(\%)} \\
\hline\hline
(1) & 1499  & 1399              & 1346    & 53.2   & 100 \\ 
(2) &  220  &  175              &  158    & 17.1   &  91 \\
(3) &   95  &   75.4            &   59.8  & 15.6   &  84 \\
(4) &   14  &   17.0            &   11.3  &  5.7   &  71 \\
(5) &    0  &   0.23$\pm$0.04   &    0.16 &  0.07  &  34 \\
\hline\hline
\multicolumn{6}{||c||}{\bf\boldmath $\sqrts=170-172$~GeV} \\\hline\hline
Cut & Data & Total Bkg. & $\qq(\gamma)$ & 4f & {\large $\epsilon$}{\small(\%)} \\
\hline\hline
(1) & 1403   & 1254              & 1137    & 117.3    & 100 \\ 
(2) &  223   &  164.5            &  124    &  40.5    &  90 \\
(3) &   99   &   91.2            &   52.0  &  39.2    &  83 \\
(4) &   32   &   27.4            &   10.8  &  16.6    &  70 \\
(5) &    0   &   0.37$\pm$0.05   &    0.26 &   0.11   &  32 \\
\hline\hline
\end{tabular}
\caption[]{\label{table:ah6b_cuts}\sl
         Effect of the selection criteria on data,
         background (normalised to the integrated luminosity of the data) 
         and signal simulation ($\mh=60$~GeV, $\mA=30$~GeV) at the three
         centre-of-mass
         energies for the signal 
         channel \h\A\ra\bb\bb\bb. 
         The quoted errors on the background are statistical.
}
\end{center}
\end{table}

The numbers of events passing each requirement, compared with estimates from
the background simulations normalised to the integrated luminosities, are shown
in Table~\ref{table:ah6b_cuts}.
Also shown are the detection efficiencies for simulated
samples of \ee\ra\h\A\ra\A\A\A\ra\bb\bb\bb\ with $\mh=60$~GeV and
$\mA=30$~GeV. No events pass the selection
requirements for any of the three data samples, consistent with the
background expectations of 0, 0.23, 0.37 events for the 
$130-136$, 161, $170-172$~GeV samples, respectively. 

The systematic errors on the detection efficiencies for the signal are
dominated by the statistics of the 500-event Monte Carlo samples,
typically 7\%. In addition to the statistical errors, 
there are errors due to the modelling of the selection
variables. A Monte Carlo sample of 10,000 events with $\mh=60$~GeV,
$\mA=30$~GeV for $\sqrts=171$~GeV was used to study systematic effects.  
The most significant of these effects on the efficiencies is the b-tagging
requirement, where the modelling of the b fragmentation and lifetime and the 
vertex finding algorithms can
introduce systematic effects. These effects are similar to those for
the \h\A\ra\bb\bb\ analysis. The distributions of the significances 
of the first, second
and third most significant vertices agree between data and Monte Carlo
within the limited
statistics with  those for the (\Z$/\gamma)^*$\ra\qq($\gamma$) simulation for 
the selection variables of this analysis.
The systematic errors on the efficiency
due to the b-tagging requirement were found to be 1.7\% due to the impact
parameter resolution and fragmentation uncertainties and 3.5\% due to the
b-hadron decay multiplicity uncertainty.
The systematic errors
due to the jet reconstruction, the requirement on \sqrtsp, and the track
multiplicity are 1.2\%, 1.7\% and 7.2\%, respectively. The total
systematic error on the detection efficiency, not including Monte
Carlo statistics, is thus 8.5\%. Additional systematic errors on the
predicted total numbers of events arise from the error on the integrated
luminosity (0.6\%)~\cite{lumino}  and the theoretical uncertainty on the
cross-section (1\%).

The systematic errors on the background estimates similarly include
contributions from the
modelling of the distributions used in the event selection and from Monte
Carlo statistics.
As mentioned previously, the PYTHIA
generator underestimates the number of events with five or more jets by
approximately 23\% for the overall sample. Also the high-end tail of the  
charged track multiplicity distributions could be mismodelled.
The background estimate is also subject to the
modelling of higher-order QCD processes.
In addition, the tagging of
the third most significant vertex as a b is subject to the
misidentification of jets which do not result from b quarks, as in the
\h\A\ra\bb\bb\ analysis. 
To estimate all these sources of systematic errors related to the modelling
of the Standard Model physics, the backgrounds calculated
using the HERWIG Monte Carlo were compared with those from PYTHIA.
The HERWIG Monte Carlo describes the jet rates of the data much better.
The differences between PYTHIA and HERWIG in the predicted numbers of 
background events passing all selection requirements are  
$-$0.15, $-$0.03, 0.03 events
for $\sqrts=130-136$, 161, $170-172$~GeV
and are consistent with zero within one standard deviation. 
These are less than the statistical errors on the predictions from
either Monte Carlo generator, which for PYTHIA are 100\%, 17\%, 15\% for the 
$130-136$, 161, $170-172$~GeV Monte Carlo samples, respectively.
The systematic errors
due to the b-tagging, jet reconstruction, \sqrtsp, and track
multiplicity requirements were estimated to be 3.2\%, 5.5\%, 4.4\%,   
and 13.5\%, respectively, using the same methods as were used to calculate the
systematic errors on the detection efficiencies.
Thus the systematic error on the background is dominated
by the statistical error and amounts to approximately 100\%, 23\%, 21\% for the
$130-136$, 161, $170-172$~GeV Monte Carlo samples, respectively.
Since for this channel the predicted backgrounds are very small,
no background subtraction is applied.

\section{Statistical combination of individual search channels}
\label{section:combchan}
The searches for Higgs bosons performed by OPAL have not led
to any significant signals.
The negative results in individual search channels, based on data
at various centre-of-mass energies, are statistically combined to
increase the sensitivity. A new statistical 
method~\cite{bock}, based on ``fractional event counting", is
used for that purpose. The method is used to test the predictions of 
specific models (e.g. the MSSM or Two Higgs Field Doublet Model)
for specific parameter sets by comparing them to the experimental results.

The method assigns a weight to each candidate
event for a given hypothetical Higgs mass $m$ (test mass).
The sum of the weights for all candidates is related to a probability
to be consistent with a signal-plus-background hypothesis and
a background-only hypothesis in an analytical manner~\cite{bock}.
The probabilities for signal-plus-background and for background-only
are related in turn, using Bayesian statistics, to a confidence level.

The weight, $w_{ij}(m)$, for each candidate $j$ of each search channel $i$
at the test mass $m$ is determined by the product of a channel
scaling factor, $c_i(m)$, and another factor, $f_i(m,m_{ij})$,
which is determined by the expected mass distribution at $m$ evaluated at
the candidate mass, $m_{ij}$.

The channel scaling factors $c_i$
are determined by the signal-to-background ratio:
\[
c_i(m) = \left(
           1 + \frac{B_i(m) \cdot s(m)}{S_i^{max}(m) \cdot s_i(m)}
         \right)^{-1}.
\]
The expected number of signal events in channel $i$, $s_i(m)$, is calculated
using the model prediction for the cross-section and
branching ratio, the integrated luminosity
of the data set to which the search is applied,\footnote{
Identical search channels using different data sets are considered as 
individual channels.} 
and the signal detection efficiency.
The total expected signal rate is $s(m)=\sum_i s_i(m)$.
The function $S_i(m,m_{ij})$ is the signal probability density.
Its maximum for any possible $m_{ij}$ is $S_i^{\mathrm{max}}(m)$.
The function $B_i(m)$ is the expected differential
background rate per GeV for test mass $m$.
For channels where the mass is reconstructed, $S_i^{\mathrm{max}}(m)$ 
is inversely proportional to the mass resolution.
For channels without mass reconstruction, the ratio
$B_i(m)/S_i^{\mathrm{max}}(m)$ is replaced by the total background rate.
This implies that the $c_i$ are larger for channels where the mass
is reconstructed.

The factor $f_i(m,m_{ij})$ is given by the ratio
\[
f_i(m,m_{ij}) = \frac{S_i(m,m_{ij})}{S_i^{\mathrm{max}}(m)}.
\]

The overall event weight for each candidate is thus given by:
\[
w_{ij}(m) = K \cdot c_i(m) \cdot f_i(m,m_{ij}).
\]
The factor $K$ is a normalisation constant that fixes the largest
value of $w_{ij}(m)$ to unity.

The sum of candidate weights over all channels
$w(m)=\sum_{i,j} w_{ij}(m)$ is converted 
to a confidence level $CL(m)$~\cite{bock} for this sum to be more likely due to
signal-plus-background than due to background-only.
For example for $CL=0.05$, the signal hypothesis is rejected at 95\%
confidence level.

In some channels as indicated previously, the background is statistically
subtracted from the data to enhance the search sensitivity.
In these cases, the subtracted background is first conservatively reduced
by its systematic error.
In calculating the $c_i$ the best estimate of the background is always taken.

The uncertainty on the signal detection efficiency is accounted for using the
method described in~\cite{cousins}.

There is an overlap between some search channels in selected signal and
background events.
In the case of overlap in the signal, there is the problem of assigning the
correct weight to a candidate. This situation did not not occur for the search
results of this paper, because the efficiency of a given search
channel for the topology of any other channel is taken to be zero.
Overlap in the background is only relevant when the
background is subtracted from the data in areas where the same
candidate is found simultaneously in different search channels.
This situation does not occur for the candidate events
found in the search channels used here.

\section{Model-independent and
         Two Doublet Model interpretations}
\label{section:modindep}
Model-independent limits are given for the cross-section for the
generic processes \ee\ra~S$^0$\Z\ and \ee\ra~S$^0$P$^0$, where S$^0$ and 
P$^0$ denote scalar and pseudo-scalar neutral bosons, respectively.
The limits are conveniently expressed in terms of 
scale factors, $s^2$ and $c^2$, which relate the cross-sections of these
generic processes to those of well-known SM cross-sections
(c.f.~Eqs.~(\ref{equation:xsec_zh}),~(\ref{equation:xsec_ah})):
\begin{equation}
\sigma_{\mathrm{SZ}}=s^2~\sigma^{\mathrm{SM}}_{\mathrm{HZ}},
\label{eq:s}
\end{equation}
\begin{equation}
\sigma_{\mathrm{SP}}=c^2~\bar{\lambda}~\sigma^{\mathrm{SM}}_{\nn}.
\label{eq:c}
\end{equation}

\begin{figure}[tbp]
\centerline{ \epsfig{file=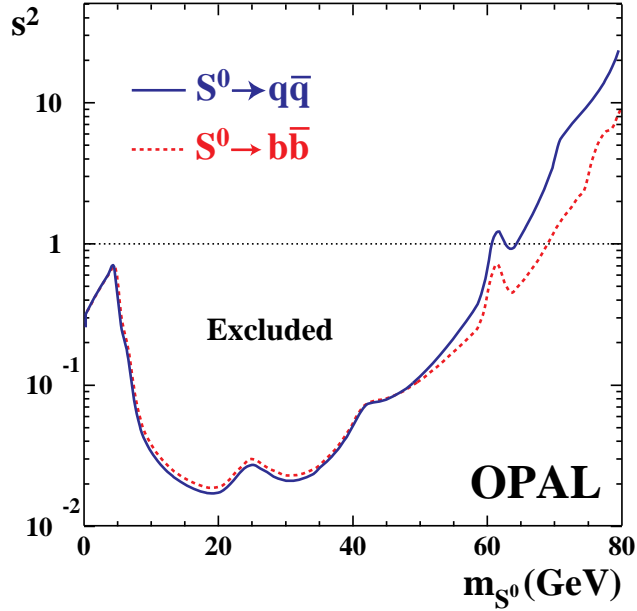,width=8.3cm} }
\caption[]{\label{modindepZh}\sl
         Upper limits at 95\% CL on $s^2$ (as defined by Eq.~(\ref{eq:s}))
         using all SM search channels and assuming the SM Higgs branching
         ratios for the S$^0$ (dashed line),
         and discarding all search channels that use b-tagging
         but assuming a hadronic branching ratio of the S$^0$ of 100\%
         (solid line).
}
\end{figure}
Figure~\ref{modindepZh} 
shows the 95\% CL upper bound for $s^2$
as a function of the S$^0$ mass, obtained from:
\[s^2=
  \frac{N_{95}^{\mathrm{SZ}}}{
       \sum~(\epsilon~{\cal L}~\sigma^{\mathrm{SM}}_{\mathrm{HZ}})},\]
where $N^{\mathrm{SZ}}_{95}$ is the 95\% CL upper limit
for the number of possible 
signal events in the data, $\epsilon$ is the signal detection efficiency
and ${\cal L}$ is the integrated luminosity.
The sum runs over the different centre-of-mass energies considered.
The dashed line is computed using all search channels and assumes
SM Higgs branching ratios for the S$^0$.
The solid line is computed assuming 100\% hadronic branching ratio for the
S$^0$ and uses only search channels that do
not employ b-tagging and is therefore more generally valid.
Below $m_{\mathrm{S}^0}\approx 5$~GeV, the direct search loses sensitivity
rapidly and the limit for $s^2$ is determined from
$\Gamma_{\mathrm{Z}^0}$ only, as described below.

\begin{figure}[tbp]
\centerline{
\mbox{\epsfig{file=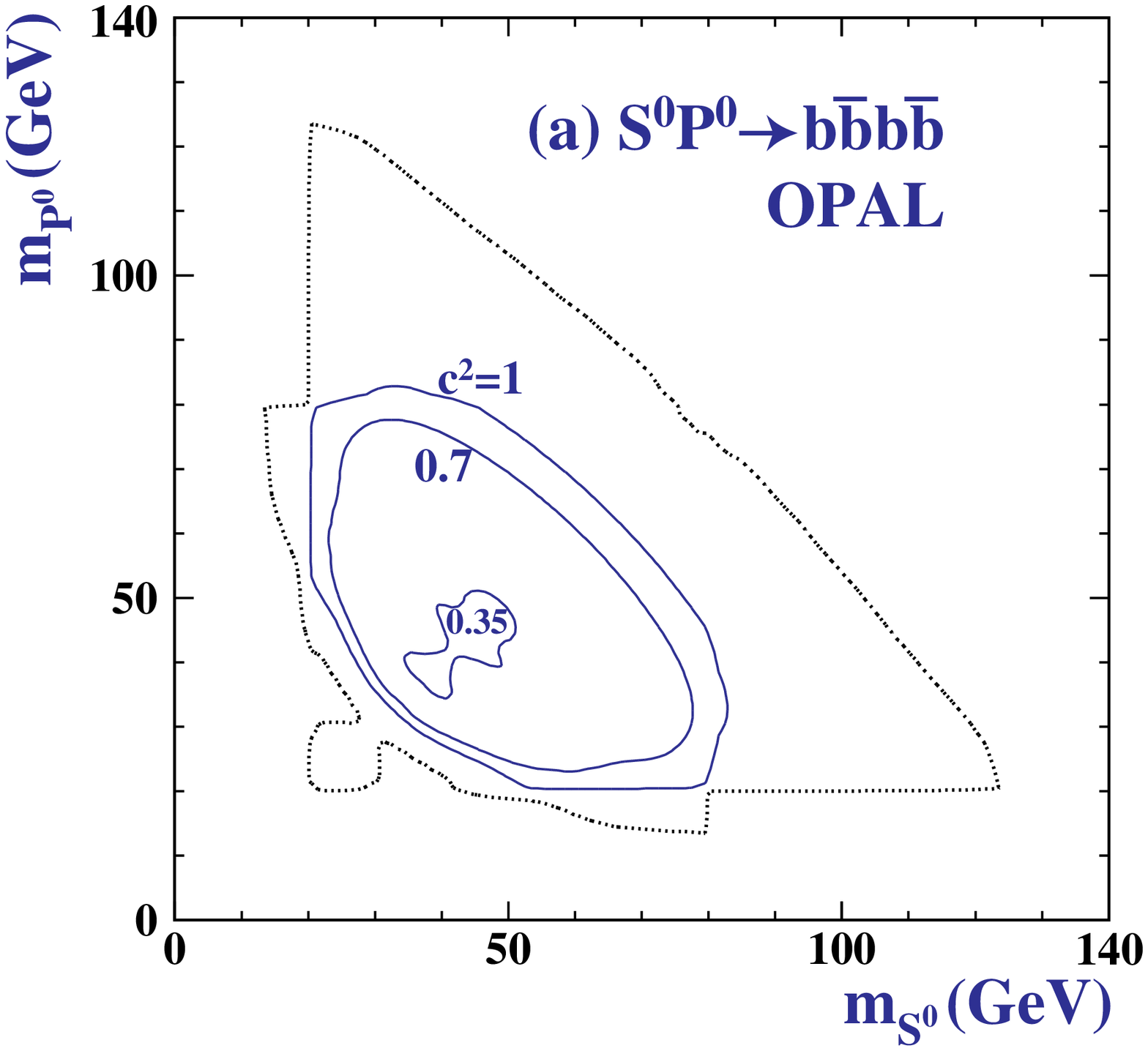,width=8.3cm}}\hfill
\mbox{\epsfig{file=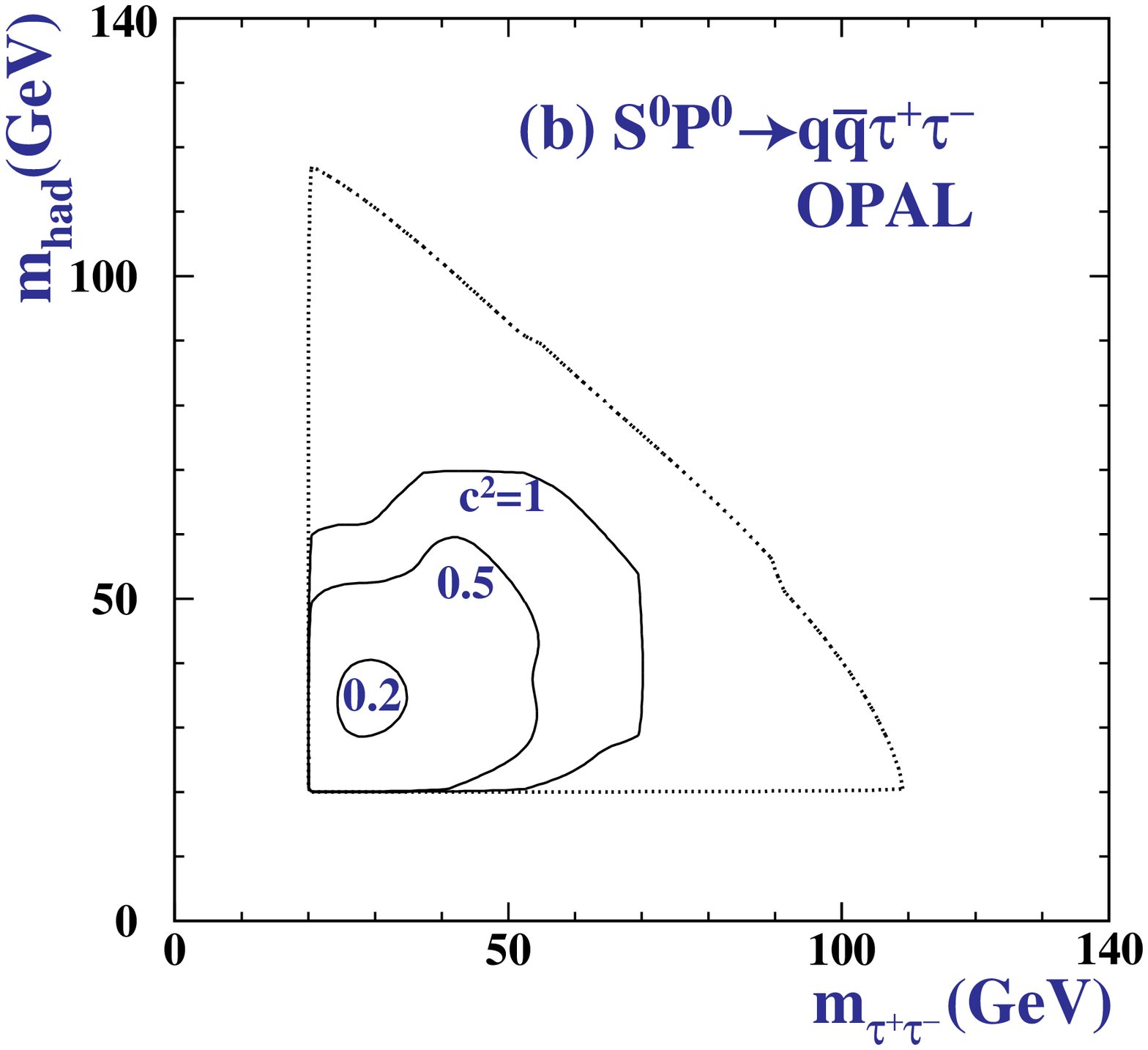,width=8.3cm}}
}
\caption[]{\label{modindephA}\sl
         Upper limits at 95\% CL for $c^2$ (see Eq.~(\ref{eq:c}))
         for:
         (a) the S$^0$P$^0$\ra\bb\bb\ search channel assuming the \bb\ branching
         ratio for both S$^0$ and P$^0$ to be 100\%, and
         (b) the S$^0$P$^0$\ra\qq\tautau\ search channel assuming a
         100\% branching ratio for this final state.
         The invariant masses of the tau-lepton pair and hadron jet-pair
         are denoted $m_{\tau\tau}$ and $m_{\mathrm{had}}$, respectively.
         The search efficiency is zero
         outside this area surrounded by the dotted line.
}
\end{figure}
Figure~\ref{modindephA} shows contours of 95\% CL
upper limits for $c^2$ in the S$^0$ and P$^0$ mass plane,
for the processes \ee\ra~S$^0$P$^0$\ra\bb\bb\ and \qq\tautau, respectively.
In both cases a 100\% branching ratio into the specific final state is
assumed.
The contours are obtained from:
\[c^2=
  \frac{N_{95}^{\mathrm{SP}}}{
       \sum~(\epsilon~{\cal L}~\bar{\lambda}~\sigma^{\mathrm{SM}}_{\nn})},\]
with $N_{95}^{\mathrm{SP}}$ being the 95\% CL upper limit for the number of
signal events in the data.
The results obtained for \bb\bb\ (Fig.~\ref{modindephA}(a)) are 
symmetric with respect to interchanging S$^0$ and P$^0$ while those obtained
for \tautau\qq\ are not. For this reason, the results for
\tautau\qq\ (Fig.~\ref{modindephA}(b)) are presented with the 
mass of the
particle decaying into \tautau\ along the abscissa and that of the particle
decaying into \qq\ along the ordinate.
The irregularities of the iso-$c^2$ contours are due to the presence
of candidate events that affect $N_{95}^{\mathrm{SP}}$.

If the decay of the \Z\ into a final state containing S$^0$ or P$^0$ is
possible, the width $\Gamma_{\mathrm{Z}}$ will be larger than when
only the known Standard Model decays are possible.
The excess width that is still possible when subtracting the predicted
Standard Model width from the measured $\Gamma_{\mathrm{Z}}$ value can be used
to place upper limits on the cross-sections of \Z\ decays into final states
with S$^0$ or P$^0$ bosons.
The additional width of the \Z\ resonance from final states not classified
as lepton pairs can be extracted from the measurement of the branching ratio
BR$(\Z\ra\ell\ell)=\Gamma(\Z\ra\ell\ell)/\Gamma(\Z\ra{\mathrm all})$.
The limits obtained from this equation are not sensitive to radiative
corrections that equally affect all final states,
because these cancel in the ratio.
The current measurement is
BR$(\Z\ra\ell\ell)=0.03366{\pm}0.00006$~\cite{pdglineshape}.
From this value and the predicted SM value obtained from~\cite{zfitter},
the difference between the measured and predicted \Z\ width is
$(-1.2~{\pm}4.4$(exp)~${\pm}1.7$(QCD)~${\pm}1.8$~(EW)$)$~MeV, resulting in an
excess width less than 7.1~MeV at 95\% CL.
The uncertainties are due to experimental errors, QCD corrections and
electroweak corrections, respectively.
This excess width of the \Z\ can still be affected by corrections,
such as vertex corrections and oblique corrections, that are different for
different final states.
However, these corrections are expected to be small for the Two Higgs Field
Doublet Model (2HDM).
In the case of a very light Higgs ($\mh<2m_{\mu}$) that remains invisible
or that decays into photons, electrons, or muons,
some of the final states Z$^*${S$^0$} with the Z$^*$ decaying to leptons could
still be classified as a lepton pair and the above limit does not apply.
However, in this case the decay-mode-independent
search limit of~\cite{sinlimit} is applied.

\begin{figure}[tbp]
\centerline{
\epsfig{file=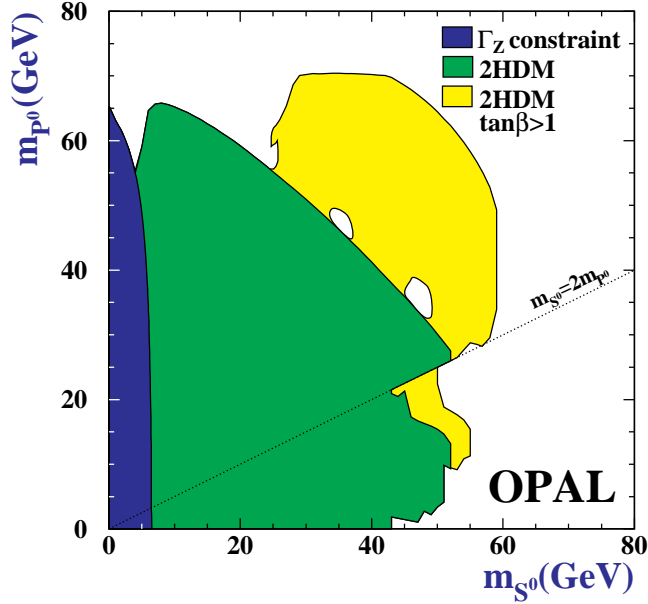,width=8.5cm}
}
\caption[]{\label{modind_gz}\sl
  Regions excluded at 95\% CL in the Type II 2HDM.
  The black region is excluded using constraints from $\Gamma_{\mathrm{Z}}$
  only.
  The dark grey region uses the direct searches for the SM Higgs in addition,
  but discarding the search channels that use b-tagging, assuming a hadronic
  branching ratio of the \h\ of 92\%.
  The light grey region is excluded for $\tanb>1$ in the 2HDM, assuming
  SM Higgs branching ratios for \h\ and \A.
}
\end{figure}
In the 2HDM the bosons S$^0$ and P$^0$ are identified with \h\ and \A,
and the couplings $s^2$ and $c^2$ are identified with \sba\ and \cba,
respectively.
The assignment of the possible excess width in $\Gamma_{\mathrm{Z}}$
to the process \Z\ra{\h}Z$^*$
yields an upper bound for
$s^2$ which depends only on the mass of \h; the assignment to
\Z\ra\h\A\ yields an upper bound for $c^2$ which depends on the
masses of both \h\ and \A.
Combining these limits, the black region in the mass plane shown in
Fig.~\ref{modind_gz} is excluded at 95\% CL
regardless of \h\ and \A\ decay modes.
In the 2HDM, disregarding the possibility of \h\ra\A\A, the most important
final states of the decays of the \h\ and \A\ bosons are \bb, \cc\ and \tautau.
The branching ratios depend on \tanb, but the hadronic
branching fraction always exceeds 92\%~\cite{hdecay}.
For $\tanb{\geq}1$ the \bb\ channel dominates 
while for $\tanb<1$ the \cc\ contribution may become the largest.

In Fig.~\ref{modind_gz} the excluded area in the (\mh,\mA) plane
is shown when the limits on $c^2$ and $s^2$ are combined.
Below the dotted line, where the \h\ra\A\A\ decay is kinematically allowed
and competes with the \h\ra\ff\ decay,
the smaller of the detection efficiencies is used.
The excluded area is therefore
valid regardless of the \h\ra\A\A\ branching ratio.
The dark grey area is excluded at 95\% CL when BR(\h\ra\qq)$\ge$92\%,
generally valid in the 2HDM.
The limit in the 2HDM for equal \h\ and \A\ masses is at
$\mh=\mA=41.0$~GeV.\footnote{
Throughout this paper numerical mass limits are quoted to 0.5~GeV precision.}
The light grey area is excluded when SM Higgs branching ratios are assumed for
\h\ and \A.
This assumption is valid in the 2HDM for $\tanb>1$.
In that case the limit for equal \h\ and \A\ masses is at $\mh=\mA=56.0$~GeV.
The holes in the exclusion of the light grey area at the edge of the
dark grey excluded region are caused by the single candidate event in the
\h\A\ra\bb\bb\ search.

\section{Interpretations of the search results in the MSSM}
\label{section:mssminterpretation}

In its most general form,
the MSSM has more than one hundred parameters. In this paper we consider
a constrained MSSM, with only five free parameters in addition to those
of the SM. The model assumes
unification of the scalar-fermion masses ($m_0$) 
at the grand unification (GUT) scale, and unification of the 
gaugino masses (which are parametrised using $M_2$, the SU(2) gaugino mass 
term at the electroweak scale) and scalar-fermion tri-linear
couplings ($A$) at the electroweak scale. 
These simplifications have practically no impact
on the MSSM Higgs phenomenology. In particular, a common
scalar-fermion mass and tri-linear coupling
is justified since only the scalar top (\sctop) sector gives important
contributions to Higgs boson masses and couplings.

Other free parameters of the model are
the supersymmetric Higgs mass parameter $\mu$, \tanb, and
the mass of the CP-odd neutral Higgs boson, \mA.
As mentioned previously, the top quark mass has a strong impact on \mh.
Therefore, it is also varied within reasonable bounds.

In scanning the MSSM parameter space, values of the above parameters are input
to the HZHA event generator~\cite{hzha,hdecay}
which is supplemented with parts of the SUSYGEN~\cite{susygen} program.
The HZHA program provides the masses 
and couplings of all Higgs bosons as well as those of the
supersymmetric partners.
It also calculates the cross-sections for \ee\ra\h\Z\ and \h\A~\cite{remt}
at each centre-of-mass energy, corrected for initial-state radiation. SUSYGEN
produces scalar-fermion masses on the electroweak scale, starting from the same
input parameters.

For the above parameters, the following ranges are considered:
\begin{itemize}
\item
$m_0$:~~0 to 1000~GeV. 
The masses of physical scalar-fermions are obtained in SUSYGEN
by running $m_0$ from the GUT scale down to the electroweak scale 
using the relevant renormalisation group equations.
\item
$M_2$:~~0 to 2000~GeV. The U(1) and SU(3) gaugino mass terms, $M_1$ and $M_3$, 
are calculated from $M_2$ using the ratios of the corresponding
coupling constants,
$M_1:M_2:M_3=\alpha_1:\alpha_2:\alpha_3$.
\item
$A$:~~$-$2.5${\cdot}m_0$ to $2.5{\cdot}m_0$.
This range is chosen to include all possible \sctop\ mixings.
\item
$\mu$:~~$-$1000 to 1000~GeV.
\item
\mA:~~5 to 160~GeV.
Beyond this range, values in 5~GeV slices around
$\mA=250$, 400, 1000 and 2000~GeV are also explored.
Masses below 5~GeV are not considered since
in that domain the decays of \A\ are uncertain.
As discussed in the previous section and shown in Fig.~\ref{modind_gz},
for $\mh<43.0$~GeV, small values of \mA\ are excluded on more general grounds.
Note that the parameter \mA\ which is used as input to HZHA and
SUSYGEN is the ``running mass", while the mass of the CP-odd Higgs
boson \A, which is used to express the results, is the physical mass.
\item
\tanb:~~0.7 to 50. This spans the theoretically favoured
range, $1<\tanb<m_{\rm t}/m_{\rm b}$, but also includes values less than
unity, which are not ruled out by theory.
\item
$m_{\rm t}$:~~165, 175 and 185~GeV. The range of values includes
approximately two standard deviations of
the measured top quark mass~\cite{mtop}.
\end{itemize}
Most parameters are scanned by dividing their 
ranges into bins of variable size and choosing the values in each bin
at random (``stratified scanning").
Exceptions are the parameter $m_{\rm t}$, for which three discrete
values are used, and the values of \mA\ greater than 160~GeV, for which
5~GeV bands around the values are used.

The following parameter scans are considered in order of increasing complexity:
\begin{description}
\item[(A)]
The parameters $m_0$ and $M_2$ are fixed at 1~TeV,
$\mu$ is fixed at $-$100~GeV and $m_{\rm t}=175$~GeV.
The parameter $A$ is fixed to 0 (minimal \sctop\ mixing) or
$\sqrt{6}$~TeV (maximal \sctop\ mixing.)
The parameters \mA\ and \tanb\ are varied as described above.
This simple scan serves as a {\em benchmark} corresponding
to the scans proposed in~\cite{lep2higgs}.
\item[(B)]
We also consider two scans which correspond to cases
of {\em minimal and maximal mixing} in the \sctop\ sector,
inducing small and large corrections to the \h\ boson mass, respectively.
In these scans the parameters $m_0$, $M_2$, \mA, \tanb, and $m_{\rm t}$
are varied independently, $\mu$ is constrained to $-0.1 m_{\mathrm{Q}}$ and
the parameter $A$ is set to
\begin{itemize}
\item
$A=0$ for minimal mixing, and 
\item
$A=\sqrt{6}~m_{\mathrm{Q}}$ for maximal mixing.
\end{itemize}
The mass $m_{\mathrm{Q}}$
of the ``left-up" scalar~quark at the electroweak scale
is uniquely determined in terms of $m_0$ and the other input
parameters~\cite{susygen}.
\item[(C)]
In the most {\em general scan} considered here,
all parameters described previously are varied
independently in the ranges that are listed above.
\end{description}

The number of parameter sets considered in scan (A) is about 50,000,
that of the scan (B) is approximately 1,000,000, and
that of scan (C) is close to 6,000,000.

Before comparing the above parameter sets to the data, these are subject to
a selection on the basis of theoretical and experimental considerations.
Only those sets are selected that provide a $\sctop_1$
mass larger than the lightest neutralino mass.
Additional experimental constraints are applied, which do not
relate directly to the searches described here.
A parameter set is rejected if the 
sum of the partial decay widths of \Z\ra\h\Zs\ and \Z\ra\h\A\ is
more than 7.1~MeV
(see the discussion of the constraint from $\Gamma_{\mathrm{Z}}$
in Section~\ref{section:modindep}).
Parameter sets giving rise to chargino or neutralino
masses~\cite{lep2neutralino}, or stop masses~\cite{lep2stop} excluded by
OPAL searches at LEP2 are also discarded.

It has been shown~\cite{frere} that
large values of $A$ and $\mu$ may give rise to non-zero vacuum
expectation values for the \sctop\ fields
which break charge and colour symmetry.
Criteria for charge- and colour-breaking (CCB) minima of
the MSSM Lagrangian have been determined~\cite{casas}, but these may be
substantially modified if also ``tunneling'' from the CCB minimum to the
electroweak minimum is taken into account.
A calculation that includes the effect of tunneling is available for one
specific scenario out of a number of distinct possibilities~\cite{kusenko}, but
a complete treatment of CCB criteria is not yet available to our knowledge.
A simple approximate criterion to avoid CCB minima is~\cite{frere}:
\begin{equation}
\label{eq:ccb}
A^2 + 3\mu^2 < x \; (\mstopL^2 + \mstopR^2),
\end{equation}
where \mstopL\ and \mstopR\ denote the left- and right-handed
scalar top masses and $x{\approx}3$.
For the specific calculation that includes tunneling this bound was shown
to be modified to $x{\approx}7.5$~\cite{kusenko}.
For the general scan (C) of the MSSM parameter space results will be shown
without applying CCB criteria,
and for CCB criteria applied with $x=3$ and $x=7.5$.

The experimental exclusion limits, at the 95\% CL, are presented below,
separately for the scans (A), (B), and (C).
In each scan the total predicted number of events from
all search channels is calculated using cross-sections, branching ratios,
luminosities and search efficiencies for all different MSSM parameter sets.
From this expected signal prediction and the number of observed events a
confidence level is calculated according to the prescription in
Section~\ref{section:combchan}.
For scans (A) and (B) the 99\% CL exclusion is also shown,
to indicate the sensitivity to the chosen exclusion confidence level.
The results are presented for each scan in four figures:
(a) in the (\mh,~\mA) plane for $\tanb>1$,
(b) in the same plane for $\tanb>0.7$,
(c) in the (\mh,~\tanb) plane, and (d) in the (\mA,~\tanb) plane.
For scans (A) and (B) the lower limits for the minimal and maximal mixing
cases differ by only small amounts and only the weaker of the two
exclusion limits is given.
The theoretically accessible area corresponds to the larger one,
for maximal stop mixing.
The theoretically inaccessible areas are shown in the figures as speckled.

\begin{figure}[p]
\centerline{
\epsfig{file=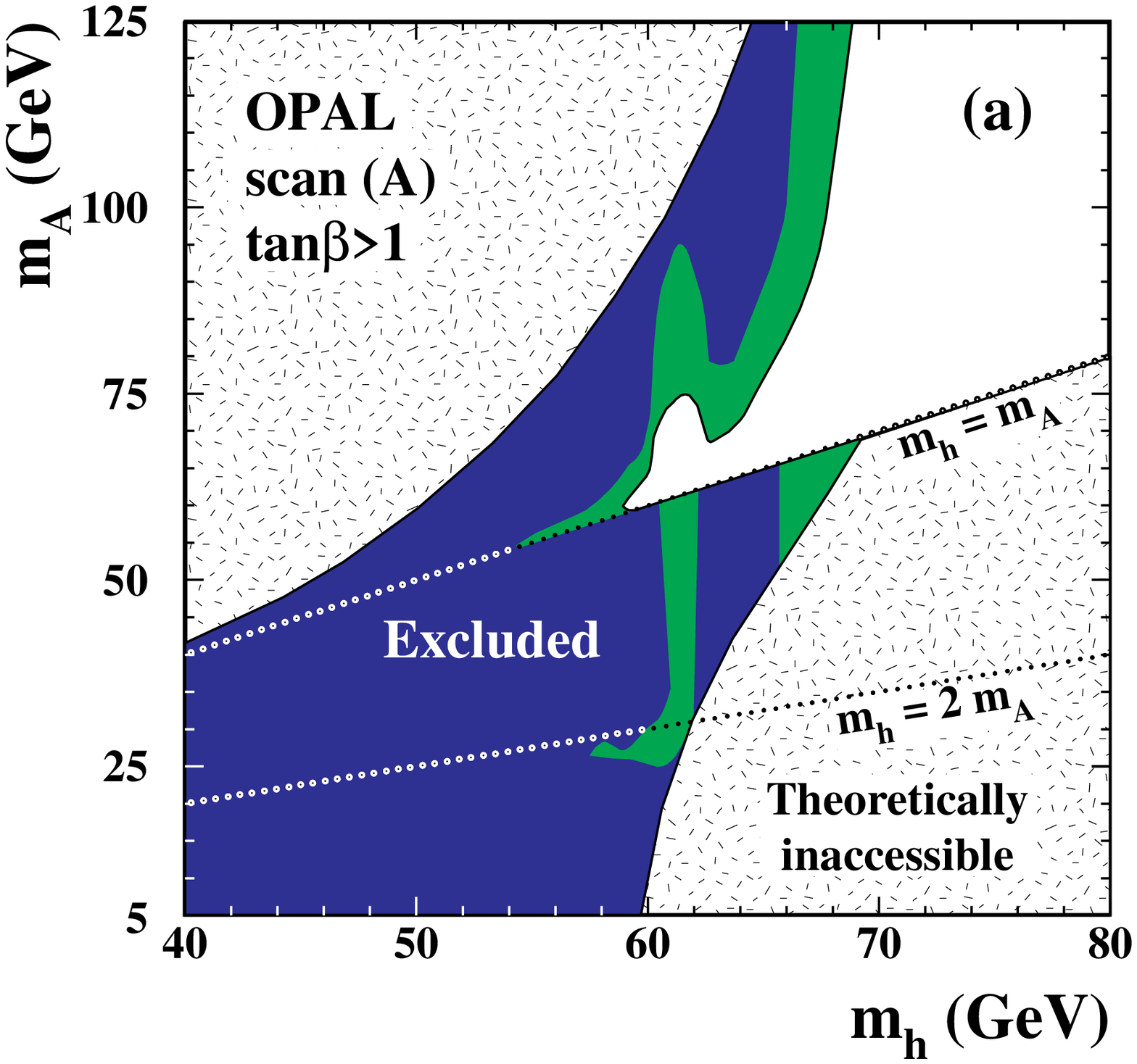,width=8.5cm}\hfill
\epsfig{file=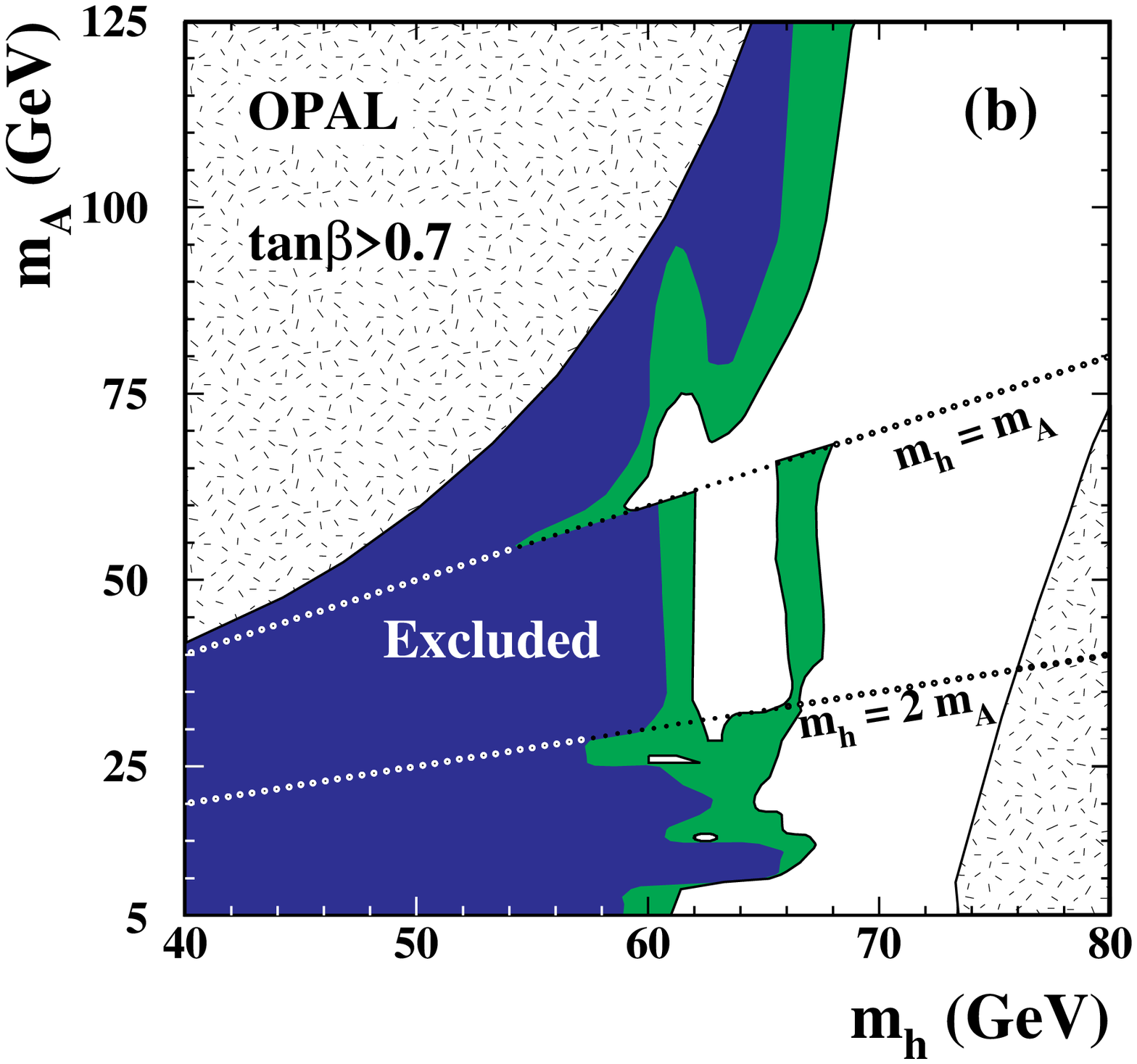,width=8.5cm}}
\centerline{
\epsfig{file=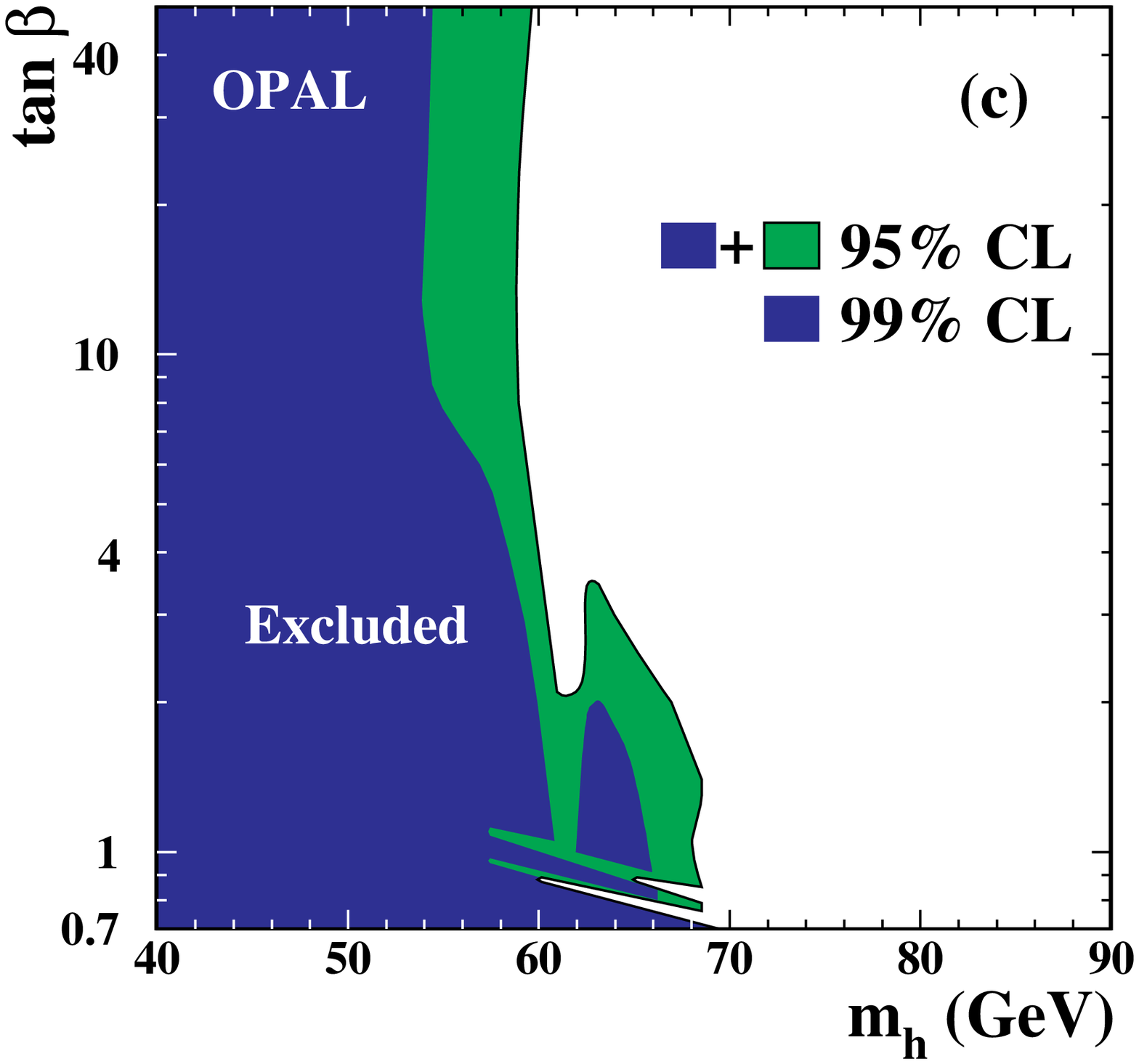,width=8.5cm}\hfill
\epsfig{file=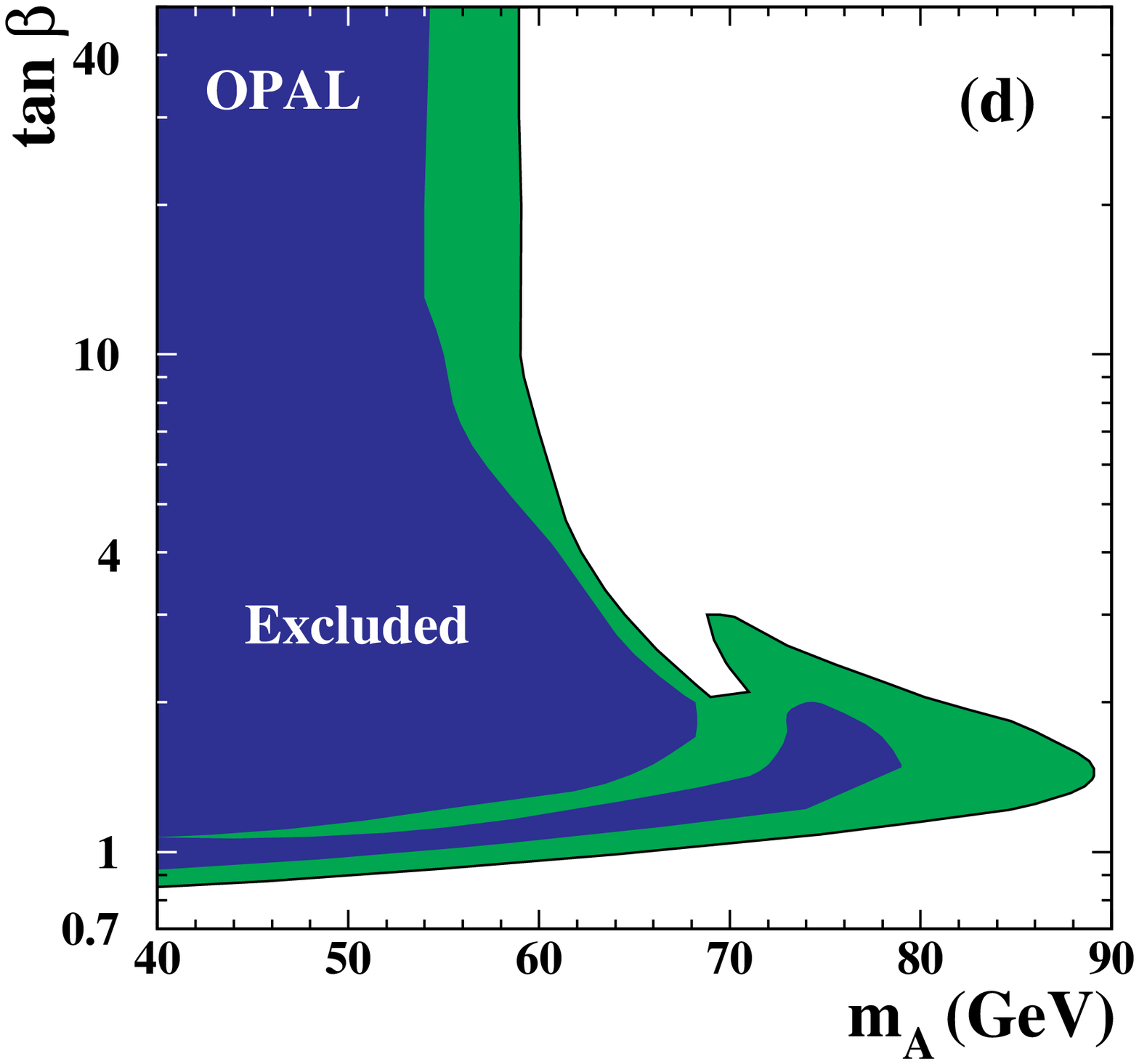,width=8.5cm}}
\caption[]{\label{figure:mssm-yb}\sl
         The MSSM exclusion for scan (A) described in the text of
         Section~\ref{section:mssminterpretation}.
         Excluded regions are shown for
         (a) the (\mh,~\mA) plane for $\tanb>1$,
         (b) the (\mh,~\mA) plane for $\tanb>0.7$,
         (c) the (\mh,~\tanb) plane, and
         (d) the (\mA,~\tanb) plane.
         The black areas are excluded at 99\% CL. The black plus the dark grey
         areas are excluded at 95\% CL.
         The speckled areas in (a) and (b) are theoretically inaccessible.
}
\end{figure}
The results for scan (A) are shown in Fig.~\ref{figure:mssm-yb}.
The structure near $\mh=60$~GeV is caused by the candidate in the analysis for
the LEP1 leptonic channel at $\mh=61.2$~GeV.
From Fig.~\ref{figure:mssm-yb}(a), for $\tanb>1$,
95\% CL lower limits can be obtained for $\mh>59.0$~GeV and $\mA>59.5$~GeV.
When the \tanb\ range is enlarged to $\tanb>0.7$ (Fig.~\ref{figure:mssm-yb}(b)),
the lower limit of \mh\ is not affected, but no lower limit
on \mA\ can be given.
Figures~\ref{figure:mssm-yb}(a) and (b) show the region $\mA>5$~GeV.
For $\mA<5$~GeV and $\mh>43$~GeV the value of \sba\ is always very close to
unity in scan (A). Therefore, in this region only the \Z\h\ production
process is relevant for the exclusion, and the limit depends neither
on \mA\ nor on the \A\ decay modes.
Hence, the limit for \mh\ at $\mA=5$~GeV is also valid for $\mA<5$~GeV.
Figure~\ref{figure:mssm-yb}(c) shows the projection onto the (\mh,~\tanb)
plane. Large values of \tanb\ correspond to \mh$\approx$\mA.
In the (\mA,~\tanb) projection of Fig.~\ref{figure:mssm-yb}(d)
the strongest limit for \mA\ of 89.0~GeV at 95\% CL is obtained at
\tanb$\approx$1.4.  However, at 99\% CL this limit drops considerably.

\begin{figure}[p]
\centerline{
\epsfig{file=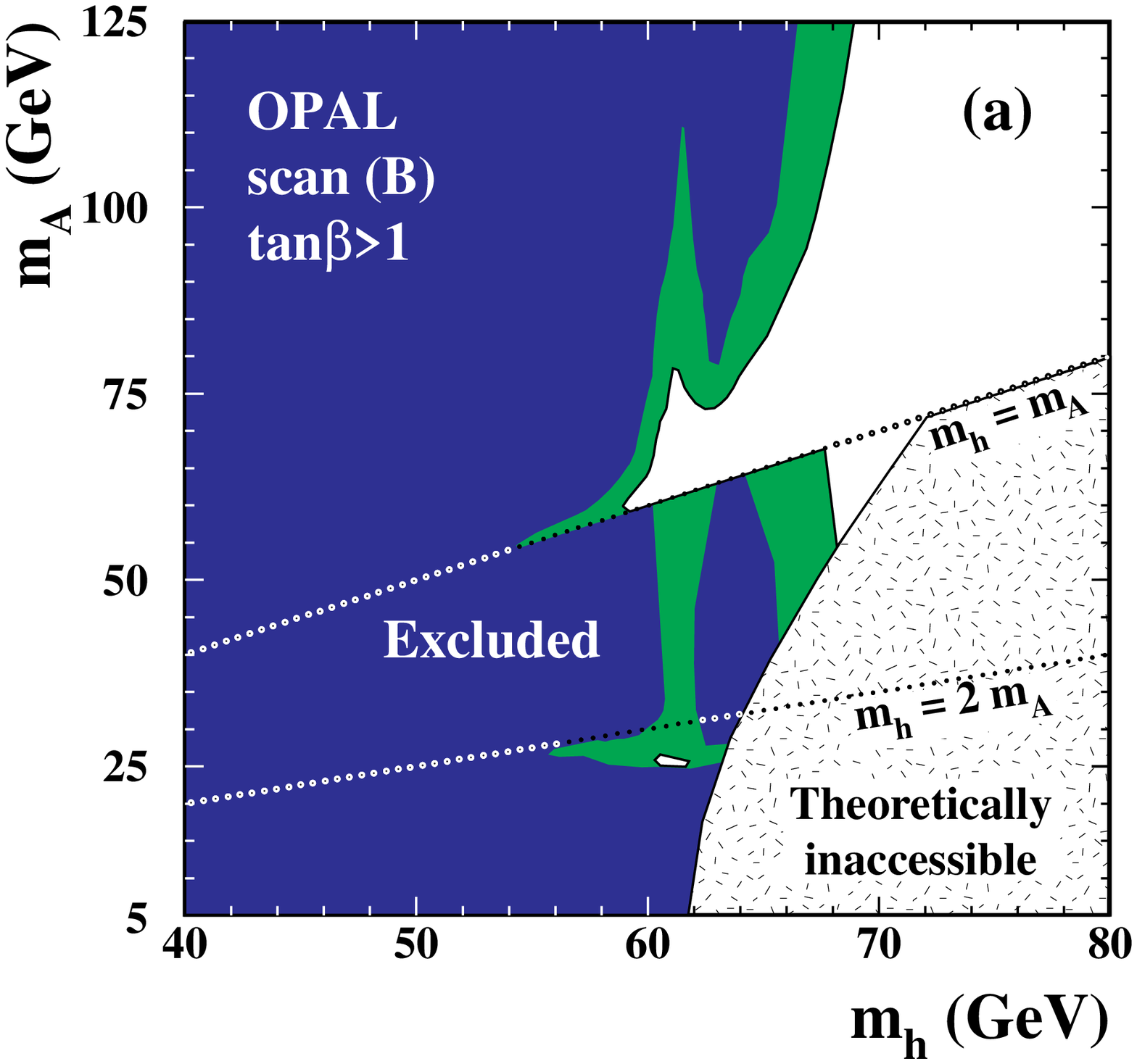,width=8.5cm}\hfill
\epsfig{file=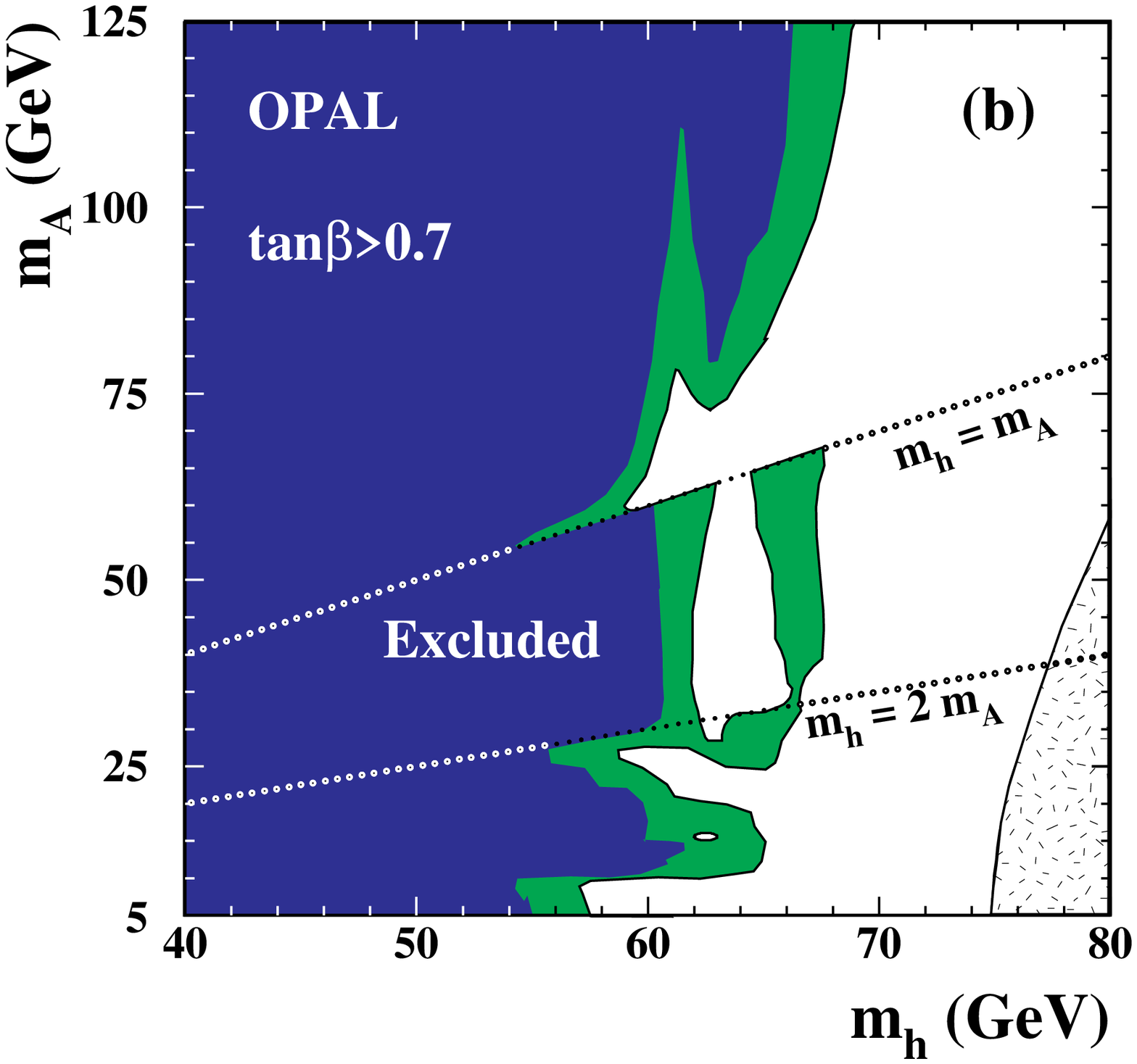,width=8.5cm}}
\centerline{
\epsfig{file=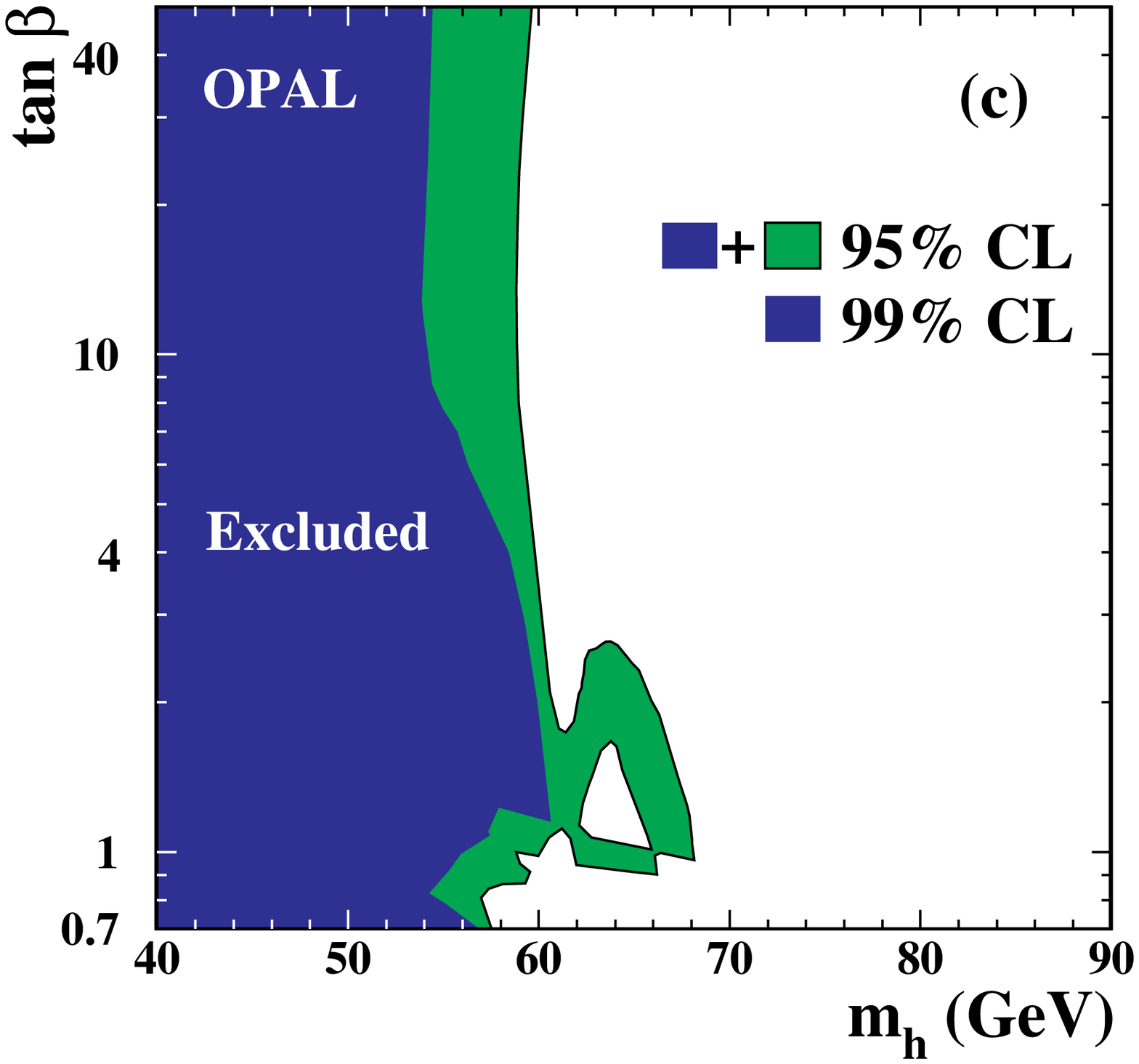,width=8.5cm}\hfill
\epsfig{file=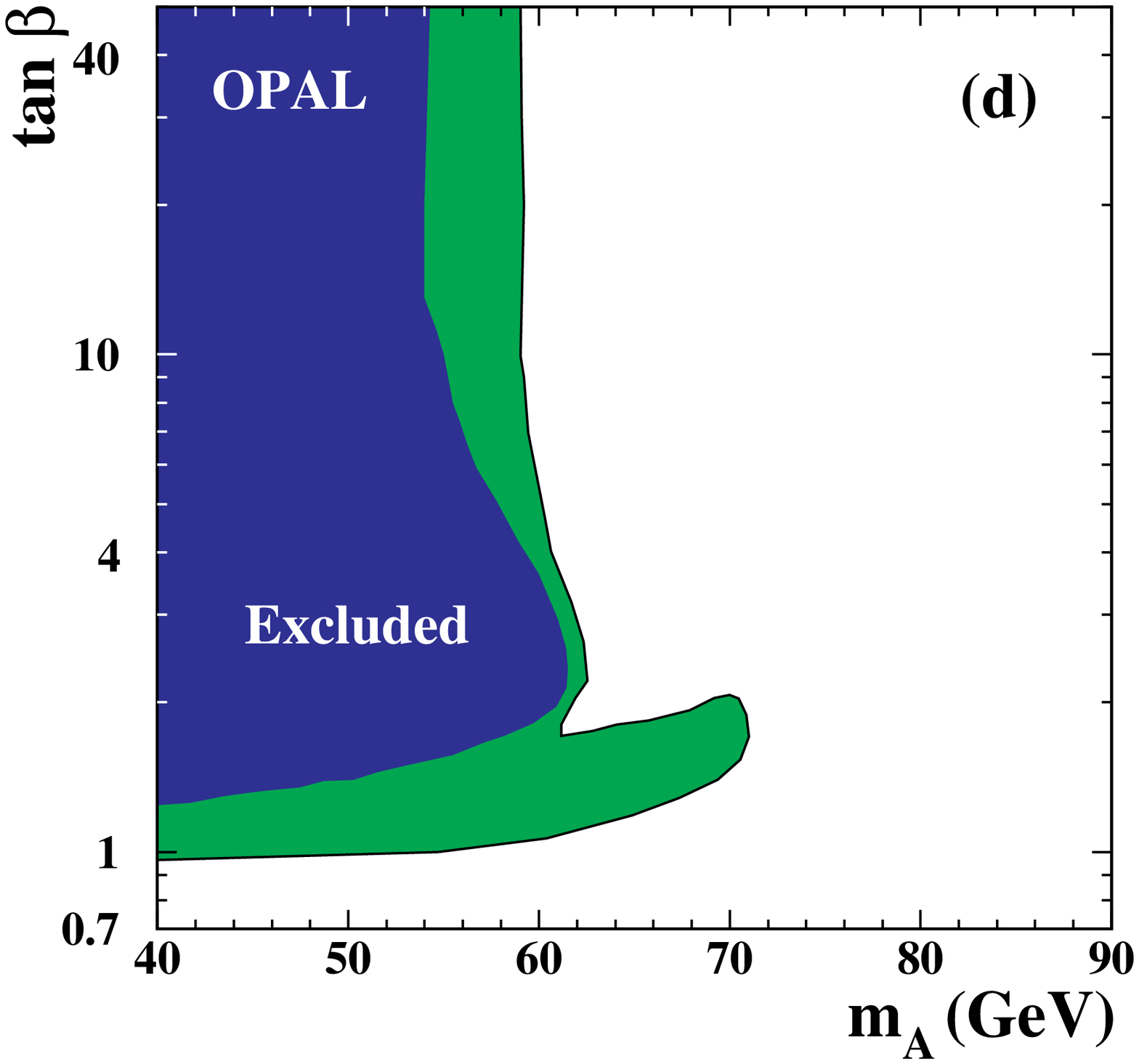,width=8.5cm}}
\caption[]{\label{figure:mssm-mm}\sl
         The MSSM exclusion for the scan (B) described in the text of
         Section~\ref{section:mssminterpretation}.
         Excluded regions are shown for
         (a) the (\mh,~\mA) plane for $\tanb>1$,
         (b) the (\mh,~\mA) plane for $\tanb>0.7$,
         (c) the (\mh,~\tanb) plane, and
         (d) the (\mA,~\tanb) plane.
         The black areas are excluded at 99\% CL. The black plus the dark grey
         areas are excluded at 95\% CL.
         The speckled areas in (a) and (b) are theoretically inaccessible.
}
\end{figure}
Figure~\ref{figure:mssm-mm} shows the results for scan (B).
Differences with respect to scan (A) are due to
the possibility of having lower $\mstop$ values.
This leads in general to modified couplings and in particular, for some
parameter sets, to a strongly enhanced branching ratio for $\h\ra\mathrm{gg}$.
The wider range of $\mstop$ in conjunction
with $m_{\mathrm{t}}=185$~GeV leads to larger theoretically accessible
regions.
Despite these modifications, many essential features such as the
limit for \mh\ at very high \mA, the limits for
\mh\ and \mA\ near $\mh=\mA$ and the absolute lower limit on \mh\ remain
unchanged.
The absolute lower limit on \mA\ for scan (B) is $\mA>23.0$~GeV at 95\% CL,
but the range $26.5<\mA<55.0$~GeV is excluded at 95\% CL.
The small unexcluded ``island'' at $60.5<\mh<61.5$~GeV
and $23.0<\mA<26.5$~GeV is due to the simultaneous effect of
a large branching ratio for $\h\ra\A\A$ and the LEP1 leptonic channel candidate.
Because the theoretically allowed region at low \mA\ and high \mh\ is
increased in scan (B) with respect to that in scan (A),
there is an additional unexcluded triangular
region near $\mh=68$~GeV with a minimum value at
$\mA=55.0$~GeV at 95\% CL.

\begin{figure}[p]
\centerline{
\epsfig{file=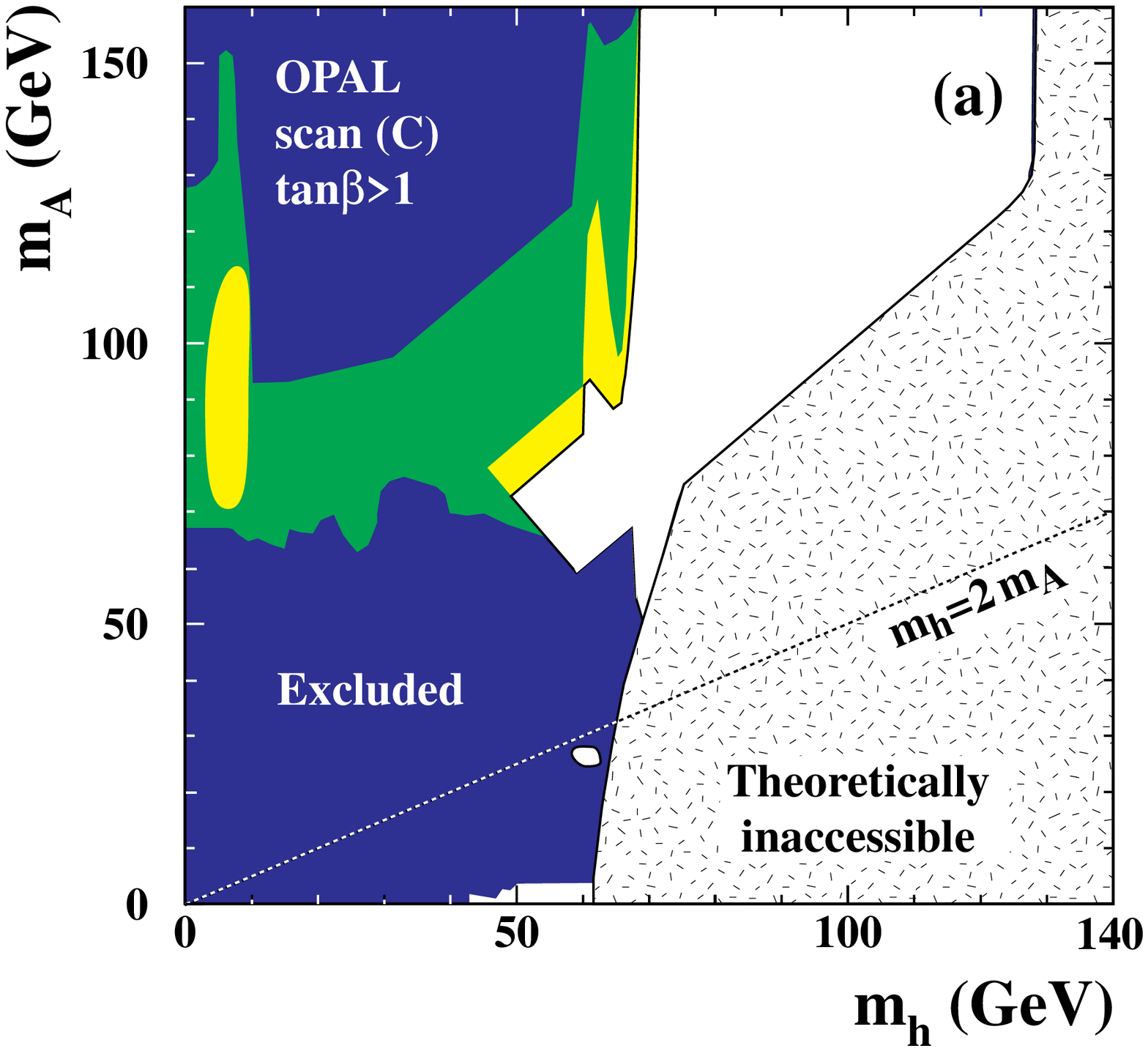,width=8.5cm}\hfill
\epsfig{file=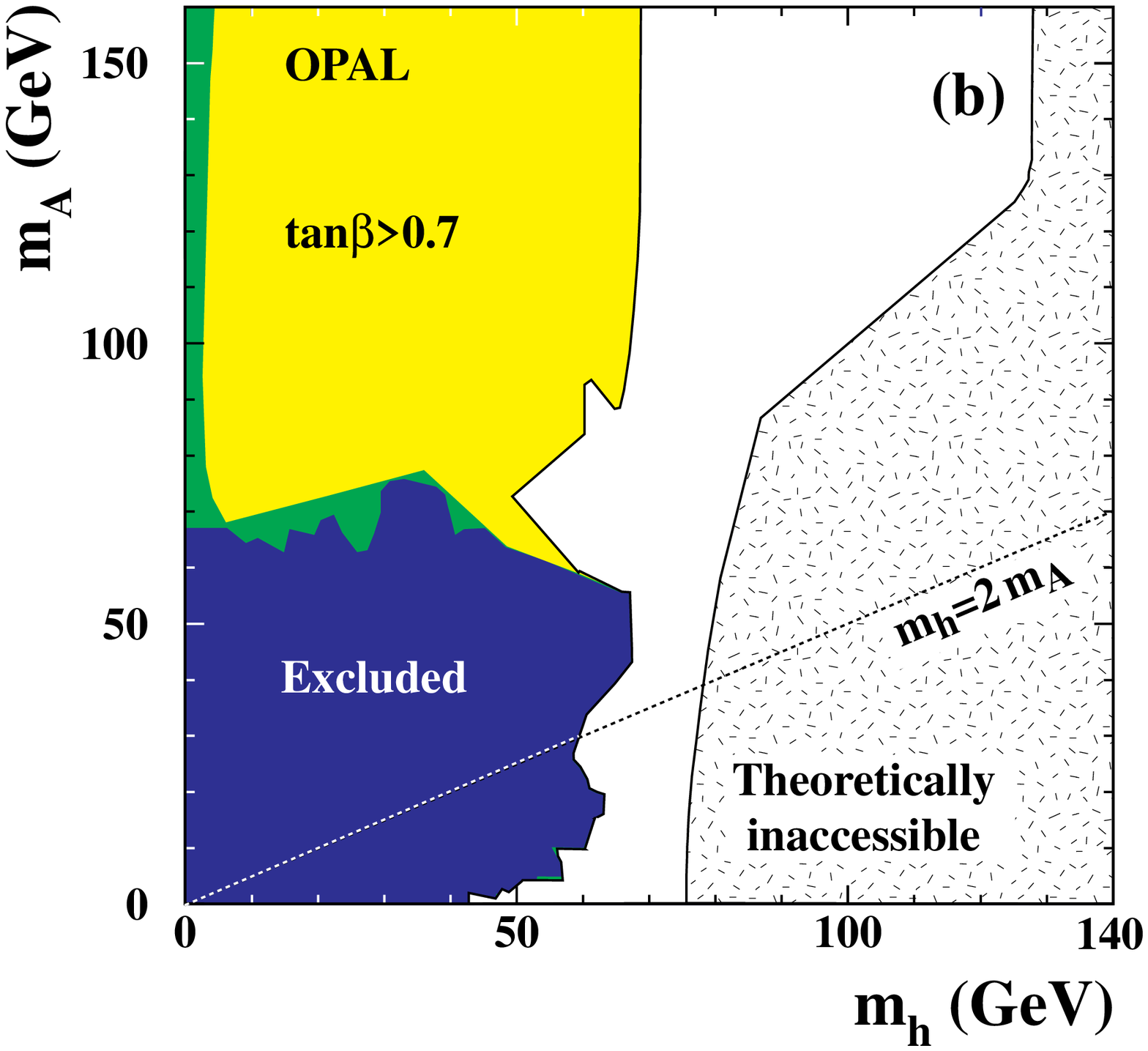,width=8.5cm}
}
\centerline{
\epsfig{file=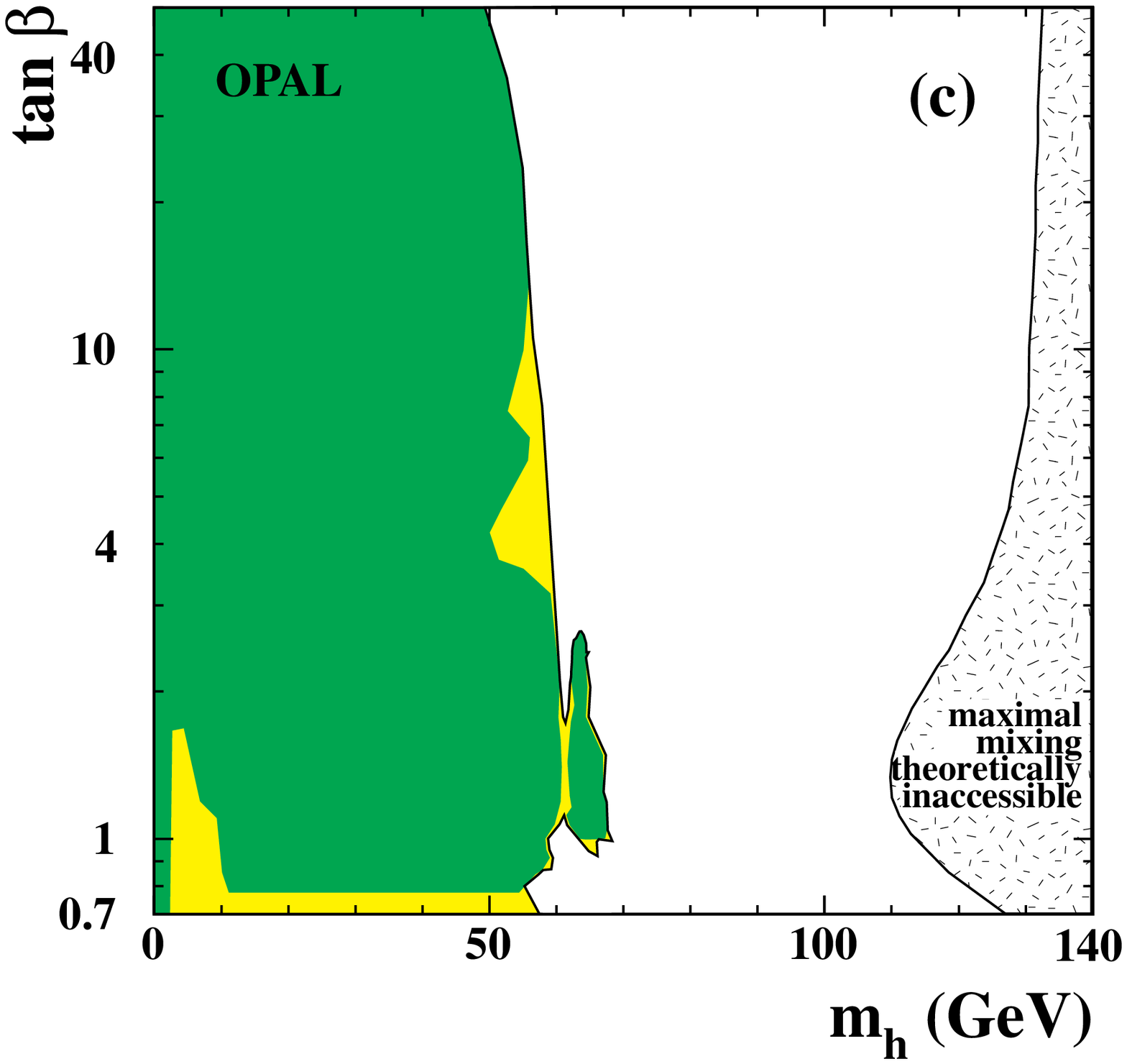,width=8.5cm}\hfill
\epsfig{file=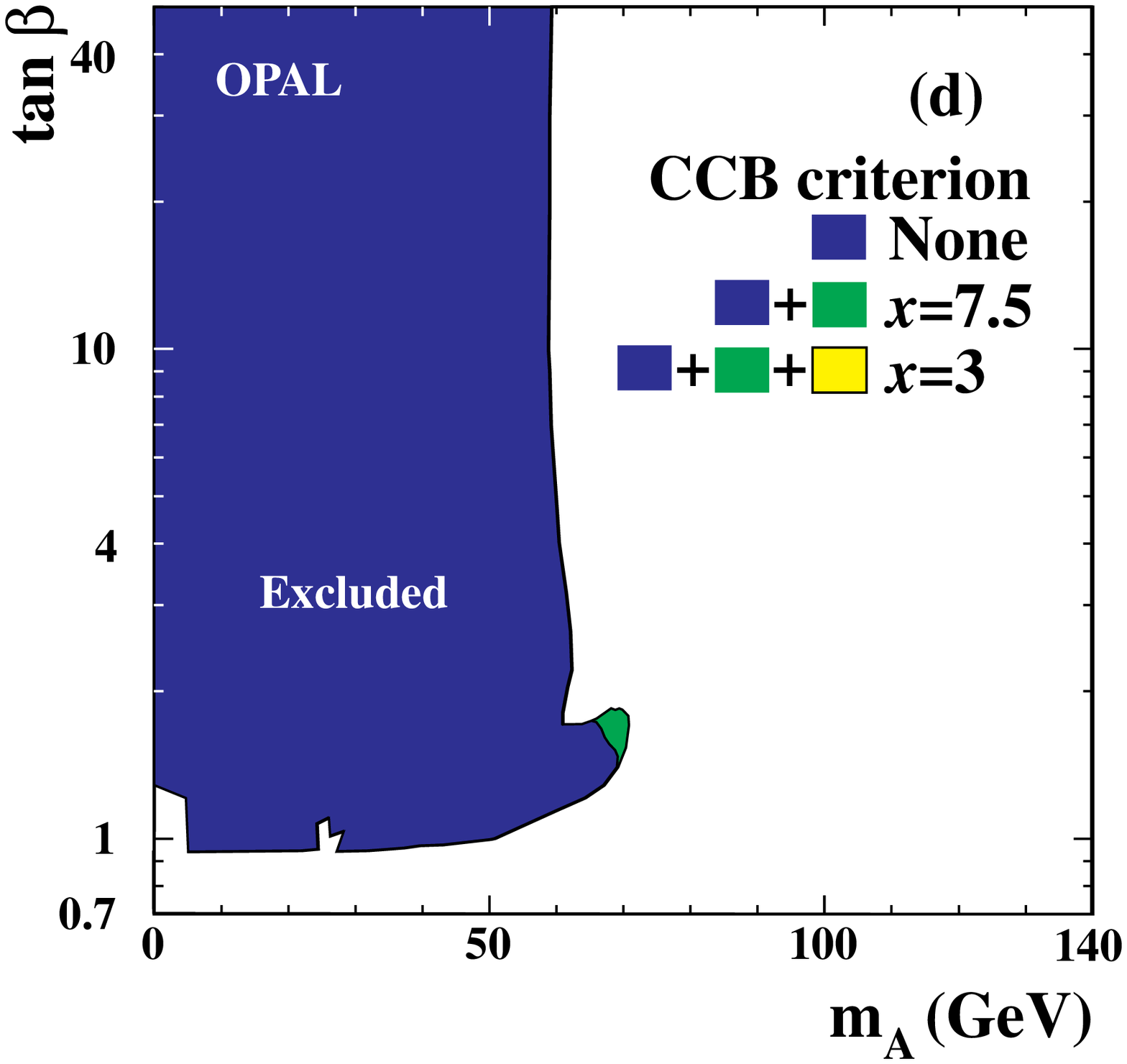,width=8.5cm}
}
\caption[]{\label{figure:mssm-gen}\sl
         The MSSM exclusion for scan (C) described in the text of
         Section~\ref{section:mssminterpretation}.
         Excluded regions are shown for
         (a) the (\mh,~\mA) plane for $\tanb>1$,
         (b) the (\mh,~\mA) plane for $\tanb>0.7$,
         (c) the (\mh,~\tanb) plane, and
         (d) the (\mA,~\tanb) plane.
         All exclusion limits are at 95\% CL.
         The black areas are excluded without applying any CCB criterion.
         When the CCB criterion of Eq.~\ref{eq:ccb} is applied with $x=7.5$
         the dark grey areas are excluded in addition. For this CCB criterion
         with $x=3$, the black, dark grey and light grey areas are all excluded.
         The speckled areas in (a), (b) and (c) are theoretically
         inaccessible.
}
\end{figure}
For the results of scan (C), shown in Fig.~\ref{figure:mssm-gen},
all exclusions are at 95\% CL.
For this scan the exclusion regions are shown for three cases:
Without the application of any CCB criterion in black,
for the CCB criterion of Eq.~\ref{eq:ccb} with $x=7.5$ in black and dark grey,
and with $x=3$ in black, dark grey and light grey.
The situation changes dramatically with respect to scans (A) and (B)
due to the appearance of unexcluded parameter sets with small values of \mh.
These points are characterised by a large mass difference $\mA-\mh$
and small \sba, hence the \h\Z\ production is 
suppressed (cf. Eq.~(\ref{equation:xsec_zh})).
The \h\A\ production is kinematically allowed at LEP2 energies in most of the
affected region, but the cross-section is small, and the current
integrated luminosity is not sufficient to exclude these parameter sets.
The existence of such parameter sets has already been pointed out
in~\cite{hempflingrosiek}.

If no CCB criterion is applied (the black area only is excluded), no absolute
lower limit for \mh\ can be given (Fig.~\ref{figure:mssm-gen}(a) and (b)).
For the CCB criterion applied with $x=7.5$,
the unexcluded area near $\mh=5$~GeV is due to
the relatively weak limit on \sba\ for this mass range,
as shown in Fig.~\ref{modindepZh}.
For $x=3$, the lower limit on \mh\ is 50.0~GeV at 95\% CL for $\mA>5$~GeV
in Fig.~\ref{figure:mssm-gen}(a), and it can be seen in
Fig.~\ref{figure:mssm-gen}(c) that this limit corresponds to $\tanb=50$.
Such high values of \tanb\ always result in \mh$\approx$\mA\ when the soft
SUSY-breaking masses are high ($>1$~TeV.)
However, as can be seen from the figures, in a general scan large values of
\tanb\ are not excluded for large mass differences $|\mA-\mh|$,
due to large higher-order corrections involving a low-mass stop.
For $\mA<5$~GeV the model-independent limit applies (cf.\ Fig.~\ref{modind_gz}).
This leads, for the $x=3$ CCB criterion, to $\mh>43.0$~GeV at 95\% CL.
Figure~\ref{figure:mssm-gen}(c) shows that there is no exclusion at all
at 95\% CL in the (\mh,~\tanb) plane
for the general scan if no CCB criterion is applied.
This is due to the fact that for any (\mh,~\tanb) combination the other
parameters can be chosen to simultaneously generate a large \mA\ and
a small \sba.
However, many of these specific parameter sets are discarded by applying
a CCB criterion.
In contrast, Fig.~\ref{figure:mssm-gen}(d), which shows the $(\mA,\tanb)$
projection, is only marginally different from
Fig.~\ref{figure:mssm-yb}(d) and~\ref{figure:mssm-mm}(d) of scans (A) and (B).
The exclusion limits on \mA\ are only slightly affected.

The results in this section suggest that the MSSM parameter
bounds, and in particular the limit on \mh,
derived from the benchmark scan (A) and minimal/maximal mixing scan (B)
are not valid when a more general scan (C) of the parameter space is performed.

\section{Summary}
Searches for neutral Higgs bosons presented here have not revealed any
significant excess beyond the background expectation from SM processes.
Evidence for both the \ee\ra\h\Z\ and \ee\ra\h\A\ production processes has been
searched for, allowing also for the decay \h\ra\A\A, when kinematically
possible.
Limits on these processes have been placed in a model-independent manner and
within the framework of the 2HDM and the MSSM.
These new limits substantially improve those previously published by OPAL.

In the model-independent approach, limits have been placed on the product of
cross-section and branching ratio for
\h\Z\ production assuming SM branching ratios and fully hadronic final
states, and for \h\A\ production with \bb\bb\ and \tautau\qq\ final states.

In the 2HDM interpretation, limits have been placed in the $(\mh,\mA)$ plane
both for the case of any value of \tanb\ and for that of $\tanb>1$.
Along the \h-\A\ mass diagonal 95\% CL lower limits are set at
$\mh\!=\!\mA>41.0$~GeV, independently of the value of \tanb\ and
at $\mh\!=\!\mA>56.0$~GeV for $\tanb>1$.

In the MSSM, three different scans over the model parameters
have been performed.
For the simplest, benchmark scan, in which all parameters except \mA\ and
\tanb\ are fixed,
a lower limit at 95\% CL on $\mh>59.0$~GeV and $\mA>59.5$~GeV is derived
for $\tanb>1$.
For $\tanb>0.7$, the limit on \mh\ remains at 59.0~GeV,
but no limit for \mA\ is obtained.

For the MSSM parameter scan with minimal and maximal \sctop\ mixing, where
only the $\mu$ and $A$ parameters are constrained and all other parameters
are left free, the 95\% CL excluded area has slightly less extent
than for the benchmark scan.

In the general scan of the MSSM parameter space,
parameter sets arise that cannot be excluded at 95\% CL.
All of these correspond to small, but non-negligible,
cross-sections for the process \ee\ra\h\A,
and will either be observed or excluded
as the integrated luminosity of the data increases.
A fraction of those parameter sets can be excluded
if requirements are made to avoid charge- and colour-breaking vacua.

\section*{Acknowledgements}
We particularly wish to thank the SL Division for the efficient operation
of the LEP accelerator at all energies
 and for
their continuing close cooperation with
our experimental group.  We thank our colleagues from CEA, DAPNIA/SPP,
CE-Saclay for their efforts over the years on the time-of-flight and trigger
systems which we continue to use.  In addition to the support staff at our own
institutions we are pleased to acknowledge the  \\
Department of Energy, USA, \\
National Science Foundation, USA, \\
Particle Physics and Astronomy Research Council, UK, \\
Natural Sciences and Engineering Research Council, Canada, \\
Israel Science Foundation, administered by the Israel
Academy of Science and Humanities, \\
Minerva Gesellschaft, \\
Benoziyo Center for High Energy Physics,\\
Japanese Ministry of Education, Science and Culture (the
Monbusho) and a grant under the Monbusho International
Science Research Program,\\
German Israeli Bi-national Science Foundation (GIF), \\
Bundesministerium f\"ur Bildung, Wissenschaft,
Forschung und Technologie, Germany, \\
National Research Council of Canada, \\
Research Corporation, USA,\\
Hungarian Foundation for Scientific Research, OTKA T-016660, 
T023793 and OTKA F-023259.\\

\end{document}